\documentclass[11pt]{article}
\usepackage{latexsym,amsmath,amssymb,graphics}
\usepackage[cp1251]{inputenc}  
\numberwithin{equation}{section}

\newtheorem{theorem}{Theorem}
\newcommand{\bt}{\begin{theorem}}
\newcommand{\et}{\end{theorem}}

\newtheorem{dr}{Definition}
\newcommand{\bdr}{\begin{dr}}
\newcommand{\edr}{\end{dr}}
\makeatletter\@addtoreset{dr}{section}

\newtheorem{lema}{Lemma}

\def\bl{\begin{lema}}
\def\el{\end{lema}}

\newtheorem{tver}{Assertion}

\def\bt{\begin{tver}}
\def\et{\end{tver}}

\def\pl
{\lower 3pt\hbox{$\displaystyle+ \atop\raise 5pt\hbox{$\displaystyle.$}$}}

\title{Group classification of nonlinear wave equations}

\author{ V. Lahno\\ \small State Pedagogical
University, 2 Ostrogradskogo Street, 36000 Poltava, Ukraine \thanks{e-mail:
laggo@poltava.bank.gov.ua.} \\
R.~Zhdanov
\\
\small Institute of Mathematics of NAS of Ukraine,  3 Tereshchenkivska Street,
01601 Kyiv, Ukraine\thanks{Corresponding author, e-mail: renat@imath.kiev.ua} \\
O.~Magda\\
\small Institute of Mathematics of NAS of Ukraine,  3 Tereshchenkivska Street,
01601 Kyiv, Ukraine}
\date{}
\begin{document}
\maketitle

\begin{abstract}
We perform complete group classification of the general class of quasi linear
wave equations in two variables. This class may be seen as a broad generalization
of the nonlinear d'Alembert, Liouville, sin/sinh-Gordon and Tzitzeica equations. 
In this way we derived a number of new genuinely nonlinear invariant models 
with high symmetry properties. In particular, we obtain four classes of 
nonlinear wave equations admitting five-dimensional invariance groups. 
Applying the symmetry reduction technique we construct multi-parameter 
families of exact solutions of those wave equations.
\end{abstract}
 
\section*{Introduction}

More than century ago Sophus Lie introduced the concept of
continuous transformation group into mathematical physics and
mechanics. His initial motivation was to develop a theory of
integration of ordinary differential equations enabling to answer
the basic questions, like, why some equations are integrable and
others are not. His fundamental results obtained on this way,  can
be seen as a far reaching generalization of the Galua's and Abel's
theory of solubility of algebraic equations by radicals. Since that
time the Lie’s theory of continuous transformation groups has become
applicable to an astonishingly wide range of mathematical and
physical problems.

It was Lie who was the first to utilize group properties of
differential equations for constructing of their exact solutions. In
particular, he computed the maximal invariance group of the
one-dimensional heat conductivity equation and applied this symmetry
to construct its explicit solutions. Saying it the modern way, he
performed symmetry reduction of the heat equation. Since late 1970s
symmetry reduction becomes one of the most popular tools for solving
nonlinear partial differential equations (PDEs).

By now symmetry properties of the majority of fundamental equations
of mathematical and theoretical physics are well known. It turns out
that for the most part these equations admit wide symmetry groups.
Especially this is the case for linear PDEs and it is this rich
symmetry that enables developing a variety of efficient methods for
mathematical analysis of linear differential equations. However,
linear equations give mathematical description of physical, chemical
or biological processes in a first approximation only. To provide a
more detailed and precise description a mathematical model has to
incorporate nonlinear terms. Note that some important equations (for
example, the Yang-Mills equations) of theoretical physics are
essentially nonlinear in a sense that they have no linearized
version.

Hyperbolic type second-order nonlinear PDEs in two independent
variables play a fundamental role in modern mathematical physics.
Equations of this type are utilized to describe different types of
wave propagation. They are used in differential geometry, in various
fields of hydro- and gas dynamics, chemical technology, super
conductivity, crystal dislocation to mention only a few applications
areas. Surprisingly the list of equations utilized is rather narrow.
In fact, it is comprised by the Liouville, sine/sinh-Gordon,
Goursat, d’Alembert and Tzitzeica equations and a couple of others.
Popularity of these very models has a natural group-theoretical
interpretation, namely, all of them have non-trivial Lie or
Lie-B\"acklund symmetry. By his very reason some of them are
integrable by the inverse problem methods (see,
e.g.,\cite{magda69}--\cite{m86}) or are linearizable
\cite{m87}--\cite{m92}) and completely integrable \cite{m88,m91}.

In this connection it seems a very important problem to select from the
reasonably extensive class of nonlinear hyperbolic type PDEs those enjoying
the best symmetry properties. Saying 'reasonably extensive" we mean
that this class should contain the above enumerated equations
as particular cases on the one hand, and on the other it should
contain a wide variety of new invariant models of potential interest for
applications. The list of the so obtained invariant equations will
contain candidates for realistic nonlinear mathematical models of the
physical and chemical processes enumerated above.

The history of group classification methods goes back to Lie itself.
Probably, the very first paper on this subject is \cite{magda5}, where
Lie proves that a linear two-dimensional second-order PDE may admit at
most a three-parameter invariance group (apart from the trivial
infinite-parameter symmetry group, which is due to linearity).

The modern formulation of the problem of group classification of
PDEs was suggested by Ovsyan\-ni\-kov in \cite{z26}. He developed a
regular method (we will refer to it as to Lie-Ovsyannikov method)
for classifying differential equations with non-trivial symmetry and
performed complete group classification of the nonlinear heat
conductivity equation. In a number of subsequent publications more
general types of nonlinear heat equations were classified (review of
these results can be found in \cite{magda41}).

However, even a very quick analysis of the papers on group classification of
PDEs reveals that the overwhelming majority of them deal with equations whose
arbitrary elements (functions) depend on one variable only. There is a deep
reason for this fact, which is that Lie-Ovsyannikov method becomes inefficient
for PDEs containing arbitrary functions of several variables. To achieve a
complete classification one either needs to specify the transformation group
realization or somehow restrict arbitrariness of functions contained in the
equation under study.

Recently, we developed an efficient approach enabling to overcome this
difficulty for low dimensional PDEs \cite{z18,magda16}. Utilizing it
approach we have obtained a final solution of the problem of group
classification of the general quasi-linear heat conductivity equations
in two independent variables.

In this paper we apply the approach mentioned above to perform group
classification of the most general quasi-linear hyperbolic type PDE in
two independent variables.

\section{ Group classification algorithm }
\setcounter{section}{1}
\setcounter{equation}{0}

We begin this section by formulating the problem to be solved.
Then we present a brief review of the already known results.
Finally we give a short description of our approach to group
classification of PDEs (for the detailed account of the necessary
facts, see (\cite{magda16})).

While classifying a given class of differential equations into
subclasses, one can use different classifying features, like linearity,
order, the number of independent or dependent variables, etc. In group
analysis of differential equations the principal classifying features are
symmetry properties of equations under study. This means that classification
objects are equations together with their symmetry groups. This point of view
is based on the well-known fact that any PDE admits a (possibly trivial) Lie
transformation group. And what is more, any transformation group corresponds
to a class of PDEs, which are invariant under this group. So to perform group
classification of a class of PDEs means describing all possible (inequivalent)
pairs (PDE, maximal invariance group), where PDE should belong to the class
of equations under consideration.

We perform group classification of the following class of
quasi-linear wave equations
\begin{equation} \label {1.1}
u_{tt} = u_{xx} + F(t,x,u,u_x).
\end{equation}
Here $F$ is an arbitrary smooth function, $u = u(t,x).$ Hereafter we
adopt notations $u_t = \frac{\partial u}{\partial t}, \
u_x = \frac{\partial_u}{\partial_x}, u_{tt} = \frac{\partial^2 u}
{\partial t^2}, \ldots.$

Our aim is describing {\it all} equations of the form (\ref{1.1}) that admit
nontrivial symmetry groups. The challenge of this task is in the word {\it all}.
If, for example, we fix the desired invariance group, then the classification
problem becomes a University textbook exercise on Lie group analysis. A slightly
 more cumbersome (but still tractable with the standard Lie-Ovsyannikov approach)
is the problem of group classification of equation with arbitrary functions of,
at most, one variable.

As equations invariant under similar Lie groups are identical within
the group-theoretic framework, it makes sense to consider non
similar transformation groups \cite{magda1,magda2} only. The
important example of similar Lie groups is provided by Lie
transformation groups obtained one from another by a suitable change
of variables. Consequently, equations obtained one from another by a
change of variables have similar symmetry groups and cannot be
distinguished within the group-theoretical viewpoint. That is why,
we perform group classification of (\ref{1.1}) within a change of
variables preserving the class of PDEs (\ref{1.1}).

The problem of group classification of linear hyperbolic type equation
\begin{equation} \label{1.2}
u_{tx} +A(t,x) u_t +B(t,x) u_x +C(t,x) u=0
\end{equation}
with $u = u(t,x)$, was solved by Lie \cite{magda5} (see, also, \cite{magda6}).
In view of this fact, we consider only those equations of the form (\ref{1.1})
which are not (locally) equivalent to the linear equation (\ref{1.2}).

As we have already mentioned in Introduction, the Lie-Ovsyannikov method of
group classification of differential equations has been suggested in \cite{z26}.
Utilizing this method enabled solving group classification problem for a
number of important one-dimensional nonlinear wave equations:
\begin{eqnarray*}
u_{tt} = u_{xx} +F(u); && \cite{magda32}-\cite{magda9} \\
u_{tt} = [f(u)u_{x}]_x; && \cite{magda10}-\cite{magda11} \\
u_{tt} = f(u_x) u_{xx}; && \cite{magda11,magda12} \\
u_{tt} = F(u_x) u_{xx} + H(u_x); && \cite{magda31}\\
u_{tt} = F(u_{xx}); && \cite{magda11}\\
u_{tt} = u^m_x u_{xx}+f(u); && \cite{m79}\\
u_{tt}+f(u) u_t = (g(u) u_x)_x + h(u) u_x. && \cite{m80}
\end{eqnarray*}

Analysis of the above list shows immediately that arbitrary elements
($=$ arbitrary functions) depend on one variable, at most. This is not
coincidental, indeed, the Lie-Ovsyannikov approach works smoothly for the
case when the arbitrary elements are functions of one variable. The reason
for this is that the obtained system of determining equations is still
over-determined and can be effectively solved within the same ideology used
while computing maximal symmetry group of PDEs containing no arbitrary elements.

The situation becomes much more complicated for the case when arbitrary
elements are functions of two (or more) arguments. By this very reason
group classification of nonlinear wave equations
\begin{eqnarray*}
u_{tt} = (f(x,u) u_x)_x ; && \cite{magda13}\\
u_{tt} +\lambda u_{xx} = g(u, u_x) ; && \cite{magda36,magda9a}\\
u_{tt} = [f(u) u_x +g(x,u)]_x; && \cite{magda44}-\cite{magda14}\\
u_{tt}=f(x,u_x) u_{xx} +g(x, u_x) && \cite{magda15}
\end{eqnarray*}
is not complete.

We suggest new efficient approach to the problem of group classification
of low dimensional PDEs in \cite{z18,magda16}. This approach is based on
Lie-Ovsyannikov infinitesimal method and classification results for abstract
finite-dimensional Lie algebras. It enables us to obtain the complete solution
of group classification problem for the general heat equation with a nonlinear
source
$$ u_t = u_{xx} +F(t,x,u,u_x).$$
Later on, we perform complete group classification of the most general
quasi-linear evolution equation \cite{m81}--\cite{m83}
$$ u_t = f(t,x,u,u_x) u_{xx} + g(t,x,u,u_x).$$
We use the above approach to achieve a complete solution of the group
classification problem for the class of equations (\ref{1.1}).

Our approach to group classification of the class of PDEs (\ref{1.1})
consists of the following steps (for more details, see \cite{m83}):
\begin{enumerate}
\renewcommand{\theenumi}{\Roman{enumi}}
\item Using the infinitesimal Lie method we derive the system of determining
equations for coefficients of the first-order operator that generates
symmetry group of equation (\ref{1.1}) (Note that the determining equations
which explicitly depend on the function $F$ and its derivatives
are called classifying equations). Integrating equations that do not
depend on $F$ we obtain the form of the most general infinitesimal
operator admitted by equation (\ref{1.1}) under arbitrary $F$.
Another task of this step is calculating the equivalence group ${\cal E}$
of the class of PDEs (\ref{1.1}).

\item We construct all realizations of Lie algebras $A_n$ of the
dimension $n \leq 3$ in the class of operators obtained at the
first step within the equivalence relation defined by transformations
from the equivalence group ${\cal E}$. Inserting the so
obtained operators into classifying equations we select those
realizations that can be symmetry algebras of a differential equation
of the form (\ref{1.1}).

\item We perform extension of the realizations constructed at the previous
step to realizations of higher dimensional ($ n > 3$) Lie algebras.
Since extending symmetry algebras results in reducing arbitrariness
of the function $F$, at some point this function will
contain either arbitrary functions of at most one variable or arbitrary
constants. At this point, we apply the standard classification method,
which is due to Lie and Ovsyannikov, to derive the maximal
symmetry group of the equation under study thus completing its
group classification.
\end{enumerate}

As a result of performing the above enumerated steps we get the
complete list of inequivalent equations of the form (\ref{1.1})
together with their maximal (in Lie's sense) symmetry algebras.

We say that the group classification problem is completely solved
when it is proved that
\begin{enumerate}
\renewcommand{\labelenumi}{\arabic{enumi} )}
\item The constructed symmetry algebras are maximal invariance algebras
of the equations under consideration;
\item The list of invariant equations contains only inequivalent
ones, namely, no equation can be transformed into another one from the list
by a transformation from the equivalence group ${\cal E}$.
\end{enumerate}

\section{Preliminary group classification of equation (1.1)}
\setcounter{section}{2}
\setcounter{equation}{0}

According to the above algorithm we are looking for infinitesimal operator of
symmetry group of equation (\ref{1.1}) in the form
\begin{equation} \label{1.3}
Q=\tau(t,x,u)\partial_t+\xi(t,x,u)\partial_x+\eta(t,x,u)\partial_u,
\end{equation}
where $\tau,\xi,\eta$ are smooth functions defined on an open domain
$\Omega$ of the space $V= {\mathbb R}^2\times {\mathbb R}^1$ of
independent ${\mathbb R}^2=\langle t,x\rangle $ and dependent
${\mathbb R}^1=\langle u \rangle  = u(t,x)$ variables.

Operator (\ref{1.3}) generates one-parameter invariance group of
(\ref{1.1}) iff its coefficients $\tau, \xi, \eta, \epsilon$
satisfy the equation (Lie's invariance criterion)
\begin{equation} \label{1.4}
\varphi^{tt} -\varphi^{xx} -\tau F_t -\xi F_x -\eta F_u -\varphi^x F_{u_x}
\Biggl \vert_{(\ref{1.1})} = 0,
\end{equation}
where
\begin{eqnarray*}
\varphi^t &=& D_t (\eta) -u_t D_t (\tau) -u_{x} D_t (\xi), \\
\varphi^x &=& D_x (\eta) -u_t D_x (\tau) -u_{x} D_x (\xi), \\
\varphi^{tt} &=& D_t (\varphi^t) -u_{tt} D_t (\tau) -u_{tx} D_t (\xi), \\
\varphi^{xx} &=& D_x (\varphi^x) -u_{tx} D_x (\tau) -u_{xx} D_x (\xi)
\end{eqnarray*}
and $D_t, D_x$ are operators of total differentiation with respect
to the variables $t,x$. As customary, by writing $\Biggl
\vert_{(\ref{1.1})}$ we mean that one needs to replace $u_{tt}$ and
its differential consequences with the expression $u_{xx} + F$ and
its differential consequences in (\ref{1.4}).

After a simple algebra we represent (\ref{1.4}) in the form of system
of four PDEs:
\begin{eqnarray} \label{1.5}
(1) && \xi_u = \tau_u= \eta_{uu}=0, \nonumber\\
(2) && \tau_t-\xi_x=0, \ \ \xi_t -\tau_x =0, \nonumber \\
(3) && 2 \eta_{tu}+\tau_x F_{u_x} =0, \\
(4) && \eta_{tt} -\eta_{xx} -2 u_x \eta_{xu} +[\eta_u -2 \tau_t] F -\tau F_t
-\xi F_x\nonumber \\
&& -\eta F_u -[\eta_x +u_x (\eta_u -\xi_x)] F_{u_x} =0. \nonumber
\end{eqnarray}

It follows immediately from (1) that $\tau = \tau(t,x), \ \xi = \xi(t,x),
\eta = h(t,x) u+r(t,x).$ In the sequel we differentiate between
the cases $F_{u_x u_x} \not =0$ and $F_{u_x u_x} =0.$

\underline{Case $F_{u_x u_x} \not =0$.} It follows from (3) that $\tau_x = h_t
=0$. Taking into account this fact and equation (2) we obtain $\tau = \lambda t
+\lambda_1, \xi = \lambda x+\lambda_2, \ h =h(x)$, where $ \lambda, \lambda_1,
\lambda_2$ are arbitrary real constants.

\underline{Case $F_{u_x u_x}  =0$.} If this is the case, then $F = g(t,x,u) u_x
+f(t,x, u)$, where $f$ and $g$ are arbitrary smooth functions.

Given the condition $g_u \not =0$, it follows from (3) that $\tau_x = h_t =0$.
So that taking into account equation (2) we arrive at the expressions $\tau,
\xi, h$ given in the previous case.

If $g_u =0$, then $f_{uu} \not =0$ (otherwise equation (\ref{1.1}) becomes
linear).

Let $g\equiv 0$. It follows from (3) that $\eta = h(x) u + r(t,x)$. Equation (4)
now reads as
\begin{eqnarray*}
&& r_{tt} -r_{xx} -2 h' u_x +[h -2\tau_t] f - \tau f_t -\xi f_x -[hu +r] f_u =0.
\end{eqnarray*}
As functions $\tau, \xi, h, r, f$ do not depend on $u_x$, $h' =0$. Hence $\eta =
mu +r(t,x) $, where $m$ is an arbitrary real constant.

Furthermore, if $g = g(t,x) \not =0,$ then it is straightforward to verify that
system of equations ((3), (4)) is equivalent to the following equations:
\begin{eqnarray*}
&& 2 h_t = -\tau_x g, \ \
2 h_x = -\tau_t g -\tau g_t -\xi g_x, \\
&& (h_{tt} -h_{xx}) u +r_{tt} -r_{xx} +f[h-2 \tau_t] \ \
- \tau f_t -\xi f_x -[hu+r]f_u -(h_x u+r_x) g = 0.
\end{eqnarray*}
Integrating (2) yields $\tau = \varphi (\theta ) +\psi(\sigma),$ $\xi =
-\varphi(\theta) +\psi(\sigma),$ where $\varphi, \psi$ are arbitrary smooth
functions of $\theta = t-x, \sigma = t+x$ and we arrive at the following
theorem.

\begin{theorem} \label{theo1.2.1}
Provided $F_{u_x u_x} \not =0,$ the maximal invariance group of equation
(\ref{1.1}) is generated by the following infinitesimal operator:
\begin{equation} \label{1.6}
Q = (\lambda t+\lambda_1) \partial_t+(\lambda x +\lambda_2) \partial_x+[h(x) u
+r(t,x)] \partial_u,
\end{equation}
where $\lambda, \lambda_1, \lambda_2$ are real constants and $h = h(x), \ r =
r(t,x),$  $F = F(t,x,u,u_x)$ are functions obeying the constraint
\begin{eqnarray} \label{1.7}
&& r_{tt}-r_{xx}-\frac{d^2h}{dx^2}u-2\frac{dh}{dx}u_x+
(h-2\lambda)\,F\nonumber\\
&& \quad -(\lambda t+\lambda_1)\,F_t -(\lambda x+\lambda_2)\,F_x -(h
u+r)\,F_u\\
&& \quad -(r_x+\frac{dh}{dx}u+(h-\lambda)u_x)\,F_{u_x} =0.\nonumber
\end{eqnarray}

If $F = g(t,x,u) u_x + f(t,x,u), \ g_u \not =0,$ then the maximal invariance
group of equation (\ref{1.1}) is generated by infinitesimal operator
(\ref{1.6}), where $\lambda, \lambda_1, \lambda_2$ are real constants $h, r, g,
f$ are functions satisfying system of two equations
\begin{eqnarray}\label{1.8}
&& -2h'-\lambda g = (\lambda t+\lambda_1) g_t +(\lambda x+\lambda_2) g_x +(hu
+r) g_u,\nonumber\\
&& -h''u +r_{tt}-r_{xx}+(h-2 \lambda)f =(\lambda t +\lambda_1) f_t +(\lambda x
+\lambda_2) f_x\\
&& \ \ \ \ \ \ +(hu +r) f_u + g(h'u+r_x).\nonumber
\end{eqnarray}
Next, if $F = g(t,x) u_x + f(t,x,u), \ q \not \equiv 0, \ f_{uu} \not =0,$
then the infinitesimal operator of the invariance group of equation
(\ref{1.1}) reads as
$$
Q = \tau(t,x) \partial_t+\xi(t,x) \partial_x+(h(t,x) u+r(t,x)) \partial_u,
$$
where $\tau, \xi, h, r, g, f$ are functions satisfying system of PDEs
\begin{eqnarray*}
&& \tau_t -\xi_x=0, \ \ \xi_t -\tau_x =0, \\
&& 2 h_t = -\tau_x g, \ \ 2 h_x = -\tau_t g-\tau g_t -\xi g_x, \\
&& (h_{tt} -h_{xx}) u +r_{tt} -r_{xx} +f(h-2 \tau_t) -\tau f_t \\
&& -\xi f_x -(hu+r) f_u -(h_x u+r_x) g =0.
\end{eqnarray*}
Finally, if $F = f(t,x,u), \ f_{uu} \not =0,$ then the maximal invariance group
of equation (\ref{1.1}) is generated by infinitesimal operator
$$
Q = [\varphi(\theta) + \psi(\sigma)]
\partial_t-[\varphi(\theta)-\psi(\sigma)] \partial_x+[ku+r(t,x)] \partial_u,
$$
where $ k \in {\mathbb R}, \ \theta = t-x, \ \sigma = t+x$ and
functions $\varphi, \psi, r, f$ and constant $k$ satisfy the following
equation:
\begin{eqnarray*}
&& r_{tt} -r_{xx} +[k-2 \varphi'-2\psi']f-(\varphi+\psi) f_t +\\
&& + (\varphi-\psi) f_x -(ku +r) f_u =0, \ \ \varphi' = \frac{d \varphi}{d
\theta}, \ \ \psi' =\frac{d \psi}{d \theta}.
\end{eqnarray*}
\end{theorem}

Summing up the above considerations we conclude that the problem of group
classification of equation (\ref{1.1}) reduces to the one of classifying
equations of more specific forms
\begin{eqnarray}
u_{tt} &=& u_{xx} + F(t,x,u,u_x), \ \ F_{u_x u_x} \not =0; \label{1.9} \\
u_{tt} &=& u_{xx} + g(t,x,u) u_x +f(t,x,u), \ \ g_u \not =0; \label{1.10} \\
u_{tt} &=& u_{xx} + g(t,x) u_x +f(t,x,u), \ \ g \not =0, \ f_{uu} \not =0;
\label{1.11} \\
u_{tt} &=& u_{xx} + f(t,x,u), \ \ f_{uu} \not =0. \label{1.12}
\end{eqnarray}
Consider the last two equations. By the change of variables
$$ \bar t = t-x, \ \ \bar x = t+x, \ \ u = v( \bar  t, \bar x)$$
we reduce them to equations
\begin{eqnarray}\label{1.13}
v_{\bar t \bar x} &=& \frac{1}{4} f(\bar t, \bar x, v), \nonumber \\
v_{\bar t \bar x} &=& -\frac{1}{4} g(\bar t, \bar x) (v_{\bar t} -v_{\bar x})
+\frac{1}{4} f(\bar t, \bar x, v).
\end{eqnarray}
Now, making the change of variables
$$ \tilde t = \bar t, \ \ \tilde x = \bar x, \ \ \tilde v(\tilde t, \tilde x) =
\Lambda (\bar t, \bar x) v, $$
where $\Lambda = \exp \left [-\frac{1}{4} \int g(\bar t, \bar x) d \bar x
\right]$, we transform (\ref{1.13}) to become
$$\tilde v_{\tilde t \tilde x} = \left(\frac{1}{4} g -\Lambda^{-1} \Lambda_{\bar
t} \right) \tilde v_{\tilde x}-\frac{1}{4} g \Lambda^{-1} \Lambda_{\tilde t}
\tilde v +\frac{1}{4} \Lambda^{-1} \Lambda_{\bar x} g \tilde v +\Lambda^{-1}
f.$$
Hence we conclude that the following assertion holds true.

\begin{tver} \label{tv1.2.1}
The problem of group classification of equations (\ref{1.11}), (\ref{1.12}) is
equivalent to the one of classifying equations
\begin{eqnarray}
u_{tx} &=& g(t,x) u_x +f(t,x,u), \ g_x \not =0, \ f_{uu} \not =0; \label{1.14}
\\
u_{tx} &=& f(t,x,u), \ \ f_{uu} \not =0. \label{1.15}
\end{eqnarray}
\end{tver}
Note that condition $g_x \not =0$ is essential, since otherwise (\ref{1.14})is
locally equivalent (\ref{1.15}).

Summing up, we conclude that the problem of group classification of (\ref{1.1})
reduces to classifying more particular classes of PDEs (\ref{1.9}), (\ref{1.10},
(\ref{1.14}), (\ref{1.15}). In what follows, we provide full calculation details
for equations (\ref{1.11}) and (\ref{1.12}) only. The reason is just to save
space and still be able to present all details of the algorithm.

First, we consider equations (\ref{1.10}), (\ref{1.14}), (\ref{1.15}).

\section{ Group classification of equation (2.8) }
\setcounter{section}{3}
\setcounter{equation}{0}

According to Theorem \ref{theo1.2.1} invariance group of equation (\ref{1.10})
is generated by infinitesimal operator (\ref{1.6}). And what is more, the real
constants $\lambda, \lambda_1, \lambda_2$ and functions $h, r, g, f$ satisfy
equations (\ref{1.8}). System (\ref{1.8}) is to be used to specify both
the form of functions $f$, $g$ from (\ref{1.10}) and functions $h, r$ and
constants $\lambda, \lambda_1, \lambda_2$ in (\ref{1.6}). It is called
the determining (sometimes classifying) equations.

Efficiency of the Lie method for calculation of maximal invariance group of PDE
is essentially based on the fact that routinely this system is over-determined.
This is clearly not the case, since we have only one equation for four (!)
arbitrary functions and three of the latter depend on two variables. By this
very reason the direct application of Lie approach in the Ovsyannikov's spirit
is no longer efficient when we attempt classifying PDEs with arbitrary functions
of several variables.

Next, we compute the equivalence group $\cal E$ of equation (\ref{1.10}). This group
is generated by invertible transformations of the space ${\rm V}$ preserving
the differential structure of equation (\ref{1.10}) (see, e.g., \cite{magda1}).
Saying it another way, group transformation from $\cal E$
$$ \bar t = \alpha(t,x,u), \hskip 5mm  \bar x = \beta(t,x,u), \hskip 5mm v =
U(t,x,u), \ \ \ \frac{D(\bar t,\bar x, v)}{D(t,x,u)} \not =0,$$
should reduce (\ref{1.10}) to equation of the same form
$$
v_{\bar t \bar t} = v_{\bar x\bar x} +\tilde g(\bar t, \bar x, v) v_{\bar x}
+\tilde f(\bar t, \bar x, v), \ \tilde g_{v} \not =0
$$
with possibly different $\tilde f, \tilde g$.

As proved by Ovsyannikov \cite{magda1}, it is possible to modify the Lie's
infinitesimal approach to calculate equivalence group in essentially
same way as invariance group. We omit the simple intermediate
calculations and present the final result.

\begin{tver}\label{tv1.3.1}
The maximal equivalence group $\cal E$ of equation (\ref{1.10})
reads as
\begin{equation}\label{1.16}
\bar t = kt + k_1, \ \ \bar x= \epsilon kx + k_2, \ \ v = X(x) u + Y(t,x),
\end{equation}
where $k \not =0, \ X \not =0, \ \epsilon = \pm 1, k, k_1, k_2 \in {\mathbb R}$,
and $X, Y$ are arbitrary smooth functions.
\end{tver}

This completes the first step of the algorithm.

\subsection{ Preliminary group classification of equation (2.8).}

First, select equations of the form (\ref{1.10}) admitting one-parameter
invariance groups.

\begin{lema} \label{l1.3.1}
There exist transformations (\ref{1.16}) that reduce operator (\ref{1.6}) to
one of the six forms:
\end{lema}
\begin{eqnarray}\label{1.17}
&& Q = m(t \partial_t +x \partial_x), \ \ m \not =0; \ \
 Q = \partial_t +\beta \partial_x, \ \ \beta \geq 0; \ \nonumber \\
&& Q = \partial_t +\sigma(x) u \partial_u, \ \sigma \not =0; \ \
 Q = \partial_x; \\
&& Q = \sigma(x) u \partial_u, \ \ \sigma \not =0; \ \
 Q = \theta (t,x) \partial_u, \ \ \theta \not =0. \nonumber
\end{eqnarray}

{\it Proof.} Change of variables (\ref{1.16}) reduces operator (\ref{1.6}) to
the form
\begin{equation}\label{1.18}
\tilde Q = k(\lambda t+\lambda_1) \partial_{\bar t}+\epsilon k (\lambda x
+\lambda_2) \partial_{\bar x} + [Y_t(\lambda t +\lambda_1) +(\lambda
x+\lambda_2)(X'u+Y_x)+X(hu+r)]\partial_v.
\end{equation}
If $\lambda \not =0$ in (\ref{1.6}), then putting $k_1= \lambda^{-1} \lambda_1
k, \ k_2 =\epsilon \lambda^{-1} \lambda_2 k,$ and taking as $X$, $Y \ (X \not
=0)$ integrals of system of PDEs
\begin{eqnarray*}
&& X'  (\lambda x+\lambda_2) +X h =0, \\
&& Y_t (\lambda t +\lambda_1) +Y_x(\lambda x+\lambda_2) +X r=0
\end{eqnarray*}
in (\ref{1.16}), we reduce (\ref{1.18}) to the form
$$
\tilde Q = \lambda(\bar t \partial_{\bar t} +\bar x \partial_{\bar x}).
$$

Provided $\lambda=0$ and $\lambda_1 \not =0,$ we analogously obtain
$$
\tilde Q = \partial_{\bar t} +\beta \partial_{\bar x}, \ \ \beta \geq 0; \ \
Q = \partial_{\bar t} +\sigma(\bar x) v \partial_v, \ \ \sigma \not =0.
$$

Next, if $\lambda= \lambda_1 =0, \ \lambda_2 \not =0,$ in (\ref{1.6}), then
putting $k = \epsilon \lambda^{-1}_2$, and taking as $X$, $Y$ \ $(X \not =0)$
integrals of equations
$$ \lambda_2 X' +h X =0, \hskip 10mm Y_x + r X=0,$$
we reduce operator (\ref{1.18}) to become $ \tilde Q = \partial_{\bar x}.$

Finally, the case $\lambda = \lambda_1 = \lambda_2 =0,$ gives rise to operators
$\tilde Q = \sigma (\bar x) v \partial_v, \hskip 10mm \tilde Q = \theta(\bar t,
\bar x) \partial_v.$

Rewriting the above operators in the initial variables $t, x$ completes the
proof.

\begin{theorem} \label{theo1.3.1}
  There are exactly five inequivalent equations of the form (\ref{1.10})
  that admit one-parameter transformation groups. They are listed below
  together with one-dimensional Lie algebras generating their invariance
  groups (note that we do not present the full form of invariant PDEs
  just the functions $f$ and $g$)
$$\begin{array}{l}
A^1_1 = \langle t \partial_t+x\partial_x \rangle : g=x^{-1} \tilde g(\psi, u),
\\
\hskip 10mm f = x^{-2} \tilde f(\psi, u), \psi = t x^{-1}, \tilde g_{u} \not
=0;\\[2mm]
A^2_1 = \langle \partial_t+\beta \partial_x \rangle : g = \tilde g(\eta, u), \ f
= \tilde f(\eta, u),\\[2mm]
\hskip 10mm \eta = x-\beta t , \ \ \beta \geq 0, \ \tilde g_u \not =0; \\[2mm]
A^3_1 = \langle \partial_t+\sigma(x) u \partial_u \rangle: g = -2 \sigma'
\sigma^{-1} \ln |u|+ \tilde g(\rho, x), \\[2mm]
\hskip 10mm f = (\sigma' \sigma^{-1})^2 u \ln^2 |u|-\sigma' \sigma^{-1} \tilde
g(\rho, x) u \ln |u|-\sigma^{-1}\sigma'' u \ln |u| +u \tilde f(\rho, x), \\[2mm]
\hskip 10mm \rho = u \exp({-t \sigma}), \ \ \sigma \not =0;\\[2mm]
A^4_1 = \langle \partial_x \rangle : g = \tilde g(t,u), \ f = \tilde f(t,u), \
\tilde g_u\not =0;\\[2mm]
A^5_1 = \langle \sigma(x) u \partial_u \rangle : g = -2 \sigma' \sigma^{-1} \ln
|u|+ \tilde g(t,x), \ f = (\sigma' \sigma^{-1})^2 u \ln^2 |u|\\[2mm]
\hskip 10mm -(\sigma^{-1} \sigma'' +\sigma^{-1} \sigma' \tilde g(t,x)) u \ln
|u|+ u \tilde f(t,x), \ \ \sigma' \not =0.
\end{array}$$
\end{theorem}

{\it Proof.} If equation (\ref{1.10}) admits a one-parameter invariance group,
then it is generated by operator of the form (\ref{1.6}). According to
Lemma \ref{l1.3.1}, the latter is equivalent to one of the six operators
(\ref{1.17}). That is why, all we need to do is integrating six systems of
determining equations corresponding to operators (\ref{1.8}). For
the first five operators solutions of determining equations are easily
shown to have the form given in the statement of theorem.

We consider in more detail the operator $Q = \theta(t,u) \partial_u$.
Determining equations (\ref{1.8}) for this operator reduce to the
form
$$
\theta_{tt} -\theta_{xx} = \theta f_u +\theta_x g, \ \ \ \ \theta g_u =0,
$$
whence we get $g_u =0$. Consequently, the system of determining
equations is incompatible and the corresponding invariant equation
fails to exist.

Non equivalence of the invariant equations follows from non equivalence
of the corresponding symmetry operators.

The theorem is proved.

\begin{lema}\label{l1.3.2}
 There are no realizations of semi-simple Lie algebras generated by
 operators of the form (\ref{1.6}).
\end{lema}

{\it Proof.} To prove the lemma it suffices to check that there are no
realizations of the lowest order simple Lie algebras by operators
(\ref{1.6}). The commutation relations defining those read as \cite{magda61}:
$$
\begin{array}{l}
so(3) = \langle e_1, e_2, e_3 \rangle: \ [e_1, e_2] = e_3, \ \
[e_1, e_3] = -e_2, \ \ [e_2, e_3] = e_1; \\[2mm]
sl(2,{\mathbb R}) = \langle e_1, e_2, e_3 \rangle : \ [e_1, e_2] = 2 e_2, \ \
[e_1, e_3] = -2 e_3, \ [e_2, e_3] = e_1.
\end{array}
$$

We start by noting that one of the basis operators $e_1, e_2, e_3$ can be
reduced to one of the five operators (\ref{1.17}) (see, Lemma \ref{l1.3.1}).

We consider in detail the case of operator
\begin{equation} \label{1.19}
t \partial_t + x \partial_x
\end{equation}
only, since the remaining operators are treated in a similar way.

Let the basis operator $e_1$ of the algebras $so(3)$ and $sl(2,{\mathbb R})$
be of the form (\ref{1.19}). Computing the commutator of $e_1$ and $Q$ of the
form (\ref{1.6}) yields the relation
$$
[e_1, Q] = -\lambda_1 \partial_{t} -\lambda_2 \partial_x +
[x h' u + x r_x + t r_t ]\partial_u.
$$
To satisfy the first two commutation relations for each of the algebras
under study, the basis operators $e_2, e_3$ are to be of the form
$$
\alpha_1 \partial_t +\alpha_2 \partial_x +(\gamma (x) u +\mu(t,x)) \partial_u,
$$
where $\alpha_1, \alpha_2 \in {\mathbb R}$, $ \gamma$ and $\mu$ are smooth
functions. It is straightforward to verify that these operators cannot
satisfy the third commutation relation for either algebra $sl(2,{\mathbb R})$
and $so(3)$.

The lemma is proved.

\begin{theorem} \label{theo1.3.2}
There are no nonlinear equations (\ref{1.10}) whose invariance algebras are
isomorphic to semi-simple Lie algebras or contain those as sub-algebras.
\end{theorem}
{\it Proof.} Suppose the inverse. Let (\ref{1.10}) be an equation whose
invariance algebra contain sub-algebra that is semi-simple Lie algebra
$L$. Then by properties of semi-simple Lie algebras, there exist linear
combinations of the basis elements of $L$ forming the basis of either
$so(3)$ or $sl(2,{\mathbb R}).$ However, due to Lemma \ref{l1.3.2} there are no
realizations of the algebras $so(3)$, $sl(2,{\mathbb R})$ by operators
(\ref{1.6}). We arrive at the contradiction which proves the theorem.

It follows from Theorem \ref{theo1.3.2} and Levi-Maltsev theorem
(see, e.g., \cite{magda61, magda19}) that nonlinear equations
(\ref{1.10}) can admit invariance algebras of the dimension higher
than one, provided, (1) those are isomorphic to real solvable Lie
algebras, or (2) their finite dimensional sub-algebras are real and
solvable. Using this fact and also the concept of compositional row
for solvable Lie algebras we may perform hierarchical classification
of invariant equations starting from the lowest dimensional solvable
Lie algebras and increasing dimension by one till we exhaust all
possible invariant equations. We start by considering
two-dimen\-si\-on\-al solvable Lie algebras.

There exist two inequivalent two-dimensional solvable Lie algebras
\cite{magda19, magda20}
\begin{eqnarray*}
&& A_{2.1} = \langle e_1, e_2 \rangle \ : \ [e_1, e_2] =0; \\
&& A_{2.2} = \langle e_1, e_2 \rangle \ : \ [e_1, e_2] = e_2.
 \end{eqnarray*}
To construct all possible realizations of the above algebras we take as the first
basis element one of the realizations of one-dimensional invariance algebras
obtained above. The second operator is looked for in the form (\ref{1.6}). In
the case of commutative algebra $A_{2.1}$ there is no difference between
operators $e_1$ and $e_2$, while for the algebra $A_{2.2}$ those two
operators need separate analysis. We give full computation details for the
case, when one of the basis elements is of the form $A^1_1$ given in
Theorem \ref{theo1.3.1}.

\underline{Algebra $A_{2.1}$.} Let operator $e_1$ be of the form (\ref{1.19})
and operator $e_2$ read as (\ref{1.6}). Then it follows from the relation
$[e_1, e_2 ] =0$ that $ \lambda_1 = \lambda_2 = x h' =0, \ \
t r_t +x r_x =0.$ Consequently, we may choose the basis elements of the
algebra realization in the form $\langle t \partial_t +x \partial_x,
(mu + r(\psi)) \partial_u \rangle,$ where $m \in {\mathbb R}, \
\psi = t x^{-1}.$ Provided $m =0,$ the operator $e_2$ becomes $r(\psi)
\partial_u$. As established earlier, this realization does not satisfy
the determining equations. Hence, $m \not =0.$ Making the change
of variables
$$
\bar t = t, \ \ \bar x = x, \ \ \ v = u +m^{-1} r(\psi)
$$
reduces the basis operators in question to the form $\bar t \partial_{\bar t}
+\bar x \partial_{\bar x},$ $mv \partial_v.$ That is why we can restrict
our considerations to the realization $ \langle t \partial_t +x \partial_x,
u \partial_u \rangle.$

The second determining equation from (\ref{1.8}) after being written for
the operator $u \partial_u$ takes the form $u g_u =0,$ whence it follows
that this realization does not satisfy the determining equations. So the
realization $A^1_1$ cannot be extended to a realization of the
two-dimensional algebra $A_{2.1}$.

\underline{Algebra $A_{2.2}.$} If operator $e_1$ is of the form (\ref{1.19}),
then it follows from $[e_1, e_2] = e_2$ that $\lambda = \lambda_1 = \lambda_2 =0,$
$x h' =h,$ $t r_t +x r_x =r.$

Next, if $e_2$ reads as (\ref{1.19}), then we get from $[e_1, e_2] = e_2 $ the
erroneous equality $1=0$.

So the only possible case is when $e_2 = (mxu+xr(\psi)) \partial_u,$
$m \not =0,$ \ $\psi=tx^{-1}$, which gives rise to the following
realization of the algebra $A_{2.2}$:\ \ $ \langle t \partial_t +
x \partial_x, xu \partial_u \rangle.$ This algebra is indeed invariance
algebra of an equation from the class (\ref{1.10}) and the functions $f$
and $g$ read as:
$g = -2 x^{-1} \ln |u| +x^{-1} \tilde g(\psi),\ \
f = x^{-2} u \ln^2 |u| -x^{-2} \tilde g(\psi) u \ln |u| +
x^{-2} u \tilde f(\psi), \ \psi = tx^{-1}.
$

Analysis of the remaining realizations of one-dimensional Lie
algebras yields ten inequivalent $A_{2.1}$- and $A_{2.2}$-invariant
equations (see the assertions below). What is more, the obtained
(two-dimensional) algebras are maximal symmetry algebras of the
corresponding equations.

\begin{theorem} \label{theo1.3.3}
There are, at most, four inequivalent $A_{2.1}$-invariant nonlinear equations
(\ref{1.10}). Below we list the realizations of $A_{2.1}$ and the corresponding
expressions for $f$ and $g$.
\begin{eqnarray*}
1)&& \langle \partial_t, \sigma(x) u \partial_u, \rangle : g =
-2 \sigma' \sigma^{-1} \ln |u|,\\[2mm]
&& f = (\sigma' \sigma^{-1})^2 u \ln^2 |u| -\sigma^{-1} \sigma'
u \ln |u| +u \tilde f(x), \ \ \sigma' \not =0; \\[2mm]
2) && \langle \partial_t, \partial_x \rangle : g = \tilde g(u), \
f = \tilde f(u), \ \tilde g_u \not =0;\\[2mm]
3) && \langle \partial_x, \partial_t+u \partial_u \rangle : g = \tilde g(\omega), \
f = \exp(t) \tilde f (\omega), \ \omega = \exp({-t}), \ \tilde g_{\omega}\not =0;\\[2mm]
4) && \langle \sigma(x) u \partial_u,  \partial_t -\frac{1}{2} k \sigma(x) \psi(x)
u \partial_u \rangle : g = -2 \sigma' \sigma^{-1} \ln |u| +kt+ \tilde g(x), \\[2mm]
&& f = (\sigma' \sigma^{-1})^2 u \ln^2 |u| -\sigma^{-1} \sigma'' u \ln |u| -\sigma^{-1}
\sigma'(kt +\tilde g(x)) u \ln |u| \\[2mm]
&& +u\left[\frac{1}{2} k \sigma' \sigma^{-1} t +\frac{1}{4} k^2 t^2 +\frac{1}{2} k
\tilde g(x) + \tilde f(x)\right],\\[2mm]
&& k \not =0, \ \sigma' \not =0, \ \psi = \int \sigma^{-1} dx.
\end{eqnarray*}
\end{theorem}

\begin{theorem} \label{theo1.3.4}
 There exist, at most, six inequivalent $A_{2.2}$-invariant nonlinear
equations (\ref{1.10}). Below we list the realizations of $A_{2.1}$ and the corresponding expressions for $f$ and $g$.
$$\begin{array}{l}
1) \ \langle t \partial_t +x \partial_x, k^{-1} |x|^k u \partial_u \rangle :
g = x^{-1} (-2 k \ln|u| +\tilde g(\psi)),\\[2mm]
\hskip 10mm f = x^{-2} u (-k^2 \ln^2 |u| +k \tilde g(\psi) \ln |u| +k(k-1)
\ln |u| +\tilde f(\psi)), \\[2mm]
\hskip 10mm  k \not =0, \ \psi = t x^{-1};\\[2mm]
2) \ \langle \partial_t +\beta \partial_x, \ \exp({\beta^{-1}x}) u
\partial_u \rangle :
g = -2 \beta^{-1} \ln |u| +\tilde g(\eta), \\[2mm]
\hskip 10mm  f = \beta^{-2} u \ln^2 |u| -(\beta^{-2} +
\beta^{-1} \tilde g(\eta))u \ln |u|
+u \tilde f(\eta), \\[2mm]
\hskip 10mm \beta>0, \ \eta = x-\beta t;\\[2mm]
3) \ \langle -t \partial_t -x \partial_x, \partial_t +
\beta \partial_x \rangle :
g = \eta^{-1} \tilde g(u), \ \ f = \eta^{-2} \tilde f(u), \
\beta \geq 0,\\[2mm]
\hskip 10mm \eta = x -\beta t, \ \tilde g_u \not =0;\\[2mm]
4) \ \langle -t \partial_t -x \partial_x, \partial_t +
m x^{-1} u \partial_u \rangle :
g = x^{-1} (2m \psi +\tilde g(\omega)), \\[2mm]
\hskip 10mm f=x^{-1} [-2 m \psi u -2m \psi -2 -\tilde g(\omega)+ \exp({m \psi})
\tilde g(\omega)],\\[2mm]
\hskip 10mm m>0, \ \omega = u \exp({-m \psi}), \ \psi = t x^{-1}, \
\tilde g_\omega \not =0;\\[2mm]
5) \ \langle \partial_x, e^x u \partial_u \rangle : g = -2 \ln |u| +\tilde g(t), \
f = u \ln^2 |u| -\\[2mm]
\hskip 10mm - u \ln |u| (1+\tilde g(t)) +u \tilde f(t);\\[2mm]
6) \ \langle -t \partial_t -x \partial_x,\partial_x \rangle : g = t^{-1}
\tilde g(u), \ f = t^{-2} \tilde f(u), \ \tilde g_u \not =0.
\end{array}$$
\end{theorem}

\subsection{ Completing group classification of (\ref{1.10}). }

As the invariant equations obtained in the previous subsection
contain arbitrary functions of, at most, one variable, we can use
the standard Lie-Ovsyannikov approach to complete group classification of
(\ref{1.10}). We give the computation details for the case of the
first $A_{2.1}$-invariant equation, the remaining cases are treated
in a similar way.

Putting $g=-2 \sigma' \sigma^{-1} \ln |u|, f=(\sigma' \sigma^{-1}) u \ln^2 |u|
-\sigma^{-1} \sigma'' u \ln |u|+ u \tilde f(x),$  $ \sigma = \sigma(x), \
\sigma' \not =0$ we rewrite the first determining equation to become:
$$
-2 h' +2 \lambda \sigma' \sigma^{-1} \ln |u| = -2 (\lambda x+\lambda_2)(\sigma'
\sigma^{-1})'_x \ln |u| -2h \sigma' \sigma^{-1} -2 r \sigma' \sigma^{-1} u^{-1}.
$$
As $h = f(x), \sigma = \sigma(x), \ r= r(t,x), \lambda, \lambda_2 \in
{\mathbb R},$ the above relation is equivalent to the following ones:
$$
h' = \sigma' \sigma^{-1} h, \ \ r=0, \ \
\lambda \sigma' \sigma^{-1}=-(\lambda x+\lambda_2) (\sigma' \sigma^{-1})'.
$$
If $\sigma$ is an arbitrary function, then $\lambda = \lambda_2 =r =0,$
$h = C \sigma, \ C\in {\mathbb R}$ and we get $\langle \partial_t, \sigma(x)
u \partial_u \rangle$ as the maximal symmetry algebra. Hence, the extension
of symmetry algebra is only possible when the function $\psi = \sigma' \sigma^{-1}$
is a (non-vanishing identically) solution of equation
$$ (\alpha x +\beta) \psi' +\alpha \psi =0, \ \ \alpha, \beta \in R, \
|\alpha|+|\beta| \not =0.$$
If $\alpha \not =0,$ then at the expense of displacements by $x$ we can get
$\beta =0$, so that $\psi = m x^{-1}, \ m \not =0.$ Integrating the remaining
determining equations we get
$$
g = -2 m x^{-1} \ln |u|,\ \
f = m x^{-2} [ mu \ln^2 |u| -(m-1) u \ln |u| +nu], \ \
m \not =0, \ m,n \in {\mathbb R}.
$$
The maximal invariance algebra of the obtained equation is the three-dimensional
Lie algebra $ \langle \partial_t$, $|x|^m u \partial_u, t \partial_t +
x \partial_x \rangle$ isomorphic to $A_{3.7}.$

Next, if $\alpha =0,$ then $\beta \not =0$ and $\psi =m, \ m \not =0.$ If this is the
case, we have
$$g = \ln|u|, \ f = \frac{1}{4} u \ln^2 |u| -\frac{1}{4} u \ln |u| +nu, \ n \in R.$$
The maximal invariance algebra of the above equation reads as
$$ \langle \partial_t, \partial_x, \exp\left({-\frac{1}{2}x}\right) u \partial_u \rangle$$
and is isomorphic to $A_{3.2}.$

Similarly we prove that the list of inequivalent equations of the form (\ref{1.10})
admitting three-dimensional symmetry algebras is exhausted by the equations given below.
Note that the presented algebras are maximal. This means, in particular, that maximal
symmetry algebra of equation (\ref{1.10}) is, at most, three-dimensional.

{$A_{3.2}$-invariant equations}
$$\begin{array}{l}
1) \ u_{tt} = u_{xx} +u_x \ln |u| +\frac{1}{4} u \ln^2 |u|-\frac{1}{4} u \ln |u| +nu \
(n \in {\mathbb R}) :
 \ \langle \partial_t, \partial_x, \exp\left({-\frac{1}{2}x}\right) u \partial_u \rangle; \\[2mm]
2) \ u_{tt} = u_{xx} +m[\ln|u| -t] u_{x} +\frac{m^2}{4} u  [(\ln |u| -t)(\ln |u| -t-1)]
+nu ( m>0,\\[2mm]
 \hskip 10mm  n \in R):\langle \partial_x, \partial_t+u \partial_u,
 \exp\left({-\frac{1}{2}mx}\right) u\partial_u \rangle.
\end{array}$$

{$A_{3.4}$-invariant equations}
$$\begin{array}{l}
1) \ u_{tt} = u_{xx} +x^{-1} [2 \ln |u| +m x^{-1} t +n] u_x + x^{-2} u \ln |u| \\[2mm]
\hskip 10mm +(mx^{-1} t +n -2) x^{-2} u \ln |u| +\frac{1}{4} m^2 x^{-4} t^2 u +\frac{1}{2}
m (n-3) x^{-3} t u+ p x^{-2} u \\[2mm]
\hskip 10mm ( \ m \not =0, \ n,p \in {\mathbb R}): \langle t\partial_t+x \partial_x, x^{-1}u
\partial_u, \partial_t-\frac{m}{2} x^{-1} \ln |x| u \partial_u \rangle.
\end{array}$$

{$A_{3.5}$-invariant equations}
$$\begin{array}{l}
1) \ u_{tt} = u_{xx} +|u|^m u_x +n |u|^{1+2m} \ (m \not =0, \ n \in {\mathbb R}):
\langle \partial_t,\partial_x,  t\partial_t+ x \partial_x -m^{-1} u
\partial_u \rangle; \\[2mm]
2) \ u_{tt} = u_{xx} +e^u u_x +n e^{2u} \ (n \in {\mathbb R}):
\langle \partial_t, \partial_x, t \partial_t +x \partial_x -\partial_u \rangle ;\\[2mm]
3) \ u_{tt} = u_{xx} -x^{-1} [2 \ln |u| -m x^{-1} t -n] u_x +x^{-2} u \ln^2 |u| \\[2mm]
\hskip 10mm - x^{-2} (m x^{-1} t +n) u \ln |u|
+u x^{-2} \left[ \frac{m}{4} x^{-2} t^2 +\frac{m}{2} (n-1) x^{-1} t +p\right] \\[2mm]
\hskip 10mm  (m, n, p \in {\mathbb R}): \langle t \partial_t +x \partial_x, xu \partial_u,
\partial_t +\frac{m}{4} x^{-1} u \partial_u \rangle.
\end{array}$$

{$A_{3.7}$-invariant equations}
$$
\begin{array}{l}
1) \ u_{tt} = u_{xx} -2 m x^{-1} u_x \ln |u| +mx^{-2} [mu \ln^2 |u| -(m-1) u
\ln |u|+nu] \\[2mm]
\hskip 10mm  (m \not =0,1; n \in {\mathbb R}):  \langle \partial_t,
|x|^m u \partial_u, t \partial_t +x \partial_x \rangle; \\[2mm]
2) \ u_{tt} = u_{xx}-x^{-1} [2k+\ln |u| -m x^{-1} t -n]u_x +
k^2 x^{-2} u \ln^2 |u| \\[2mm]
\hskip 10mm -k x^{-2} [mtx^{-1} +k +n -1] u \ln |u| +
\frac{1}{2} m (k-2+n) t x^{-3} u\\[2mm]
\hskip 10mm +\frac{1}{4} m^2 t^2 x^{-4} u +p x^{-2} u \ \
( |k| \not =0,1; m \not =0, n,p\in {\mathbb R}):\\[2mm]
\hskip 10mm \langle t\partial_t+x \partial_x, |x|^k u \partial_u, \partial_t+
\frac{m}{2(1+k)} x^{-1} u \partial_u \rangle.
\end{array}
$$

This completes group classification of nonlinear equations (\ref{1.10}).

\section{ Group classification of equation (2.12) }
\setcounter{section}{4}
\setcounter{equation}{0}

Omitting calculation details we present below the determining equations
for symmetry operators admitted by equation (\ref{1.14}).

\begin{tver} \label{tv1.4.1}
The maximal invariance group of PDE (\ref{1.14}) is generated by the
infinitesimal operator
\begin{equation} \label{1.20}
Q = \tau(t) \partial_t+\xi(x) \partial_x+[h(t)u +r(t,x)] \partial_u,
\end{equation}
where $\tau,\xi, h,r,f,g$ are smooth functions satisfying the conditions
\begin{eqnarray} \label{1.21}
&& r_{tx} +f[h-\tau_t-\xi_x] = g r_x +\tau f_t +\xi f_x +[hu+r]f_u , \\
&& h_t = \tau_t g+\tau g_t +\xi g_x. \nonumber
\end{eqnarray}
\end{tver}

\begin{tver} \label{tv1.4.2}
The equivalence group $\cal E$ of (\ref{1.14}) is formed by the
following transformations of the space ${\rm V}$:
\begin{eqnarray} \label{1.22}
(1) && \bar t = T(t) , \ \bar x = X(x), \ v = U(t)u +Y(t,x), \ \ t'X'U \not =0;
\\
(2) && \bar t = T(x) , \ \bar x = X(t), \ v = \Psi(x)\Phi(t,x)u +Y(t,x), \ \
t'X'\Psi \not =0, \nonumber\\
&& \Phi(t,x) = \exp[-\int g(t,x) dt], \ g_x \not=0. \nonumber
\end{eqnarray}
\end{tver}

As the direct verification shows, given arbitrary functions $g$ and $f$,
it follows from (\ref{1.21}) that $\tau=h=\xi=r=0$. So that in the generic case
the maximal invariance group of (\ref{1.14}) is the trivial group of identical
transformations.

We begin classification of (\ref{1.14}) by constructing equations that
admit one-dimensi\-on\-al symmetry algebras.

\begin{lema} \label{l1.4.1}
There exist transformations (\ref{1.22}) reducing operator (\ref{1.20})
to one of the seven canonical forms given below
\begin{eqnarray} \label{1.23}
Q&=& t \partial_t+x\partial_x; \ \ Q = \partial_t; \ \ Q =
\partial_x+tu\partial_u; \nonumber \\
Q&=& \partial_x+\epsilon u \partial_u, \ \epsilon =0,1; \ Q = tu \partial_u, \\
Q&=& u \partial_u, \ \ Q = r(t,x) \partial_u, \ r \not =0. \nonumber
\end{eqnarray}
\end{lema}
{\it Proof.} Transformations (\ref{1.22}) reduce operator $Q$ (\ref{1.20})
to become
\begin{equation} \label{1.24}
\tilde Q = \tau T' \partial_{\bar t} +\xi X' \partial_{\bar x} +[(\tau U' +Uh)
u +\tau Y_t+\xi Y_x + Ur] \partial_v.
\end{equation}
Provided $\sigma \cdot \xi \not =0$, we can choose as $T, X, U, Y$ non vanishing identically
solutions of equations
$$ \tau T' = T, \ \ \xi X'= X, \ \ \tau U' + h U=0, \ \ \tau Y_t +\xi Y_x +Ur =0$$
thus getting operator $\tilde Q$ (\ref{1.24}) in the form
$ \tilde Q = \bar t\partial_{\bar t} +\bar x \partial_{\bar x}.$
If $\tau \not =0,$ a $\xi =0,$ then taking as $T, U, Y$ solutions of equations
$$ \tau T' =1, \ \tau U' +h U=0 \ (U \not =0), \ \tau Y_t + U r = 0$$
reduces operator (\ref{1.20}) to become
$\tilde Q = \partial_{\bar t}.$
If $\tau =0, \ \xi \not =0,$ then under $h' \not =0$ we get operator
$\tilde Q = \partial_{\bar x} + \bar t v \partial_v$. Next, if $h'=0,$
we arrive at the operator
$\tilde Q = \partial_{\bar x} +\epsilon v \partial_v,$ where either
$\epsilon =0$ or $\epsilon = 1.$

Finally, the case $\tau = \xi =0,$ gives rise to the following operators
$ \tilde Q = \bar t v\partial_v, \ \ \tilde Q =v \partial_v, \ \ \tilde Q =
r(\bar t, \bar x) \partial_v.$
After rewriting these in the initial variables we get the
operators listed in the statement of lemma. The lemma is proved.

\begin{theorem} \label{theo1.4.1}
There exist, at most, three inequivalent nonlinear equations (\ref{1.14})
that admit one-dimen\-si\-on\-al invariance algebras. The form of functions
$f, g$ and the corresponding symmetry algebras are given below.
\begin{eqnarray*}
A^1_1 &=& \langle t \partial_t +x \partial_x \rangle : g = t^{-1} \tilde
g(\omega), \ f = t^{-2} f(u, \omega), \ \ \omega = tx^{-1}, \ \
\tilde g_\omega \not =0, \ \ f_{uu} \not =0; \\
A^2_1& =& \langle \partial_t \rangle: g = \tilde g(x), \ f = \tilde f(x,u), \
\tilde g' \not =0, \ \tilde f_{uu} \not =0; \\
A^3_1 &=& \langle \partial_x +tu \partial_u \rangle : g=x+\tilde g(t), f =
e^{tx} \tilde f(t,\omega), \ \omega = e^{-tx} u, \ \tilde f_{\omega \omega} \not
=0.
\end{eqnarray*}
\end{theorem}
{\it Proof.} If equation (\ref{1.14}) admits one-parameter transformation
group, then the latter is generated by infinitesimal operator (\ref{1.20}).
According to Lemma \ref{l1.4.1} there exist equivalence transformations
(\ref{1.22}) reducing this operator to one of the seven canonical
operators (\ref{1.23}). With this fact in hand, we turn to solving the
determining equations (\ref{1.21}) for each of those operators. The first
three operators yield invariant equations and corresponding symmetry
algebras given in the statement of theorem. The next two operators give rise to
inconsistent equations.

Finally, the remaining operators yield that the functions $f$ and $g$
are linear in $u$, which means that the corresponding invariant
equations are linear.

It is straightforward to verify that for the case of arbitrary
functions $\tilde f$, $\tilde g$, the corresponding one-dimensional algebras
are maximal invariance algebras.

The theorem is proved.

We proceed now to analyzing equations (\ref{1.14}) admitting two-dimensional
symmetry algebras.

\begin{theorem} \label{theo1.4.2}
There exist, at most,  three inequivalent nonlinear equations
(\ref{1.14}) that admit two-dimen\-si\-on\-al symmetry algebras, all of
them being $A_{2.2}$-invariant equations. The forms of functions
$f$ and $g$ and the corresponding realizations of the Lie algebra
$A_{2.2}$ are given below
\begin{eqnarray*}
A^1_{2.2} &=& \langle t \partial_t +x \partial_x, t^2 \partial_t +x^2
\partial_x+ mut \partial_u \rangle \ (m \in {\mathbb R} ): \\
&& g = [mt+(k-m)x ]t^{-1} (t-x)^{-1}, \ k\not =0, \\
&& f = |t-x|^{m-2} |x|^{-m} \tilde f (\omega), \\
&& \omega = u |t-x|^{-m} |x|^m , \tilde f_{\omega \omega} \not =0; \\
A^2_{2.2} &=& \langle  t \partial_t +x \partial_x, t^2 \partial_t +mtu
\partial_u \rangle \ (m \in {\mathbb R} ): \\
&& g = t^{-2} [kx +mt], \ k \not =0, \ f = |t|^{m-2} |x|^{-m} \tilde f(\omega),
\\
&& \omega = |t|^{-m} |x|^m u, \ \tilde f_{\omega \omega} \not =0; \\
A^3_{2.2} &=& \langle t \partial_t +x \partial_x, x^2 \partial_x +tu \partial_u
\rangle : \\
&& g = (tx)^{-1} (mx-t) \ (m \in {\mathbb R} ), \ f=x^{-2} \exp({-tx^{-1}}) \tilde f(\omega),
\\
&& \omega = u \exp({tx^{-1}}), \ \tilde f_{\omega \omega} \not =0.
\end{eqnarray*}
\end{theorem}

To prove the theorem we need to extend realizations  $A^i_1 \
(i=1,2,3)$ to realizations of the algebras $A_{2.1}, A_{2.2}$
by operators (\ref{1.20}). We skip calculation details.

Note that if the functions $\tilde f$ are arbitrary, then the invariance
algebras given in the statement of Theorem \ref{theo1.4.2} are maximal.

Now we can complete the group classification presented in
Theorem \ref{theo1.4.2} with the use of Lie-Ovsyan\-ni\-kov classification
routine.

We consider in some detail the case of $A^1_{2.2}$-invariant
equations (the remaining cases are treated in a similar way).
The second determining equation from (\ref{1.21}) reads now as
\begin{equation} \label{1.25}
(t-x)^2 h_t = t^{-1} \tau_t [m(t-x)^2 +kx(t-x)] \ \
+\tau [-t^{-2} m(t-x)^2 -2 k t^{-1} x+kt^{-2} x^2] +k \xi.
\end{equation}
Differentiating right- and left-hand sides of (\ref{1.25}) twice
by $x$ yields
$$
h_t = (m-k) (t^{-1} \tau_t -t^{-2} \tau) +k \xi''.
$$
Hence we get $\xi'''= 0$ and
\begin{eqnarray*}
\xi&=& \lambda_1 x^2 +\lambda_2 x +\lambda_3 ,\ \ \lambda_1, \lambda_2,
\lambda_3 \in {\mathbb R}, \\
h&=& (m-k) t^{-1} \tau+\lambda_1 kt+\lambda_4 ,\ \
\lambda_4 \in {\mathbb R}.
\end{eqnarray*}
With account of the above results we obtain from (\ref{1.25}) that $\tau =
\lambda_1 t^2 +\lambda_2 t+\lambda_3.$ So it follows from (\ref{1.25}) that
the coefficients of infinitesimal operator (\ref{1.20}), which generates
symmetry group of $A^1_{2.2} $-invariant equation, read necessarily as
\begin{eqnarray*}
\tau &=& \lambda_1 t^2 +\lambda_2 t +\lambda_3, \\
\xi &=& \lambda_1 x^2 +\lambda_2 x +\lambda_3, \\
h&=& m \lambda_1 t +(m-k) \lambda_3 t^{-1} +(m-k) \lambda_2 +\lambda_4, \ \
\lambda_1, \lambda_2, \lambda_3, \lambda_4 \in {\mathbb R}.
\end{eqnarray*}
Consequently, the first determining equation from (\ref{1.21}) takes the form
\begin{eqnarray}\label{1.26}
&& \{ x^{-1} (t-x)^{-1} [((m-k)\lambda_2 +\lambda_4)(tx-x^2) -(m-k) \lambda_3
t^{-1} x^2 \nonumber \\
&&- k \lambda_3 x +m \lambda_3 t] \omega +r|t-x|^{-m} |x|^m \} \tilde
f_{\omega}\nonumber \\
&& -x^{-1}(t-x)^{-1} [((m-k) \lambda_2 +\lambda_4 )(tx-x^2) \\
&& -(m-k)\lambda_3 t^{-1} x^2 -k \lambda_3 x+m\lambda_3 t]\tilde f\nonumber\\
&&= |t-x|^{-m+2} |x|^m [r_{tx} -t^{-1} (m+kx(t-x)^{-1} r_x].\nonumber
\end{eqnarray}
It follows from (\ref{1.26}) that if the functions$\tilde f$
are arbitrary, then the maximal invariance algebra of the equation under
study coincide with the realization $A^1_{2.2}.$ What is more, an extension
of the invariance algebra is only possible when the function $\tilde f$ obey
the following equation:
\begin{equation} \label{1.27}
(a \omega +b) \tilde f_{\omega} -a \tilde f =c,
\end{equation}
where $a,b,c \in  R , |a|+|b| \not =0.$ On the other hand, it follows from
(\ref{1.27})
$$(a\omega +b) \tilde f_{\omega\omega} =0,$$
whence $f_{\omega\omega} =0.$ We arrive at the contradiction, which
proves that there are no extension of the realization $A^1_{2.2}$ in
question to the higher dimensional invariance algebra of the
equation (\ref{1.14}). Analy\-zing $A^2_{2.2}$- and
$A^2_{2.3}$-invariant equations we arrive at the same conclusion.

Consequently, there are no nonlinear equations of the form (\ref{1.14})
whose maximal invariance algebras are solvable Lie algebras of the
dimension higher than two. Next, as the algebra $sl(2, {\mathbb R})$
contains two-dimensional subalgebra isomorphic to $A_{2.2}$, there
are no nonlinear equations (\ref{1.14}), whose invariance algebras
are either isomorphic to $sl(2, {\mathbb R})$ or contain it as a
subalgebra. Finally, we verified that there are no realizations
of the algebra $so(3)$ by operators (\ref{1.20}).

Summing up the above reasonings we formulate the following assertion.
\begin{theorem} \label{theo1.4.3}
A nonlinear equation (\ref{1.14}) having non-trivial symmetry properties
is equivalent to one of the equations listed in Theorems \ref{theo1.4.1} and
\ref{theo1.4.2}.
\end{theorem}

This completes group classification of the class of nonlinear PDEs
(\ref{1.14}).

\section{Group classification of equation (2.13)}
\setcounter{section}{5}
\setcounter{equation}{0}

As earlier, we present the results of the first step
of our group classification algorithm skipping derivation details.

\begin{tver} \label{tv1.5.1}
Invariance group of equation (\ref{1.15}) is generated by infinitesimal
operator
\begin{equation} \label{1.28}
Q = \tau (t ) \partial_t+\xi(x) \partial_x+(ku+r(t,x))\partial_u,
\end{equation}
where $k$ is a constant $\tau, \xi, r, f$ are functions satisfying the relation
\begin{equation} \label{1.29}
r_{tx} +[k-\tau'-\xi']f=\tau f_t +\xi f_x +[ku+r]f_u.
\end{equation}
\end{tver}

\begin{tver} \label{tv1.5.2}
Equivalence group $\cal E$ of the class of equations (\ref{1.15}) is
formed by the following trans\-for\-ma\-ti\-ons:
\begin{eqnarray} \label{1.30}
(1) && \bar t = T(t), \ \ \bar x = X(x), \ \ v = mu +Y(t,x), \nonumber \\
(2) && \bar t = T(x), \ \ \bar x = X(t), \ \ v=mu +Y(t,x), \ \ T' X' m \not =0.
\end{eqnarray}
\end{tver}

Note that given arbitrary $f$, it follows from (\ref{1.29}) that
$\tau = \xi = k = r = 0$, i.e., the group admitted is
trivial. To obtain equations with nontrivial symmetry we need to specify
properly the function $f$. To this end we perform classification
of equations under study admitting one-dimensional invariance algebras.

\begin{lema} \label{l1.5.1}
There exist transformations from the group $\cal E$ (\ref{1.30}) that reduce
(\ref{1.28}) to one of the four canonical forms:
\begin{eqnarray*}
Q&=& \partial_t +\partial_x+\epsilon u \partial_u \ \ (\epsilon =0,1):\\
Q&=& \partial_t +\epsilon u \partial_u \ \ (\epsilon =0,1); \\
Q&=& u \partial_u, \ Q = g(t,x) \partial_u \ (g \not =0).
\end{eqnarray*}
\end{lema}
{\it Proof.} Utilizing transformations (1) from (\ref{1.30}) we reduce the
operator $Q$ to one of the following forms:
\begin{eqnarray*}
Q&=& \partial_{\bar t} +\partial_{\bar x}+\epsilon v \partial_v \ \ (\epsilon
=0,1):\\
Q&=& \partial_{\bar t} +\epsilon v \partial_v \ \ (\epsilon =0,1); \\
Q&=& \partial_{\bar x} +\epsilon v \partial_v \ \ (\epsilon =0,1); \\
Q&=& v \partial_v, \ Q = g({\bar t} ,\bar x) \partial_v \ (g \not =0).
\end{eqnarray*}
Next, we note that the change of variables
$\tilde t = \bar x , \ \ \tilde x = \bar t, \ \ \tilde v = v,$
which is of the form (2) from (\ref{1.30}), transforms the second
operator into the third one.

Finally, rewriting the obtained operators in the initial variables completes
the proof.

\begin{theorem} \label{theo1.5.1}
There exist exactly two nonlinear equations of the form (\ref{1.15})
admitting one-dimensional invariance algebras. The corresponding
expressions for function $f$ and invariance algebras are given below.

\begin{eqnarray*}
A^1_1 &=& \langle \partial_t +\partial_x +\epsilon u \partial_u \rangle  \
(\epsilon=0,1): \ f = e^{\epsilon t} \tilde f(\theta, \omega), \ \
\theta = t-x, \ \omega = e^{-\epsilon t} u; \tilde f_{\omega \omega} \not
=0;\\
A^2_1 &=& \langle \partial_t +\epsilon u \partial_u \rangle \
(\epsilon =0,1): f = e^{\epsilon t} \tilde f (x, \omega), \ \
\omega = e^{-\epsilon t} u, \ \tilde f_{\omega \omega} \not =0.
\end{eqnarray*}
\end{theorem}
To prove the theorem, it suffices to select those operators from the list of
Lemma \ref{l1.5.1} that can be invariance algebra of nonlinear equation of the
form (\ref{1.15}). To this end we need to solve equation (\ref{1.29}) for each
of the operators in question.

The first two operators yield $A^1_1$- and $A^2_1$-invariant equations.
The last two operators gives rise to linear invariant equations (\ref{1.15}),
which are not taken into account.

What is more, if the function $\tilde f$ is arbitrary, then the algebras
$A^1_1$ and $A^2_1$ are maximal invariance algebras of the corresponding
equations.

Next, we classify nonlinear equations admitting symmetry algebras of the
dimension higher than one. We begin by considering equations whose
invariance algebras contain semi-simple subalgebras. It turns out,
that the class of operators (\ref{1.28}) contain no realizations of
the algebra $so(3)$. Furthermore, it contains four inequivalent
realizations of the algebra $sl(2, {\mathbb R})$ given below.
\begin{eqnarray*}
(1) && \langle \partial_t, \frac{1}{2} e^{2t} \partial_t, -\frac{1}{2} e^{-2t}
\partial_t \rangle; \\
(2) && \langle \partial_t, \frac{1}{2} e^{2t}(\partial_t+\partial_u),
-\frac{1}{2} e^{-2t} (\partial_t-\partial_u) \rangle; \\
(3) && \langle \partial_t ,\frac{1}{2} e^{2t}(\partial_t+x \partial_u),
-\frac{1}{2} e^{-2t}(\partial_t-x \partial_u) \rangle ;\\
(4) && \langle \partial_t+\partial_x, \frac{1}{2} e^{2t} \partial_t+\frac{1}{2}
e^{2x} \partial_x, -\frac{1}{2} e^{-2t} \partial_t-\frac{1}{2} e^{-2x}
\partial_x+\epsilon[e^{-2x}-e^{-2t}]\partial_u \rangle,
\ \ \epsilon =0,1.
\end{eqnarray*}

Before analyzing $sl(2, {\mathbb R})$-invariant equations let us
briefly review the group properties of the Liouville equation
\begin{equation} \label{1.31}
u_{tx} = \lambda e^u, \ \ \lambda \not = 0.
\end{equation}
It is a common knowledge that the maximal invariance group of this equation
is the infinite-parameter group generated by the following infinitesimal
operator \cite{magda3}:
$$ Q = h(t) \partial_t +g(x) \partial_x-(h'+g') \partial_u,$$
where $h$ and $g$ are arbitrary smooth functions. Note that due to this fact
the Liouville equation can be linearized by a (non-local) change of
variables (see, e.g., \cite{magda41,magda42,m78}).

After a simple algebra we obtain that realizations (1), (3), (4) with
$\epsilon = 1$ cannot be invariance algebras of nonlinear equation of the form
(\ref{1.15}). Realization (2) is the invariance algebra of equation
$$ u_{tx} = \tilde f(x) e^{-2 u}, \ \ \tilde f \not =0$$
which reduces to equation (\ref{1.31}) via the change of variables
$$ t = t, \ \  \ x=x, \  \ \ u =-\frac{1}{2} (v -\ln|\tilde f|), \ \  \ \
v=v(t,x).$$

Finally making use of the change of variables
$$ \bar t = e^{-2 t}, \ \ \bar x = e^{-2 x}, \ \ v=u$$
we rewrite (4) under $\epsilon=0$ to become
$$ \langle \partial_t+\partial_x, t \partial_t+x \partial_x, t^2
\partial_t+x^2\partial_x \rangle.$$
The corresponding invariant equation reads as
\begin{equation}
 \label{1.32}
u_{tx} = (t-x)^{-2} \tilde f(u), \ \ \tilde f_{uu} \not =0.
\end{equation}
If the function $\tilde f$ is arbitrary, then the above presented
realization is the maximal invariance algebra of the equation under
study. Using Lie-Ovsyannikov algorithm we establish that extension
of symmetry is only possible when $\tilde f = \lambda e^u +2.$ However,
the corresponding equation is reduced to the Liouville equation
by the change of variables
$$ t = t, \ x=x, \ \ u = v(t,x) +2 \ln |t-x|.$$

Thus the only inequivalent nonlinear equations (\ref{1.15}) whose invariance
algebras contain semi-simple subalgebras are given in (\ref{1.31}) and
(\ref{1.32}), where $\tilde f$ is an arbitrary smooth function of $u$.

To complete group classification of equation (\ref{1.15}) we need to
describe equations whose invariance algebras are solvable Lie
algebras of the dimension higher than one. We begin with those realizations
of two-dimensional Lie algebras $A_{2.1}$, $A_{2.2}$, which can be admitted
by nonlinear equations (\ref{1.15}).

It turns out that the class of operators (\ref{1.28}) contains within
the equivalence relation only one realization of the algebra $A_{2.1}$
which meets the invariance requirements, namely,
$$
\langle \partial_t + \epsilon_1 u \partial_u, \partial_x +\epsilon_2
u\partial_u \rangle \ (\epsilon_1 =0,1; \ \ \epsilon_2 = 0,1).
$$
The corresponding invariant equation reads as
\begin{equation} \label{1.33}
u_{tx}= \exp(\epsilon_1 t+\epsilon_2 x) \tilde f(\omega), \ \ \omega = u
\exp(-\epsilon_1 t-\epsilon_2 x).
\end{equation}
Analysis of equation (\ref{1.33}) with arbitrary $f(\omega)$ shows that under $\epsilon_1
+\epsilon_2 \not =0$ the above realization is its maximal invariance
algebra. Provided $\epsilon_1 = \epsilon_2 =0,$ the equation takes the form
\begin{equation} \label{1.34}
u_{tx} = f(u)
\end{equation}
and its maximal invariance algebra is the three-dimensional Lie algebra of
the operators
$$ \langle \partial_t,\partial_x, t \partial_t-x \partial_x \rangle,$$
which is isomorphic to $A_{3.6}.$

It is a common knowledge (see, e.g., \cite{magda32,magda43,magda9}) that
(\ref{1.34}) admits higher symmetry if it is equivalent either to the
Liouville equation (\ref{1.31}), or to the nonlinear d'Alembert
equation
\begin{equation} \label{1.35}
u_{tx} = \lambda |u|^{n+1}, \ \ \lambda \not =0, \ \ n \not =0,-1.
\end{equation}
The maximal invariance algebra of (\ref{1.35}) is the four-dimensional
Lie algebra of the operators
$$\langle t \partial_t-\frac{1}{n} u \partial_u, x \partial_x -\frac{1}{n} u
\partial_u, \partial_t, \partial_x \rangle. $$
It is isomorphic to the Lie algebra $A_{2.2} \oplus A_{2.2}.$

Extension of symmetry algebra of equation (\ref{1.33}) with $\epsilon_1=1, \epsilon_2 =0,$
is only possible when:
\begin{eqnarray}
u_{tx}& =& \lambda e^{-mt} |u|^{m+1}, \ \lambda \not =0, \ m \not =0,-1;
\label{1.36} \\
u_{tx} &=& \lambda e^{t}\exp(u e^{-t}), \ \lambda \not =0. \label{1.37}
\end{eqnarray}
The maximal invariance algebra of (\ref{1.36}) is the four-dimensional
Lie algebra of operators
$$\langle \partial_t+u \partial_u, e^{mt}  \partial_t, \partial_x,
x\partial_x-\frac{1}{m} u \partial_u \rangle, $$
which is isomorphic to $A_{2.2} \oplus A_{2.2}.$ Note that the change of variables
$$\bar t = e^{-mt}, \ \bar x = x, \ u = v(\bar t, \bar x)$$
reduces the above equation to the form (\ref{1.35}).

The maximal invariance algebra of (\ref{1.37}) is spanned by the operators
$$\langle \partial_t +u \partial_u, \partial_x, x \partial_x -e^t \partial_u
\rangle,$$
and is isomorphic to $A_1 \oplus A_{2.2}.$

Analysis of $A_{2.2}$-invariant equations yields the following results. The class
of operators (\ref{1.28}) contains six inequivalent realizations of the algebra
$A_{2.2}$ which meet the invariance requirements
\begin{eqnarray}\label{1.38}
(1) && \langle - t \partial_t+x \partial_u, \partial_t \rangle ; \nonumber \\
(2) && \langle -t \partial_t -x \partial_x, \partial_t+\partial_x \rangle
;\nonumber \\
(3) && \langle -t \partial_t-x \partial_x +u \partial_u, \partial_t+\partial_x
\rangle ; \\
(4) && \langle -t \partial_t+\partial_u, \partial_t \rangle ; \nonumber \\
(5) && \langle -t \partial_t -x \partial_x -u \partial_u, \partial_t \rangle;
\nonumber \\
(6) && \langle  -t \partial_t -x \partial_x, \partial_t \rangle .\nonumber
\end{eqnarray}

Equation invariant under realization (1) reads as
\begin{equation} \label{1.39}
u_{tx} = \exp(x^{-1} u).
\end{equation}
Its maximal symmetry algebra is the three-dimensional Lie algebra of
operators
$$ \langle -t \partial_t +x \partial_u, \partial_t, x \partial_x+
u \partial_u \rangle
$$
isomorphic to $A_{2.2} \oplus A_1.$ Note that the change of variables
$$\bar t=x, \ \ \bar x=e^t, \ \ u = v(\bar t, \bar x)$$
reduces  (\ref{1.39}) to the form (\ref{1.37}).

Equation invariant under the second realization of $A_{2.2}$
is of the form (\ref{1.32}). It has already been studied while
describing $sl(2, {\mathbb R})$-invariant equations.

Realizations (3) and (4) give no new invariant equations as well.

New invariant equation are obtained with the use of the fifth realization
from (\ref{1.37}). It has the form
$$ u_{tx} = x^{-1} \tilde f(\omega), \ \ \omega = x^{-1}u.$$
If the function $\tilde f$ is arbitrary, then the realization in
question is maximal invariance algebra of the above equation. Further
extension of symmetry properties is only possible if $\tilde f(\omega)
= \lambda |\omega|^{m+1}$, which gives the following invariant equation:
$$u_{tx} = \lambda |x|^{-m-2} |u|^{m+1}, \ \ \lambda \not =0, \ \ m \not =0, -1,
-2.$$
Its maximal symmetry algebra is the three-dimensional Lie algebra having the basis
$$ \langle \partial_t, t \partial_t+x \partial_x+u \partial_u,  x
\partial_x+\frac{m+1}{m} u \partial_u \rangle.$$
This algebra is isomorphic to $A_{2.2} \oplus A_1.$

We sum up the above results in the following assertion.

\begin{theorem} \label{theo1.5.2}
The Liouville equation
$ u_{tx} = \lambda e^u, \ \ \lambda \not=0,$
has the highest symmetry among equations (\ref{1.15}). Its maximal
invariance algebra is infinite-dimensi\-on\-al and is spanned by
the following infinite set of basis operators:
$$ Q = h(t) \partial_t+g(x) \partial_x-(h'(t)+g'(x)) \partial_u,$$
where $h$ and $g$ are arbitrary smooth functions. Next, there exist
exactly nine inequivalent equations of the form (\ref{1.15}), whose
maximal invariance algebras have dimension higher that one. Those
equations and their invariance algebras are given in Table 1.
\end{theorem}

\begin{center}
 Table I.\ {\it Invariant equations (\ref{2.13})}
\end{center}
\vskip 5mm
\begin{tabular}{|c|c|c|c|}\hline
Number & Function $f$& Symmetry operators& Invariance algebra \\
& & & type \\ \hline
1& $e^t \tilde f(\omega), $ &$\partial_t+u \partial_u, \partial_x$&$A_{2.1}$ \\
& $\omega = u e^{-t} , \tilde f_{\omega \omega} \not =0$ & & \\ \hline
2& $e^{t+x} \tilde f(\omega), $ &$\partial_t+u \partial_u,$&$A_{2.1}$ \\
& $\omega = u e^{-t-x} , \tilde f_{\omega \omega} \not =0$ &$ \partial_x+u
\partial_u$ & \\ \hline
3& $(t-x)^{-3} \tilde f(\omega), $ &$-t\partial_t-x \partial_x+u
\partial_u,$&$A_{2.2}$ \\
& $\omega =(t-x) u, \tilde f_{\omega \omega} \not =0$ &$ \partial_t+
\partial_x$ & \\ \hline
4& $x^{-1} \tilde f(\omega), $ &$-t\partial_t-x \partial_x-u
\partial_u,$&$A_{2.2}$ \\
& $\omega =x^{-1} u, \tilde f_{\omega \omega} \not =0$ &$ \partial_t$ & \\
\hline
5& $(t-x)^{-2} \tilde f(u), $ &$\partial_t+ \partial_x,$&$sl (2, R )$ \\
& $ \tilde f_{u u} \not =0$ &$ t\partial_t+ x \partial_x,$ & \\
& & $t^2 \partial_t +x^2 \partial_x$& \\ \hline
6& $\exp({x^{-1}u})$ &$-t\partial_t+x \partial_u,$&$A_{2.2}\oplus A_1$ \\
&  &$ \partial_t,x \partial_x+u \partial_u$ & \\ \hline
7 &$\lambda |x|^{-m-2} |u|^{m+1},$ &$ \partial_t, t \partial_t -\frac{1}{m} u
\partial_u,$ &$A_{2.2} \oplus A_1$ \\
& $\lambda \not =0, m\not =0, -,1 -2$ &$ x \partial_x +\frac{m+1}{m} u
\partial_u$ & \\ \hline
8 &$ \tilde f(u), \tilde f_{u u } \not =0$ &$ \partial_t, \partial_x, -t
\partial_t -x \partial_x $&$ A_{3.6}$ \\ \hline
9 &$\lambda |u|^{n+1}, \lambda \not =0, n\not =0, -1$ &$ t\partial_t-\frac{1}{n}
u \partial_u$ &$A_{2.2} \oplus A_{2.2}$ \\
& &$ x \partial_x -\frac{1}{n} u \partial_u$ & \\
& &$  \partial_t, \partial_x$ &  \\ \hline
\end{tabular}

\section{Group classification of equation (2.7)}
\setcounter{section}{6}
\setcounter{equation}{0}

The first step of the algorithm of group classification of (\ref{1.9})
$$ u_{tt} = u_{xx} +F(t,x,u,u_x), \ \ F_{u_x u_x} \not =0$$
has been partially performed in the second chapter. It follows from
Theorem \ref{theo1.2.1} that the invariance group of equation (\ref{1.9})
is generated by infinitesimal operator (\ref{1.6}).
What is more, the real constants $\lambda,\lambda_1,\lambda_2$ and real-valued
functions $h = h(x),r = r(t,x),$  $F = F(t,x,u,u_x)$ obey the relation (\ref{1.7}).
The equivalence group of the class of equations (\ref{1.9}) is formed by
transformations (\ref{1.16}).

The above enumerated facts enable using results of group classification
of equation (\ref{1.10}) in order to classify equation (\ref{1.9}). In
particular, using Lemmas \ref{l1.3.1} and \ref{l1.3.2} it is straightforward
to verify that the following assertions hold true.

\begin{theorem} \label{theo2.1.1}
There are, at most, seven inequivalent classes of
nonlinear equations (\ref{1.9}) invariant under the one-dimensional
Lie algebras.
\end{theorem}
Below we give the full list of the invariant equations and the
corresponding invariance algebras.
\begin{eqnarray*}
A^1_1&=&\langle t\partial_t+x \partial_x\rangle:\quad F = t^{-2}
G(\xi, u, \omega),\ \xi = tx^{-1},\ \omega = x u_x; \\[1mm]
A^2_1&=&\langle\partial_t+k\partial_x\rangle \ (k>0):\quad
F=G(\eta,u,u_x),\ \eta=x-kt; \\[1mm]
A^3_1&=&\langle\partial_x\rangle:\quad F=G(t,u,u_x); \\[1mm]
A^4_1&=&\langle\partial_t\rangle:\quad F=G(x,u,u_x); \\[1mm]
A^5_1 &=&\langle\partial_t+f(x)u\partial_u\rangle \
(f\not=0):\\[1mm]
&& \quad F=-tf''u+t^2(f')^2u  -2tf'u_x+e^{tf}G(x,v,\omega),\\[1mm]
&&  v=e^{-tf}u,\ \ \omega=u^{-1}u_x-f'f^{-1}\ln|u|; \\[1mm]
A^6_1&=&\langle f(x)u\partial_u\rangle \ (f\not=0):\quad
F=-f^{-1}f''u\ln|u|\\[1mm]
&& \quad-2f^{-1}f'u_x\ln|u|+f^{-2}(f')^2u\ln^2|u|+uG(t,x,\omega),\\[1mm]
&& \omega=u^{-1}u_x-f'f^{-1}\ln|u|;\\[1mm]
A^7_1&=&\langle f(t,x)\partial_u\rangle \ (f\not=0):\quad
F=f^{-1}(f_{tt}-f_{xx})u+G(t,x,\omega),\\[1mm]
&& \omega=u_x-f^{-1}f_xu.
\end{eqnarray*}
Note that if the functions $F$ and $G$ are arbitrary, then the given
algebras are maximal (in Lie's sense) symmetry algebras of the respective
equations.

\begin{theorem} \label{theo2.1.2}
An equation of the form (\ref{1.9}) cannot admit Lie
algebra which has a subalgebra having nontrivial Levi factor.
\end{theorem}

With account of the above facts we conclude that nonlinear
equations (\ref{1.9}) admit a symmetry algebra of the dimension higher than
one only if the latter is a solvable real Lie algebra. That is why, we turn
to classifying equations (\ref{1.9}) whose invariance algebras are
two-dimensional solvable Lie algebras.

As the calculations are similar to those performed in the third section,
we present the final result only. Namely, we give the form of invariant
equations and the corresponding realizations of the two-dimensional
invariance algebras.

\noindent
{\bf  I. $A_{2.1}$-invariant equations}
\begin{eqnarray*}
A^1_{2.1}&=&\langle t\partial_t+x\partial_x,\ u\partial_u
\rangle:\quad F=x^{-2}uG(\xi, \omega),\\[1mm]
 && \xi=tx^{-1},\ \omega=u^{-1}xu_x;\\[1mm]
A^2_{2.1}&=&\langle t\partial_t+x\partial_x,\
\sigma(\xi)\partial_u\rangle\ (\sigma\not=0,\ \xi=tx^{-1}):\\[1mm]
&& F=x^{-2}[\sigma^{-1}((1-\xi^2)\sigma''-2\xi\sigma')u +G(\xi,\omega)],\\[1mm]
&& \omega=\xi\sigma'u+\sigma xu_x;
\end{eqnarray*}
\begin{eqnarray*}
A^3_{2.1}&=&\langle\partial_t+k\partial_x,\ u\partial_u\rangle\
(k>0):\quad F=uG(\eta,\omega),\\[1mm] && \eta=x-kt,\
\omega=u^{-1}u_x;\\[1mm]
A^4_{2.1}&=&\langle\partial_t+k\partial_x,\
\varphi(\eta)\partial_u\rangle\ (k>0,\ \eta = x-kt,\
\varphi\not=0):\\[1mm] &&
F=(k^2-1)\varphi''\varphi^{-1}u+G(\eta,\omega),\ \omega=\varphi
u_x-\varphi'u; \\[1mm]
A^5_{2.1}&=&\langle\partial_t+k\partial_x,\
\partial_x+u\partial_u\rangle\ (k>0): \\[1mm]
&& F = e^{ \eta}\,G(\omega, v),\ \eta=x-kt,\
\omega=ue^{-\eta},\ v=u^{-1}u_x; \\[1mm]
A^6_{2.1}&=&\langle\partial_t,\ \partial_x\rangle:\quad
F=G(u,u_x);\\[1mm]
A^7_{2.1}&=&\langle\partial_x,\ u\partial_u\rangle:\quad
F=uG(t,\omega),\ \omega=u^{-1}u_x; \\[1mm]
A^8_{2.1}&=&\langle\partial_x,\ \varphi(t)\partial_u\rangle\
(\varphi\not=0): \\[1mm] && F=\varphi^{-1}\varphi''u+G(t,u_x); \\[1mm]
A^{9}_{2.1}&=&\langle\partial_t,\ \partial_u\rangle:\quad
F = G(x,u_x);\\[1mm]
A^{10}_{2.1}&=&\langle\partial_t,\ f(x)u\partial_u\rangle\
(f\not=0):\\[1mm] && F=-u^{-1} u^2_x +u G(x,\omega);\\[1mm]
&& \omega=u^{-1}u_x-f'f^{-1}\ln|u|; \\[1mm]
A^{11}_{2.1}&=&\langle\partial_t+f(x)u\partial_u,\
g(x)u\partial_u\rangle\ (\delta=f^{-1}f'-g^{-1}g'\not =0):
\\[1mm] && F=-g^{-1}g''u\ln|u|-2 g^{-1}g'u_x\ln|u|
\\[1mm] && \quad + g^{-2}(g')^2u\ln^2|u|-2f\delta tu_x
+2f\delta g'g^{-1}tu\ln|u| \\[1mm]
&& \quad+f^2\delta^2t^2u+f(g^{-1}g''-f^{-1}f'')tu+u G(x,\omega),\\[1mm]
&& \omega=u^{-1}u_x-g'g^{-1}\ln|u|- t f\delta;\\[1mm]
A^{12}_{2.1}&=&\langle\partial_t+f(x)u\partial_u,\
e^{tf}\partial_u\rangle\ (f\not=0): \\[1mm]
&& F=[f^2-t f''+t^2(f')^2]u-2tf'u_x+e^{tf}G(x,\omega),\\[1mm] &&
\omega=e^{-tf}(u_x-tf'u);\\[1mm]
A^{13}_{2.1}&=&\langle f(x)u\partial_u,\ g(x)u\partial_u\rangle \
(\delta=f'g-g'f \not=0):\\[1mm]
&& F=-u^{-1}u^2_x-\delta^{-1}\delta' u_x \\[1mm]
&& \quad+\delta^{-1}[f''g'-g''f']u\ln|u|+u G(t,x);
\\[1mm]
A^{14}_{2.1}&=&\langle\varphi(t)\partial_u,\
\psi(t)\partial_u\rangle\ (\varphi'\psi-\varphi\psi'\not=0):\\[1mm]
&& F=\varphi^{-1}\varphi''u+G(t,x,u_x),\
\varphi''\psi-\varphi\psi''=0.
\end{eqnarray*}
{\bf II. $A_{2.2}$-invariant equations}
\begin{eqnarray*}
A^1_{2.2}&=&\langle t\partial_t+x\partial_x,\
xu\partial_u\rangle:\quad F=x^{-2}u\ln^2|u| \\[1mm]
 && \quad-2x^{-1}u_x\ln|u|+t^{-2}uG(\xi,\omega),\
\xi=tx^{-1};\\[1mm]
 && \omega=xu^{-1}u_x-\ln|u|; \\[1mm]
A^2_{2.2}&=&\langle t\partial_t+x\partial_x,\
t\varphi(\xi)\partial_u\rangle \ (\varphi\not=0,\
\xi=tx^{-1}):\\[1mm]
&& F=t^{-2}(1-\xi^2)\varphi^{-1}\xi(2\varphi'+\xi\varphi'')u
+t^{-2}G(\xi,\omega),\\[1mm]
 && \omega=x\varphi u_x+\xi\varphi'u;
\end{eqnarray*}
\begin{eqnarray*}
A^3_{2.2}&=&\langle\partial_t+k\partial_x,\
\exp({k^{-1}x})u\partial_u\rangle \ (k>0):\\[1mm]
&& F=k^{-2}u \ln^2|u|-2k^{-1}u_x\ln|u|-k^{-2} u \ln |u|\\[1mm]
&&+uG(\eta,\omega),\ \eta=x-kt,\ \omega=u^{-1}u_x-k^{-1}\ln|u|;\\[1mm]
A^4_{2.2}&=&\langle\partial_t+k\partial_x,\
e^t\varphi(\eta)\partial_u\rangle\ (\eta=x-kt,\ k>0,\ \varphi\not=0):\\[1mm]
&& F=\left((k^2-1)\varphi''\varphi^{-1}-2k\varphi'\varphi^{-1}
+1\right)u+G(\eta,\omega),\\[1mm]
 && \omega=\varphi u_x-\varphi'u,\
\varphi'=\frac{d\varphi}{d\eta};\\[1mm]
A^5_{2.2}&=&\langle -t\partial_t-x\partial_x,\
\partial_t+k\partial_x\rangle\ (k>0):\\[1mm]
&& F=\eta^{-2}G(u,\omega),\ \eta=x-kt, \ \ \omega = u_x \eta;\\[1mm]
A^6_{2.2}&=&\langle -t\partial_t-x\partial_x+mu\partial_u,\
\partial_t+k\partial_x\rangle\ (k>0,\ m\not=0):\\[1mm]
&& F=|\eta|^{-2-m}G(v,\omega),\ \eta=x-kt, \\[1mm]
&& \omega=u|\eta|^m,\ v=u_x |\eta|^{m+1};\\[1mm]
A^7_{2.2}&=&\langle\partial_x,\ e^xu\partial_u\rangle:\quad
F=u\ln^2|u|-u \ln |u| -2u_x\ln|u|\\[1mm]
&& +uG(t,\omega),\ \omega=u^{-1}u_x-\ln|u|; \\[1mm]
A^8_{2.2}&=&\langle\partial_x,\ e^x\varphi(t)\partial_u\rangle\
(\varphi\not=0):\\[1mm]
&& F=(\varphi^{-1}\varphi''-1)u+G(t,\omega),\ \omega=u_x-u;\\[1mm]
A^9_{2.2}&=&\langle-t\partial_t-x\partial_x,\
\partial_x\rangle:\quad F=t^{-2}G(u,t u_x);\\[1mm]
A^{10}_{2.2}&=&\langle -t\partial_t-x\partial_x+ku\partial_u,\
\partial_x\rangle,\ (k\not=0):\\[1mm]
&& F=|t|^{-2-k}G(v,\omega),\ v=|t|^k u,\ \omega=|t|^{k+1}u_x;\\[1mm]
A^{11}_{2.2}&=&\langle\partial_t,\ e^t\partial_u\rangle:\quad
F=u+G(x,u_x);\\[1mm]
A^{12}_{2.2}&=&\langle-t\partial_t-x\partial_x,\
\partial_t\rangle:\quad
F=x^{-2}G(u,\omega),\ \omega=xu_x;\\[1mm]
A^{13}_{2.2}&=&\langle\partial_t+f(x)u\partial_u,\
e^{(1+f)t}\partial_u\rangle\ (f\not=0):\\[1mm]
&& F=-\left(tf''-t^2(f')^2-(1+f^2)\right)u-2tf'u_x
\\[1mm]
  && +e^{tf}G(x,\omega),\
\omega=e^{-tf}\left(u_x-f'(t+f^{-1})u\right);\\[1mm]
A^{14}_{2.2}&=&\langle-t\partial_t-x\partial_x,\
\partial_t+kx^{-1}u\partial_u\rangle\ (k>0); \\[1mm]
&& F=-2ktx^{-3}u+k^2t^2x^{-4}u+2ktx^{-2}u_x\\[1mm]
 &&+ x^{-2}\exp({ktx^{-1}})
G(v,\omega),v=\exp({-kx^{-1}t})u,\\[1mm]
&& \omega=xu^{-1}u_x+\ln|u|;\\[1mm]
A^{15}_{2.2}&=&\langle k(t\partial_t+x\partial_x),\
|x|^{k^{-1}}u\partial_u\rangle\ (k\not=0,1): \\[1mm]
&& F=-k^{-2}(1-k)x^{-2}u\ln|u|-2k^{-1}x^{-1}u_x\ln|u|\\[1mm]
&& \quad+k^{-2}x^{-2}u\ln^2|u|+x^{-2}uG(v,\omega),\\[1mm]
&& v=tx^{-1},\ \omega=xu^{-1}u_x-k^{-1}\ln|u|;\\[1mm]
A^{16}_{2.2}&=&\langle k(t\partial_t+x\partial_x),\
|t|^{k^{-1}}\varphi(\xi)\partial_u\rangle\ (k\not=0,1,\
\varphi\not=0,\ \\ [1mm]
&& \xi=tx^{-1}):
F=[k^{-1}(k^{-1}-1)+2\xi(k^{-1}-\xi^2)\varphi^{-1}\varphi'\\[1mm]
 &&\quad+\xi^2(1-\xi)^2\varphi^{-1}\varphi'']t^{-2}u+
t^{-2}G(\xi,\omega),\\[1mm]
 && \omega=x\varphi u_x+\xi\varphi'u.
\end{eqnarray*}

In the above formulas $G$ stands for an arbitrary smooth function.
As usual, prime denotes the derivative of a function of one variable.

\subsection{Group classification of equation \\
 $u_{tt} = u_{xx} -u^{-1} u^2_x +A(x) u_x +B(x) u \ln |u| +u D(t,x)$}

Before analyzing equations (\ref{1.9}) admitting algebras of the
dimension higher than two we perform group classification of the
equation
\begin{equation} \label{2.5}
u_{tt} = u_{xx} -u^{-1} u^2_x +A(x) u_x +B(x) u \ln |u|+u D(t,x).
\end{equation}
Here $A(x), B(x), D(t,x)$ are arbitrary smooth functions. Note that the above
class of PDEs contains $A^{13}_{2.1}$-invariant equation. Importantly,
class (\ref{2.5}) contains a major part of equations of the form
(\ref{1.9}), whose maximal symmetry algebras have dimension three or
four. This fact is used to simplify group classification of equations
(\ref{1.9}).

\begin{lema} \label{l2.3.1}
If $A$, $B$ and $D$ are arbitrary, then the maximal invariance algebra
of PDE (\ref{2.5}) is the two-dimensional Lie algebra
equivalent to $A^{13}_{2.1}$ and (\ref{2.5}) reduces to
$A^{13}_{2.1}$-invariant equation. Next, if the maximal symmetry
algebra of an equation of the form (\ref{2.5}) is three-dimensional
(we denote it as $A_3$), then this equation is equivalent to one of
the following ones:
\begin{enumerate}
\renewcommand{\theenumi}{\Roman{enumi}}
\item $A_3 \sim A_{3.1}, \ A_3 = \langle \partial_t, f(x) u \partial_u,
\varphi(x) u \partial_u \rangle,$ \\[1mm]
$A=-\sigma^{-1} \sigma', \ B = \sigma^{-1} \rho, \ D =0, \sigma = f' \varphi -f
\varphi' \not =0,$ \\
$\rho = \varphi' f''-\varphi'' f';$

\item $A_3 \sim A_{3.1},  \ A_3 = \langle f(x) u \partial_u, \varphi(x) u
\partial_ u, \partial_t +\psi(x) u \partial_u \rangle,$\\
$A = -\sigma^{-1} \sigma', \ B = \sigma^{-1} \rho,$ \\
$ D = t \sigma^{-1} [\sigma' \psi' -\psi \rho-\sigma \psi''], $ \\
$ \sigma= f' \varphi -\varphi' f \not =0, \rho =f '' \varphi ' -\varphi'' f',$
\\
$ f' \psi -f \psi' \not =0, \ \ \varphi' \psi - \varphi \psi' \not =0; $
\item $D = x^{-2} G(\xi), \ \xi = tx^{-1}, \ G \not =0:$
\begin{enumerate}
\renewcommand{\labelenumii}{\arabic{enumii})}
\renewcommand{\theenumii}{\arabic{enumii}}
\item $A_3 \sim A_{3.2}, \ A_3 = \langle t \partial_t +x
\partial_x, u \partial_u, |x|^{1-n} u \partial_u \rangle, $
\vskip  2mm
$A = n x^{-1} \ (n \not =1), \ B=0$;
\item $A_3 \sim A_{3.3}, \ A_3 = \langle t \partial_t +x
\partial_x, u \partial_u, u \ln |x| \partial_u \rangle,$ $A =
x^{-1}, \ B =0;$
\vskip  2mm
\item $A_3 \sim A_{3.4}, A_3 = \langle t \partial_t +x \partial_x,
\sqrt{|x|} u \partial_u, \sqrt{|x|}\ln |x| u \partial_u \rangle,$
\vskip  2mm
$A=0,$ \ $ B = \frac{1}{4} x^{-2};$
\item $A_3 \sim A_{3.9}, A_3 = \langle t \partial_t +x \partial_x,
\sqrt{|x|} \cos (\frac{1}{2} \beta \ln |x|) u \partial_u, $ \vskip
2mm $\sqrt{|x|} \sin (\frac{1}{2} \beta \ln |x|) u  \partial_u
\rangle, $ $A =0, B = m x^{-2}, $ \vskip 2mm $ m>\frac{1}{4}, \
\beta = \sqrt{4m-1};$
\item $A_3 \sim A_{3.7}, A_3 = \langle t \partial_t +x \partial_x,
(\sqrt{|x|})^{1+\beta}  u \partial_u, (\sqrt{|x|})^{1-\beta}  u
\partial_u \rangle,  $
\vskip  2mm
 $ A=0, B = m x^{-2}, \ m < \frac{1}{4}, \ m \not =0, \ \beta =
 \sqrt{1-4m};$
\item $A_3 \sim A_{3.8}, A_3 = \langle t \partial_t +x \partial_x,
\cos (\sqrt{m}\ln |x|)   u \partial_u,  $
\vskip  2mm
$\sin (\sqrt{m} \ln |x|)   u \partial_u \rangle, \  A=x^{-1}, B =
m x^{-2}, \ m>0;$
\item $A_3 \sim A_{3.6}, A_3 = \langle t \partial_t+x \partial_x,
|x|^{\sqrt{|m|}} u \partial_u, |x|^{-\sqrt{|m|}} u \partial_u
\rangle, $ \vskip  2mm $A = x^{-1},$\ $ B = m x^{-2}, \ m<0;$
\item $A_3 \sim A_{3.4}, A_3 = \langle t \partial_t+x\partial_x,
(\sqrt{|x|})^{1-n} u \partial_u, (\sqrt{|x|})^{1-n}$
\vskip 2mm
$\times\ln|x| u
\partial_u \rangle,\ A = nx^{-1} \ (n \not =0,1), \ B = \frac{1}{4}(n-1)^2
x^{-2};$
\item $A_3 \sim A_{3.9}, A_3 = \langle t \partial_t +x \partial_x,
(\sqrt{|x|})^{1-n} \cos(\frac{1}{2} \beta \ln |x|)  u \partial_u,$
\vskip  2mm
$ (\sqrt{|x|})^{1-n} \sin (\frac{1}{2} \beta \ln |x|)  u
\partial_u \rangle,  $ $ A=n x^{-1} \ (n \not =0,1), \  $
\vskip  2mm
$ B = m x^{-2}\ (m > \frac{1}{4}(n-1)^2), \ \beta =
\sqrt{4m-(n-1)^2};$
\item $A_3 \sim A_{3.7}, A_3 = \langle t \partial_t +x \partial_x,
(\sqrt{|x|})^{1-\beta-n}  u \partial_u, (\sqrt{|x|})^{1-n+\beta}
$
\vskip  2mm
$\times u \partial_u \rangle,\ A=nx^{-1}\ (n \not =0,1),\
B = m x^{-2} $
\vskip  2mm
$(m < \frac{1}{4}(n-1)^2,\ m \not =0), \ \beta = \sqrt{(n-1)^2-4m}.$
\end{enumerate}
 \item $D = G(t),$
\begin{enumerate}
\renewcommand{\labelenumii}{\arabic{enumii})}
\renewcommand{\theenumii}{\arabic{enumii}}
\item  $A_3 \sim A_{3.3}, \ A_3 = \langle \partial_x, u
\partial_u, xu \partial_u \rangle, $
\vskip  2mm
$A = B =0;$
\item $A_3 = A_{3.2}, \ A_3 = \langle \partial_x, u \partial_u,
e^{x} u \partial_u \rangle,$
\vskip  2mm
$A = -1, \ B =0;$
\item $A_3 \sim A_{3.8}, \ A_3 = \langle \partial_x,\cos \, (x) u
\partial_u, \sin \,(x) u \partial_u \rangle, $\vskip  2mm $A =0, \
B = 1;$
\item $A_3 \sim A_{3.6}, \ A_3 = \langle \partial_x, e^{x} u
\partial_u, e^{-x} u \partial_u \rangle,$ \vskip  2mm $A = 0, \ B
= -1;$
\item  $A_3 \sim A_{3.4}, \ A_3 = \langle \partial_x,
\exp\left({\frac{1}{2}x}\right) u \partial_u, \exp\left({\frac{1}{2}x}\right) xu \partial_u
\rangle, $\vskip  2mm $A = -1, \ B = \frac{1}{4} ;$
\item $A_3 \sim A_{3.7}, \ A_3 = \langle \partial_x,
\exp\left({\frac{1}{2}(1+\beta)x}\right) u \partial_u,
\exp\left({\frac{1}{2}(1-\beta)x}\right)
u \partial_u \rangle, $\vskip  2mm $A = -1, B = m \
(m<\frac{1}{4}),  \ m \not =0, \beta = \sqrt{1-4m};$
\item $A_3 \sim A_{3.9}, \ A_3 = \langle \partial_x,
\exp\left({\frac{1}{2}x}\right) \cos (\frac{1}{2}\beta x) u \partial_u,
\exp\left({\frac{1}{2}x}\right)\sin(\frac{1}{2}\beta x)u \partial_u \rangle,
$\vskip  2mm $A = -1, B = m \ (m>\frac{1}{4} ),\ \beta = \sqrt{4m
- 1};$
\end{enumerate}
 \item $D = G(\eta), \ \eta = x-kt, \ k>0,$
\begin{enumerate}
\renewcommand{\labelenumii}{\arabic{enumii})}
\renewcommand{\theenumii}{\arabic{enumii}}
\item  $A_3 \sim A_{3.3}, \ A_3 = \langle \partial_t +k\partial_x,
u \partial_u, xu \partial_u \rangle, $\vskip  2mm $A = B =0;$
\item $A_3 = A_{3.2}, \ A_3 = \langle \partial_t +k\partial_x, u
\partial_u, e^{x} u \partial_u \rangle,$ \vskip  2mm $A = -1,\  B
=0;$
\item $A_3 \sim A_{3.8}, \ A_3 = \langle \partial_t +k\partial_x,
\cos \, (x) u \partial_u, \sin \,(x) u \partial_u \rangle, $\vskip
2mm $A =0, \ B = 1;$
\item $A_3 \sim A_{3.6}, \ A_3 = \langle \partial_t +k\partial_x,
e^{x} u \partial_u, e^{-x} u \partial_u \rangle,$ \vskip  2mm $A =
n, \ B =  -1;$
\item  $A_3 \sim A_{3.4}, \ A_3 = \langle \partial_t +k\partial_x,
\exp\left({\frac{1}{2}x}\right) u \partial_u,
\exp\left({\frac{1}{2}x}\right) xu \partial_u
\rangle, $\vskip  2mm $A = -1,\  B = \frac{1}{4};$
\item $A_3 \sim A_{3.7}, \ A_3 = \langle \partial_t +k\partial_x,
\exp\left({\frac{1}{2}(1+\beta)x}\right) u \partial_u,
\exp\left({\frac{1}{2}(1-\beta)x}\right)
u \partial_u \rangle, $\vskip  2mm $A = -1, B = m \
(m<\frac{1}{4}),  \ m \not =0, \beta = \sqrt{1-4m};$
\item $A_3 \sim A_{3.9}, \ A_3 = \langle \partial_t +k \partial_x,
\exp\left({\frac{1}{2}x}\right) \cos (\frac{1}{2}\beta x) u \partial_u, $\vskip
2mm $\exp\left({\frac{1}{2}x}\right)\sin(\frac{1}{2}\beta x)u \partial_u \rangle,
\ A = -1, \ B = m \ (m>\frac{1}{4} )  \ \beta = \sqrt{4m - 1}.$
\end{enumerate}
\end{enumerate}
\end{lema}

\noindent {\it Proof.}\  Inserting the expression
$$
F = -u^{-1} u^2_x +A(x) u_x + B(x)u \ln |u|+u D(t,x)
$$
into classifying equation (\ref{1.7}) we get the system of
determining equations for the functions $h(x)$, $r(t,x)$ and
constants $\lambda, \lambda_1, \lambda_2$:
\begin{equation} \label{2.6}
\begin{array}{l}
r=0,\ \
(\lambda x +\lambda_2) A' +\lambda A=0, \\[2mm]
(\lambda x +\lambda_2) B' +2\lambda B=0, \ \
h''+Ah'+Bh=-(\lambda t +\lambda_1) D_t -(\lambda x+\lambda_2) D_x
-2 \lambda D.
\end{array}
\end{equation}

First, consider the case of arbitrary functions $A, B, D$. The left-hand side of
the fourth equation of (\ref{2.6}) depends on $x$ only. What is more, since
$D$ is arbitrary, relation $D_t\not \equiv 0$ holds. Hence it immediately
follows that the constants $\lambda, \lambda_1, \lambda_2$ must be equal to
zero. As a consequence, the fourth equation becomes linear ordinary
differential equation for the function $h(x)$
\begin{equation} \label{2.7}
h''+ Ah' + Bh =0.
\end{equation}
The general solution of the above equation reads as
$$
h = C_1 f(x) +C_2 \varphi(x), \ \ C_1, C_2 \in {\mathbb R},
$$
$f(x)$ and $\varphi(x)$ being the fundamental system of solutions of
the equation
\begin{equation} \label{2.8}
y'' + A y' + B y =0, \ \ y = y(x).
\end{equation}
Inserting this expression into (\ref{2.7}) yields
$$
A = - \sigma^{-1} \sigma',\ B = \sigma^{-1}(\varphi' f'' - f'
\varphi''),
$$
where $\sigma = \varphi f' - \varphi' f\ne 0$, which proves the first part
of lemma.

Suppose now that $D=0.$ Then, if at least one of the functions
$A$ or $B$ is arbitrary, then $\lambda = \lambda_2 = 0$ and the
function $h$ is a solution of (\ref{2.7}). This completes the proof
of the case I of the second part of the lemma statement.

Provided functions $A$ and $B$ are not arbitrary, it follows
from the second and third equations of (\ref{2.6}) that one of the following
relations
\begin{equation} \label{2.9}
\begin{array}{l l}
1)&A = B =0; \\[2mm]
2)& A = n, \ B =m, \ m,n \in {\mathbb R},\ |n|+|m| \not =0; \\[2mm]
3)& A = n x^{-1}, \ B = m x^{-2},\ m,n \in {\mathbb R}, \ |n|+|m| \not =0
\end{array}
\end{equation}
holds. With these conditions the maximal invariance algebra of (\ref{2.5}) has
the dimension higher than three. Consequently, without loss of generality we
can suggest that $D \not =0$. Integrating the equation
$$
(\lambda t +\lambda_1) D_t +(\lambda x +\lambda_2) D_x +
2 \lambda D = H(x),
$$
under $D \not =0$ yields the following (inequivalent) expressions for
the function $D(t,x):$
\begin{equation} \label{2.10}
\begin{array}{l c l}
D&=& x^{-2} G(\xi) +x^{-2} \int x H(x) \,dx, \ \xi = t x^{-1}; \\[1mm]
D&=& G(\eta) +k^{-1} \int H(x) \, dx, \eta = x-kt, \ k>0; \\[1mm]
D&=& G(t)+\int H(x) \, dx, \\[1mm]
D&=& t H(x) +\tilde H(x).
\end{array}
\end{equation}

The change of variables
\begin{equation} \label{2.11}
\overline{t} =t, \ \ \overline{x} = x, \ \ u = \theta(x)
v(\overline{t}, \overline{x}), \ \ \theta \not =0,
\end{equation}
where $\theta$ is a solution of equation
$$
\theta^{-1} \theta'' -\theta^{-2} (\theta')^2 +A \theta^{-1}
\theta'+B \ln |\theta| +\Lambda (x) =0,
$$
preserves the form of equation (\ref{2.5}). We can use this fact
to simplify the form of the function $D$. As a result, we get
\begin{equation} \label{2.12}
\begin{array}{lcl}
D&=& x^{-2} G(\xi), \ \xi = t x^{-1}; \\[2mm]
D&=& G(\eta),\ \eta = x-kt, \ k>0; \\[2mm]
D&=& G(t), \\[2mm]
D&=& t H(x).
\end{array}
\end{equation}

If the function $D$ is given by the one of the first three expressions, then
$H(x)$ $\equiv$ $0$ and $h$ satisfies (\ref{2.7}).

Given the condition $D=tH(x),$ we have
$$
h'' +A h' +B h = -\lambda_1 H,\quad
(\lambda x +\lambda_2)H' +3 \lambda H =0.
$$
So that the maximal invariance algebra of the corresponding equation
(\ref{2.5}) is three-dimensional iff $\lambda = \lambda_2 =0,$ which yields the
case II of the second part of the lemma statement.

Turn now to the case when $D=x^{-2}G(\xi),\ \xi = t x^{-1}$. Then the
function $G \not =0$ obey the equation
\begin{equation} \label{2.13}
(\lambda_2 \xi -\lambda_1)G' +2\lambda_2 G =0.
\end{equation}
If $G$ is an arbitrary function, then $\lambda_1 = \lambda_2 =0$.
In addition, we have $\lambda \not =0$ (otherwise the maximal invariance
algebra is two-dimensional). Hence we get
$$
x A' +a =0,\quad x B' +2B =0.
$$
Consequently, functions $A$ and $B$ are given by either first
or third formula from (\ref{2.9}). Analyzing these expressions
yields ten cases of the case III of the second part of the lemma
statement.

If the function $G$ is not arbitrary, then integrating (\ref{2.13})
we get
$$\begin{array}{l}
G =p, \ \ p\in {\mathbb R},  \ p \not =0; \cr
G = p(\xi -q)^{-2}, \ p \not =0, \ q\geq 0.
\end{array} $$

Given the condition $G =p$, the parameter $\lambda_2$ vanishes.
Hence in view of the requirement for the maximal algebra to
be three-dimensional, it follows that $\lambda$ vanishes as well.
This yields the case when $A$ and $B$ in (\ref{2.5}) are arbitrary
functions (the case I of the second part of the lemma statement).
If $G = p(\xi-q)^{-2},\ p \not =0,$ then $\lambda_1 = \lambda_2 q$.
Hence we conclude that the maximal invariance algebra of the corresponding
equation (\ref{2.5}) is three-dimensional iff the functions $A, B$ are
given by formulas 3) from (\ref{2.9}) (which implies that
$\lambda_1 = \lambda_2 =0$) and we get the case III of the second
part of the lemma statement. Next, if $A, B$ are given by formulas 2 from
(\ref{2.9}) (which implies that $\lambda =0, \ D = p(t-qx)^{-2}$)
and we arrive at the case IV $(q=0)$ or the case V $(q >0)$ of the
second part of the lemma statement.

Turn now to the case $D = G(\eta), \ \eta = x-kt, \ k>0.$ If these
relations hold, then
$$\lambda(\eta G' + 2G) +(\lambda_2 -k \lambda_1) G'=0.
$$
Hence it follows that if $G$ is an arbitrary function of $\eta,$
then $\lambda =0,$ $\lambda_2 = k \lambda_1$. That is why, the maximal
invariance algebra of (\ref{2.5}) is three-dimensional iff
either $A=B=0$ or $A, B$ are given by formulas 2 from
(\ref{2.9}). So that we have obtained all equations listed in the case
V of the second part of the lemma statement.

The cases when either $G = p \ (p \not =0)$ or $G = p \eta^{-2} \
(p \not =0)$ yield no new invariant equations (\ref{2.9}).

Consider now the last possible case $D = G(t)$. If this is the case,
then the equation
$$
(\lambda t +\lambda_1) G' +2 \lambda G =0
$$
holds. Hence if follows that if $G$ is an arbitrary function,
then $\lambda = \lambda_1 = 0$. So that the maximal invariance algebra
of equation (\ref{2.5}) is three-dimensional iff $A, B$ are given
by the formula 2 from (\ref{2.9}) and we have obtained all the
invariant equations from the case IV of the second part of assertion
of the lemma. If either of relations $G=p\ (p \not =0)$ or
$G = p t^{-2} \ (p \not =0)$ hold, them no new invariant equations
having three-dimensional maximal invariance algebras can be obtained.

To complete the proof of the lemma, we need to establish non equivalence of
the obtained invariant equations. To this end it suffices to
prove that there are no transformations from the group ${\cal E}$, reducing
their invariance algebras one into another.

As we already mentioned in Section 3 there exist nine non-isomorphic
three-dimensional solvable Lie algebras $A_{3.i} = \langle e_1, e_2,
e_3 \rangle \ (i=1,2, \ldots, 9)$. We analyze in some detail the case
of the algebra $A_{3.3}$. The list of invariant equations and algebras
contains three algebras which are isomorphic to $A_{3.3}$, namely,
\begin{eqnarray*}
L_1 &=& \langle t \partial_t +x \partial_x, u \partial_u, u \ln |x| \partial_u
\rangle; \\
L_2 &=& \langle \partial_x ,u \partial_u, xu \partial_u \rangle; \\
L_2&=& \langle \partial_t +k \partial_x, u \partial_u, xu \partial_u \rangle \
(k>0).
\end{eqnarray*}
Denote  the basis elements of the algebra $L_2$ as $e_1, e_2, e_3$. Suppose
that there is a transformation $\varphi$ from the group ${\cal E}$ transforming
$L_2$ into $L_3$. In other words we suppose that there exist constants
$\alpha_i, \beta_i, \delta_i \in  R \ (i=1,2,3)$ such that the relations
\begin{eqnarray*}
\varphi(e_1)&=& \sum^3_{i=1} \alpha_i {\tilde e}_i, \ \
\varphi(e_2)= \sum^3_{i=1} \beta_i {\tilde e}_i, \ \
\varphi(e_3)= \sum^3_{i=1} \delta_i {\tilde e}_i
\end{eqnarray*}
and
$$\bigtriangleup =\left |
\begin{array}{ccc}
\alpha_1&\alpha_2&\alpha_3 \cr
\beta_1&\beta_2&\beta_3 \cr
\delta_1&\delta_2&\delta_3
\end{array}
\right | \not =0$$
hold. In the above formulas
${\tilde e}_1 = \partial_{\overline{t}} +k \partial_{\overline{x}}, \ {\tilde
e}_2 = v \partial_v, \ {\tilde e}_3 = \overline{x} v \partial_v.$ Equating the
coefficients of the linearly independent operators $\partial_{\overline{t}}$,
$\partial_{\overline{x}}, \partial_{{v}}$ yields that $\alpha_1 = \beta_1 = \delta_1 =0.$ Hence we get the contradictory equation $\bigtriangleup =0$. This
means that realizations $L_2$ and $L_3$ are non-isomorphic. Analogously, we
prove that $L_1$ and $L_2$ (as well as $L_1$ and $L_3$) are non isomorphic.

The remaining algebras are considered in a similar way. The Lemma is proved.

In what follows we will use the results on classification of abstract
four-dimensional solvable real Lie algebras
$A_4 = \langle e_1, e_2, e_3, e_4 \rangle$ \cite{magda22,magda21}.
There are ten decomposable
\begin{eqnarray*}
&& 4 A_1 = 3 A_1 \oplus A_1=A_{3.1} \oplus A_1,\ \ A_{2.2} \oplus 2 A_1 =
A_{2.2} \oplus A_{2.1} = A_{3.2} \oplus A_1, \\
&& 2 A_{2.2} = A_{2.2} \oplus A_{2.2}, \ \ A_{3.i} \oplus A_1 \ (i=3,4, \ldots,
9);
\end{eqnarray*}
and ten non-decomposable four-dimensional solvable real Lie algebras
(note that we give below non-zero commutation relations only).
\begin{eqnarray*}
A_{4.1} &:& [e_2, e_4] = e_1, \ \ [e_3, e_4] =e_2; \\
A_{4.2} &:& [e_1, e_4] = q e_1, \ \ [e_2, e_4] = e_2, \\
&& [e_3, e_4] = e_2+e_3, \ \ q\not =0; \\
A_{4.3} &:& [e_1, e_4] = e_1, \ \ [e_3, e_4] = e_2; \\
A_{4.4} &:& [e_1, e_4] = e_1, \ \ [e_2, e_4] = e_1+e_2, \\
&& [e_3, e_4] = e_2 +e_3; \\
A_{4.5} &:& [e_1, e_4] = e_1, \ \ [e_2, e_4] = qe_2, \\
&& ]e_3, e_4] = p e_3, \ \ -1 \leq p \leq q\leq 1, \ \ p \cdot q \not =0; \\
A_{4.6} &:& [e_1, e_4] = q e_1, \ \ [e_2, e_4] = p e_2 -e_3, \\
&& [e_3, e_4] = e_2 +p e_3, \ \ q\not =0, \ \ p\geq 0; \\
A_{4.7} &:& [e_2, e_3] = e_1, \ \ [e_1, e_4] = 2 e_1, \\
&& [e_2, e_4] = e_2, \ \ [e_3, e_4] = e_2 +e_3; \\
A_{4.8} &:& [e_2, e_3] = e_1, \ \ [e_1, e_4] = (1+q) e_1, \\
&& [e_2, e_4] = e_2, \ \  [e_3, e_4] = q e_3, \ \ |q| \leq 1; \\
A_{4.9} &:& [e_2, e_3] = e_1, \ \ [e_1, e_4] =2 q e_1, \\
&& [e_2, e_4] = q e_2 -e_3, \ \ [e_3, e_4] = e_2 +q e_3, \ \ q \geq 0; \\
A_{4.10} &:& [e_1, e_3] = e_1, \ \ [e_2, e_3] = e_2, \\
&& [e_1, e_4] = -e_2, \ \ [e_2, e_4] = e_1.
\end{eqnarray*}
\begin{theorem} \label{theo2.3.1}
Equation $ u_{tt} = u_{xx} - u^{-1} u^2_x$ has the widest symmetry
group amongst equations of the form (\ref{2.5}). Its maximal invariance
algebra is the five-dimensional Lie algebra
$$
A^1_5 = \langle \partial_t, \partial_x, t \partial_t+x \partial_x, xu
\partial_u, u \partial_u \rangle.
$$
There are no equations of the form (\ref{2.5}) which are
inequivalent to the above equation and admit invariance algebra of
the dimension higher than four. Inequivalent equations (\ref{2.5})
admitting four-dimensional algebras are listed below together with
their symmetry algebras.
\begin{enumerate}
\renewcommand{\theenumi}{\Roman{enumi}}
\item $D=0,$
\begin{enumerate}
\renewcommand{\labelenumii}{\arabic{enumii})}
\renewcommand{\theenumii}{\arabic{enumii}}

\item $A_4 \sim A_{3.6} \oplus A_1, \ \ A_4 = \langle \partial_t, \partial_x,
u\, {\rm ch} (\beta x) \partial_u, u \sinh (\beta x) \partial_u \rangle,$
\\[2mm]
$A=0, B = -\beta^2, \ \beta \not =0;$

\item $A_4 \sim A_{3.8} \oplus A_1, \ \ A_4 = \langle \partial_t, \partial_x,
u\cos (\beta x) \partial_u, u \sin (\beta x) \partial_u \rangle,$ \\[2mm]
$A=0, B =\beta^2, \ \beta \not =0;$

\item $A_4 \sim A_{2.1} \oplus A_{2.2}, \ A_4 = \langle \partial_t, \partial_x,
u \partial_u, e^{-x} u \partial_u \rangle, A=1, B=0;$

\item $A_4 \sim A_{3.4} \oplus A_1, \ A_4 = \langle \partial_t, \partial_x,
e^{-x} u \partial_u, x e^{-x} u \partial_u \rangle,  \ A=2, B=1;$

\item $A_4 \sim A_{3.9} \oplus A_1, \ A_4 = \langle \partial_t, \partial_x, u
e^{-x} \cos (\beta x) \partial_u, u e^{-x}\sin  (\beta x) \partial_u \rangle, $
\\[2mm]
$ A=2, B =m, m>1, \ \beta= \sqrt{m-1};$

\item $A_4 \sim A_{3.7} \oplus A_1, \ A_4 = \langle \partial_t, \partial_x, u
e^{-x} {\rm ch} (\beta x) \partial_u, u e^{-x} \sinh (\beta x) \partial_u
\rangle, $ \\[2mm]
$ A=2, \ B =m, \ m>1, \ m \not =0, \beta = \sqrt{1-m}; $

\item $A_4 \sim A_{4.2}, \ A_4 = \langle \partial_t, t \partial_t+ x\partial_x,
\sqrt{|x|} u  \partial_u, u \sqrt{|x|} \ln |x|\partial_u  \rangle,$ \\ $ \  A=0,
\ B =\frac{1}{4} x^{-2}; $

\item $A_4 \sim A_{4.5}, \ A_4 = \langle \partial_t, t \partial_t+x \partial_x,
|x|^{\frac{1}{2} +\beta} u \partial_u, |x|^{\frac{1}{2} -\beta} u \partial_u
\rangle, \ A=0,$ \\[2mm]
$ B = m x^{-2} ,  \ m<\frac{1}{4}, \ \ m \not =0, \ \beta =
\sqrt{\frac{1}{4}-m}; $

\item $A_4 \sim A_{4.6}, \ A_4 = \langle \partial_t, t \partial_t+x\partial_x,$
\\[2mm]
 $ \sqrt{|x|} \cos (\beta \ln |x| ) u \partial_u, \sqrt{|x|} \sin (\beta \ln
|x|) u \partial_u \rangle, $ \\[2mm]
$ A=0, \ B = m x^{-2}, \ m > \frac{1}{4}, \  \beta = \sqrt{m-\frac{1}{4}}; $

\item $A_4 \sim A_{4.3}, \ A_4 = \langle \partial_t, t \partial_t +x \partial_x,
u \ln |x| \partial_u, u \partial_u \rangle, \ A = x^{-1}, \ B=0;$

\item $A_4 \sim A_{3.7} \oplus A_1, \ A_4 = \langle \partial_t, t \partial_t +x
\partial_x, |x|^{1-n} u \partial_u, u \partial_u \rangle,$ \\ $ \ A= n x^{-1}, \
B =0, n \not =0,1;$

\item $A_4 \sim A_{4.5}, \ A_4 = \langle  \partial_t, t  \partial_t+x
\partial_x, |x|^{\frac{1}{2}(1-n)} u  \partial_u, |x|^{\frac{1}{2}(1-n)} u \ln
|x|  \partial_u \rangle,$ \\[2mm]
$ A=nx^{-1}, \ B = \frac{1}{4} (n-1)^2 x^{-2}, \ n \not =0,1;$

\item $A_4 \sim A_{4.5}, \ A_4 = \langle  \partial_t, t  \partial_t+x
\partial_x, |x|^{\frac{1}{2}(1-n +\beta)} u  \partial_u,
|x|^{\frac{1}{2}(1-n-\beta) }u  \partial_u \rangle,$ \\[2mm]
$ A = nx^{-1}, \ B = m x^{-2}, \ m<\frac{1}{4} (n-1)^2, \ m\not =0, \ n \not
=0,$ \\
 $  \beta = \sqrt{(n-1)^2 -4m};$

\item $A_4 \sim A_{4.6}, \ A_4 = \langle  \partial_t, t  \partial_t+x
\partial_x,  |x|^{\frac{1}{2}(1-n)} \cos (\beta \ln |x| )u  \partial_u ,$ \\
$  |x|^{\frac{1}{2}(1-n)} \sin (\beta \ln |x|) u  \partial_u \rangle, $ \ \ $
A=n x^{-1}, \ B =m x^{-2},$ \\
$ m \not =0, \ n \not =0, \ m>\frac{1}{4}(n-1)^2, \ \beta =
\sqrt{m-\frac{1}{4}(n-1)^2};$
\end{enumerate}
\item $D = ktx^{-3}, \ k>0,$
\begin{enumerate}
\renewcommand{\labelenumii}{\arabic{enumii})}
\renewcommand{\theenumii}{\arabic{enumii}}

\item $A_4 \sim A_{4.1}, \ A_4 = \langle \partial_t-\frac{1}{2} k x^{-1} u
\partial_u, t \partial_t+x\partial_x, xu \partial_u, u \partial_u \rangle, \
A=B=0;$

\item $A_4 \sim A_{4.2}, \ A_4 = \langle \partial_t-\frac{4}{9} k x^{-1} u
\partial_u, t \partial_t+x \partial_x, \sqrt{|x|} u \partial_u, \sqrt{|x|} \ln
|x| u \partial_u \rangle,$ \\[2mm]
$A=0, \ B = \frac{1}{4} x^{-2};$

\item $A_4 \sim A_{4.5}, \ A_4 = \langle \partial_t-\frac{k}{m+2} x^{-1} u
\partial_u, t \partial_t+x \partial_x, |x|^{\frac{1}{2}+\beta} u \partial_u,
|x|^{\frac{1}{2}-\beta} u \partial_u \rangle,$ \\[2mm]
$A =0, \ B = m x^{-2}, \ m \not =0,-2,  \ m <\frac{1}{4}, \ \beta =
\sqrt{\frac{1}{4}-m};$

\item $A_4 \sim A_{4.2}, \ A_4 = \langle \partial_t+\frac{1}{9} k x^{-1}(1+3 \ln
|x| u) \partial_u, t \partial_t+x \partial_x, $ \\[2mm]
$ x^2 u \partial_u, x^{-1} u \partial_ u \rangle, \ \ A=0, B = -2 x^{-2};$

\item $A_4 \sim A_{4.6},  \ A_4 = \langle \partial_t-\frac{k}{m+2} x^{-1} u
\partial_u, t \partial_t+x\partial_x, \sqrt{|x|} u \cos (\beta \ln |x|)
\partial_u, $ \\ $ \sqrt{|x|} u \sin (\beta \ln |x|) \partial_u \rangle, \ A=0,
B = m x^{-2}, \ m>\frac{1}{4}, \beta = \sqrt{m-\frac{1}{4}};$

\item $A_4 \sim A_{4.3}, \ A_4 = \langle \partial_t-k x^{-1} u \partial_u, t
\partial_t+x\partial_x, u \partial_u, u \ln |x| \partial_u \rangle,$ \\
$ A = x^{-1}, B=0;$

\item $ A_4 \sim A_{3.4} \oplus A_1, \ A_4 = \langle \partial_t+kx^{-1}(1+\ln
|x| )u \partial_u, t \partial_t+x \partial_x, u \partial_u, $ \\ $ x^{-1} u
\partial_u \rangle, \ \ A=2 x^{-1}, \ B=0;$

\item $A_4 \sim A_{3.7} \oplus A_1, \ A_4 = \langle \partial_t+\frac{k}{n-2}
x^{-1} u \partial_u, t \partial_t+x\partial_x, u \partial_u, |x|^{1-n} u
\partial_u \rangle,$\\[2mm]
 $A = n x^{-1}, B = 0, \ n \not =0,1,2;$

\item $A_4 = A_{4.4}, \ A_4= \langle \partial_t-\frac{1}{2}k x^{-1} \ln^2 |x| u
\partial_u, t \partial_t+x \partial_x,$ \\
$  x^{-1} u \partial_u, x^{-1} \ln |x| u \partial_u \rangle,$ \ \
$ A=3 x^{-1}, \ B=x^{-2};$

\item $A_4 \sim A_{4.2}, \ A_4 = \langle \partial_t-\frac{4k}{(n-3)^2}x^{-1} u
\partial_u, t \partial_t+x \partial_x, |x|^{\frac{1}{2}(1-n)} u \partial_u,$ \\
$  |x|^{\frac{1}{2}(1-n)} \ln |x| u \partial_u \rangle,$ \ \
$A = nx^{-1}, \ B = \frac{1}{4}(n-1)^2 x^{-2}, \ n \not =0,3;$

\item $A_4 \sim A_{4.5}, \ A_4 = \langle t \partial_t+x\partial_x,
\partial_t-\frac{k}{2-n+m} x^{-1} u \partial_u, |x|^{\frac{1}{2}(1-n+\beta)} u
\partial_u,$ \\
$  |x|^{\frac{1}{2} (1-n-\beta)} u \partial_u \rangle,$ \ \ $ A = n x^{-1}, \ B
= m x^{-2},$ \\
$  \ n \not =0,2, \ m \not =n-2, \ m <\frac{1}{4}(n-1)^2,$ \ \ $\beta =
\sqrt{(n-1)^2-4m};$

\item $A_4 \sim A_{4.2}, \ A_4 = \langle t \partial_t+x\partial_x,
\partial_t+\frac{k}{3-n} x^{-1} \ln |x| u \partial_u, x^{-1} u \partial_u,
|x|^{2-n} u \partial_u \rangle,$ \\[2mm]
$ A=n x^{-1}, \ B = (n-2)x^{-2}, \ n \not =0, 2, 3;$

\item $A_4 \sim A_{4.6}, \  A_4 = \langle t \partial_t+x \partial_x,
\partial_t-\frac{k}{2-n+m} x^{-1} u \partial_u,$ \\
$  |x|^{\frac{1}{2}(1-n) }u \cos (\beta \ln |x|) \partial_u,
|x|^{\frac{1}{2}(1-n)} u \sin (\beta \ln |x|) \partial_u \rangle,$ \\[1mm]
$ A=nx^{-1}, \ B = mx^{-2}, n \not =0, \ m \not =0, \ m>\frac{1}{4}(n-1)^2,$ \\
$\beta = \sqrt{m-\frac{1}{4}(n-1)^2};$
\end{enumerate}
\item $D = kt, \ k>0,$
\begin{enumerate}
\renewcommand{\labelenumii}{\arabic{enumii})}
\renewcommand{\theenumii}{\arabic{enumii}}

\item $A_4 \sim A_{4.1}, \ A_4 = \langle \partial_x, \partial_t-\frac{1}{2} kx^2
u \partial_u, xu \partial_u, u \partial_u \rangle, \ A=B=0;$

\item $A_4 \sim A_{4.3}, \ A_4 = \langle \partial_x, \partial_t-k x u\partial_u,
e^{-x} u \partial_u, u\partial_u \rangle, \ A=1, B=0;$

\item $A_4 \sim A_{3.8} \oplus A_1, \ A_4 = \langle \partial_x, \partial_t- k
\beta^{-2} u \partial_u, u \cos (\beta x) \partial_u, u \sin (\beta x)
\partial_u \rangle,$ \\[2mm]
$ A=0, \ B=\beta^2, \ \beta\not=0;$

\item $A_4 \sim A_{3.6} \oplus A_1, \ A_4 =\langle \partial_x, \partial_t+k
\beta^{-2} u \partial_u, u {\rm ch} (\beta x) \partial_u, u \sinh(\beta x)
\partial_u \rangle,$ \\[2mm]
$ A=0, \ B =-\beta^2, \ \beta \not =0;$

\item $A_4 \sim A_{3.4} \oplus A_1, \ A_4 = \langle \partial_x, \partial_t-4k
u\partial_u, \exp\left({-\frac{1}{2}x}\right) u \partial_u, x
\exp\left({-\frac{1}{2}x}\right) u \partial_u \rangle,$ \\
$  A=1, \ B = \frac{1}{4};$

\item $A_4 \sim A_{3.7} \oplus A_1, \ A_4 = \langle \partial_x,
\partial_t-km^{-1} u \partial_u, \exp\left({-\frac{1}{2}(1-\beta) x}\right) u \partial_u,
\exp\left({-\frac{1}{2} (1+\beta) x}\right) u \partial_u \rangle,$ \\[2mm]
$ A=1, \ B=m, \ m<\frac{1}{4}, \ m\not =0, \ \beta = \sqrt{1-4m};$

\item $A_4 \sim A_{3.9} \oplus A_1, \ A_4 = \langle \partial_x,
\partial_t-km^{-1} u \partial_u, \exp\left({-\frac{1}{2}x}\right) \cos (\beta x) u
\partial_u,$\\
$  \exp\left({-\frac{1}{2}x}\right) \sin (\beta x) u \partial_u \rangle, $ \\[2mm]
$ A=1, \ B =m, \ m>\frac{1}{4}, \ \beta = \sqrt{m-\frac{1}{4}};$

\end{enumerate}

\item $D = k t^{-2}, \ k \not =0,$ \\[2mm]
$ A_4 \sim A_{4.8} \ (q=-1), \ A_4 = \langle \partial_x, t \partial_t+x
\partial_x, xu \partial_u, u \partial_u \rangle, \ A=B=0;$

\item $D = m(x-kt)^{-2}, \ k>0, \ m \not =0,$ \\[2mm]
$A_4 \sim A_{4.8} \ (q=-1), \ A_4 = \langle \partial_t+k \partial_x, t
\partial_t+x \partial_x, xu \partial_u, u \partial_u \rangle, \ A=B=0.$
\end{enumerate}
\end{theorem}
{\it Proof.} According to Lemma \ref{l2.3.1} to get the list of inequivalent
equations of the form (\ref{2.5}) we need to analyze the cases when either
$D=0$ or $D$ does not vanish identically and is obtained through equations
(\ref{2.12}).

If $D=0,$ then function $h$ satisfies equation (\ref{2.7}) and
functions $A, B$ are given by one of the formulas (\ref{2.9}). It
follows from (\ref{2.6}) that the highest possible dimension of an
invariance algebra admitted by (\ref{2.5}) equals to five. An
equation admitting this algebra is equivalent to the following one:
\begin{equation} \label{2.14}
u_{tt} = u_{xx} - u^{-1} u^2_x.
\end{equation}
This proves the first part of the assertion of theorem.

The remaining expressions for the functions $A, B$ from (\ref{2.9}) yield the
fourteen items of the case I of the second assertion of theorem.

The fourth expression for the function $D$ given in (\ref{2.12}) within
the equivalence relation (\ref{2.11}) boils down to either $D = ktx^{-3},\
k>0$ or $D = kt,\ k>0$. Here the function $h$ satisfy one of the equations
$$
h'' +Ah' +B h = -\lambda_1 H,
$$
where $H = k x^{-3}$ or $H=k.$ Analyzing of the corresponding expressions
for functions $A,B$ \ (\ref{2.9}) results in the cases listed in the cases
II and III.

Next, the function $D$ given by the third formula from (\ref{2.12})
simplifies to $D = k t^{-2} \ (k \not =0),$ whence we get the results
listed in the case IV of the second assertion of theorem. Similarly,
the second expression for the function $D$ gives rise to formulas
of the case V.

The first expression for $D$ from (\ref{2.12}) gives no new invariant
equations.

What is left is to prove that the so obtained invariant equations
are inequivalent. We omit the proof of this fact.

Theorem is proved.

\subsection{Nonlinear equations (\ref{1.9}) invariant under
three-dimensional Lie algebras}

The class of PDEs (\ref{1.9}) does not contain an equation whose
invariance algebra is isomorphic to a Lie algebra with a non-trivial
Levi ideal (Theorem \ref{theo2.1.2}). That is why, to complete
the second step of our classification algorithm it suffices to
consider three-dimensional solvable real Lie algebras. We begin by
considering two decomposable three-dimensional solvable Lie algebras.

Note that while classifying invariant equations (\ref{1.9}) we skip those
belonging to the class (\ref{2.5}), since the latter has already been
analyzed.

\subsubsection{ Invariance under decomposable Lie algebras }

As $A_{3.1} = 3 A_1 = A_{2.1} \oplus A_1, \ A_{3.2} = A_{2.2} \oplus A_1,$
to construct all realizations of $A_{3.1}$ it suffices to compute all
possible extensions of the (already known) realizations of the algebras
$A_{2.1} = \langle e_1, e_2 \rangle$ and $A_{2.2} =
\langle e_1, e_2 \rangle$. To this end we need to supplement the latter
by a basis operator $e_3$ of the form (\ref{1.6}) in order to satisfy
the commutation relations
\begin{equation} \label{2.15}
[e_1, e_3] = [e_2, e_3] =0.
\end{equation}
What is more, to simplify the form of $e_3$ we may use those transformations
from ${\cal E}$ that do not alter the remaining basis operators of the
corresponding two-dimensional Lie algebras.

We skip the full calculation details and give several
examples illustrating the main calculation steps needed to
extend $A_{2.1}$ to a realization of $A_{3.1}.$

Consider the realization $A^1_{2.1}$. Upon checking commutation
relations (\ref{2.15}), where $e_3$ is of form (\ref{1.6}), we get
$$\lambda_1 = \lambda_2 = r(t,x) =0, \ \ \ h=k={\rm const}.$$
Consequently, $e_3$ is the linear combination of $e_1$, $e_2$, namely,
$e_3 = \lambda e_1 +k e_2$, which is impossible by the assumption
that the algebra under study is three-dimensional. Hence we conclude
that the above realization of $A^1_{2.1}$ cannot be extended
to a realization of the algebra $A_{3.1}$.

Turn now to the realization $A^2_{2.1}$. Checking commutation
relations (\ref{2.15}), where $e_3$ is of form (\ref{1.6}) yields
the following realization of $A_{3.1}$:
$$\langle t \partial_t +x \partial_x, \sigma(\xi) \partial_u, \gamma(\xi)
\partial_u \rangle, \ \ \xi = t x^{-1},$$
where $\gamma' \sigma -\gamma \sigma'\not =0.$ However, the corresponding invariant
equation (\ref{1.9}) is linear.

Finally, consider the realization $A^3_{2.1}.$ Inserting its basis operators
and the operator $e_3$ of the form (\ref{1.6}) into (\ref{2.15}) and
solving the obtained equations gives the following realization
of $A_{3.1}:$
$$ \langle \partial_t, \partial_x, u \partial_u \rangle.$$

Inserting the obtained coefficients for $e_3$ into the classifying equation
(\ref{1.7}) we get invariant equation
$$u_{tt} = u_{xx} +u G(\omega), \ \ \omega = u^{-1} u_x,$$
where (to ensure non-linearity) we need to have $G_{\omega \omega} \not =0$.

Similar analysis of the realizations $A^i_{2.1} \ (i=4,5, \ldots, 12, 14)$
yields three new invariant equations. For two of thus obtained
$A_{3.1}$-invariant equations the corresponding three-dimensional algebras are
maximal. The other two may admit four-dimensional invariance algebras
provided arbitrary elements are properly specified.

Handling in a similar way the extensions of $A_{2.2}$ up to realizations
of $A_{3.2}$ gives ten inequivalent nonlinear equations whose maximal
invariance algebras are realizations of the three-dimensional algebra
$A_{3.2}$ and four inequivalent equations (\ref{1.9}) admitting four-dimensional
symmetry algebras.

We perform analysis of equations admitting four-dimensional algebras
in the next sub-section. Here we present the complete list of
nonlinear equations (\ref{1.9}) whose maximal symmetry algebras are
realizations of three-dimensional Lie algebras $A_{3.1}$ and
$A_{3.2}$.

\noindent
{$A_{3.1}$-invariant equations}
\begin{eqnarray*}
A^1_{3.1} & = &\langle \partial_t, \partial_x, u\partial_u \rangle: \\
&&F=u G(\omega), \ \ \omega = u^{-1} u_x; \\
A^2_{3.1} &=& \langle \partial_x, \varphi(t) \partial_u, \psi(t) \partial_u
\rangle : \\
&&\sigma = \psi' \varphi -\psi \varphi' \not =0, \ \ \sigma' =0: \\
&& F=\varphi^{-1} \varphi'' u + G(t, u_x). \\
\end{eqnarray*}

\noindent
{$A_{3.2}$-invariant equations}
\begin{eqnarray*}
A^1_{3.2} &=& \langle \partial_t, \partial_x, e^x u \partial_u \rangle :\\[2mm]
&& F = -u^{-1} u^2_x -u \ln |u| + u G(\omega), \\[2mm]
&& \omega = u^{-1} u_x -\ln |u|: \\[2mm]
A^2_{3.2} &=& \langle -t \partial_t-x \partial_x, \partial_t+k \partial_x, u
\partial_u \rangle \ (k \geq 0): \\[2mm]
&& F = u \eta^{-2} { G}(\omega), \ \ \eta = x-kt, \\[2mm]
&& \omega = \eta u^{-1} u_x; \\[2mm]
A^3_{3.2} &=& \langle -t \partial_t-x \partial_x+mu \partial_u, \partial_t+k
\partial_x, |\eta|^{-m}\partial_u \rangle \\[2mm]
&& (\eta=x -kt,  \ k=m=0\ {\rm or} \ k>0, m \in  R): \\[2mm]
&& F = m(k^2-1)(m+1) \eta^{-2} u +|\eta|^{-2-m} {G}(\omega), \\[2mm]
&& \omega = |\eta|^m (mu+\eta u_x);\\[2mm]
A^4_{3.2}&=& \langle\partial_x, e^x u \partial_u, \partial_t+mu \partial_u
\rangle \ (m>0): \\[2mm]
&& F = -u^{-1} u^2_x -u_x +u {G}(\omega), \\[2mm]
&& \omega = u^{-1} u_x -\ln |u| +mt; \\[2mm]
\end{eqnarray*}
\begin{eqnarray*}
A^5_{3.2} &=& \langle -t \partial_t -x \partial_x, \partial_x, u \partial_u
\rangle: \\[2mm]
&& F = u t^{-2} {G}(\omega), \ \ \omega = t u^{-1} u_x; \\[2mm]
A^6_{3.2} &=& \langle -t \partial_t-x \partial_x, \partial_t+k x^{-1} u
\partial_u, u \partial_u \rangle \ (k>0):\\[2mm]
&& F =2 kt x^{-2} u_x -2 kt x^{-3} u+ k^2 t^2 x^{-4}u +x^{-2} u {G}(\omega),
\\[2mm]
&& \omega = x u^{-1} u_x +kt x^{-1}; \\ [2mm]
A^7_{3.2} &=& \langle -t\partial_t - x \partial_x, \partial_t+k x^{-1} u
\partial_u, \exp({ktx^{-1}}) \partial_u \rangle \ (k>0): \\[2mm]
&& F = 2kt x^{-2} u_x +(k^2 t^2 x^{-4} -2 k t x^{-3}+ k^2 x^{-2})u +\\[2mm]
&& +x^{-2} \exp({ktx^{-1}}) {G}(\omega), \ \  \omega = \exp({-ktx^{-1}})(x u_x +ktx^{-1}
u); \\[2mm]
A^8_{3.2} &=& \langle \frac{1}{2k} (\partial_t+k \partial_x), e^{x+kt}
\partial_u, e^\eta \partial_u \rangle \ (k>0, \eta= x-kt):\\[2mm]
&& F = (k^2-1) u +G(\eta, \omega), \ \ \omega = u_x -u; \\
A^{9}_{3.2} &=& \langle \partial_t+f(x)u \partial_u, e^{(1+f(x))t}\partial_u,
f(x) e^{f(x)t} \partial_u \rangle : \\[2mm]
&& F = -(tf''-t^2 (f')^2 -(1+f)^2)u-2tf'u_x +e^{tf} {G}(x, \omega),\\[2mm]
&& \omega = e^{-tf} (u_x -f'(t+f^{-1})u), \ \ f''+2 f^2+f=0, \ \ f \not =0;
\\[2mm]
A^{10}_{3.2} &=& \langle k(t \partial_t+x \partial_x), |t|^{k^{-1}}
|\xi|^{\frac{k-1}{2k}} \partial_u, |\xi|^{\frac{k-1}{2k}} \partial_u \rangle \ \
(k \not =0;1): \\[2mm]
&& F = \Bigl [ \frac{1-k}{k} \xi^2 +\frac{1-k^2}{4 k^2} (1-\xi^2)\Bigr] t^{-2} u
+t^{-2} {G}(\xi, \omega), \\[2mm]
&& \omega = |\xi|^{\frac{k-1}{2k}}\Bigl [ xu_x+\frac{k-1}{2k} u \Bigr], \ \ \xi
= t x^{-1}.
\end{eqnarray*}

\subsubsection{Invariance under non-decomposable three-dimensional solvable
 Lie algebras}

There exist seven non-decomposable three-dimensional solvable Lie
algebras over the field of real numbers. All those algebras contain
a subalgebra which is the two-dimensional Abelian ideal.
Con\-se\-quent\-ly, we can use the results of classification of
$A_{2.1}$-invariant equations in order to describe equations
admitting non-decomposable three-dimensional solvable real Lie
algebras. We remind that equations of the form (\ref{2.5}) has
already been analyzed and therefore are not considered in the
sequel.

As an example, we perform extension of the realization $A^{10}_{2.1}$
to all possible realizations of non-decomposable three-dimensional
solvable real Lie algebras. The remaining realizations are handled in a
similar way.

It is straightforward to verify that transformations
\begin{equation}\label{2.18}
\overline{t}=\gamma t +\gamma_1, \hskip 10mm \overline{x}=\epsilon \gamma x
+\gamma_2, \hskip 10mm v=\rho(x) u+\theta(x),
\end{equation}
where $\gamma, \gamma_1, \gamma_2 \in {\mathbb R}, \ \gamma \not =0, \ \epsilon= \pm 1, \
\rho \not =0$ are equivalence transformations for the realization $A^{10}_{2.1} =
\langle \partial_t, f(x) u \partial_u \rangle \ (f \not =0).$
That is why, we may use the above transformation in order to simplify the form
of operator $e_3$. As a result we get three inequivalent expressions for $e_3$
\begin{enumerate}
\renewcommand{\labelenumi}{\arabic{enumi} )}
\item  $e_3 = t \partial_t +x \partial_x +r(t,x) \partial_u \ (r_t \not =0$ or
$r=0)$;
\item $e_3 =  \partial_x +r(t,x) \partial_u \ (r_t \not =0$ or $r=0)$;
\item  $e_3 = r (t,x) \partial_u \ (r_t \not =0$ or $r=1)$.
\end{enumerate}

Let $e_1 = \partial_t, \ e_2 = f(x) u \partial_u$ and $e_3 = t \partial_t +
x \partial_x +r(t,x) \partial_u$, then
$$
[e_1, e_3] = \partial_t +r_t \partial_u, \ \
[e_2, e_3] = -x f' u \partial_u -rf \partial_u.
$$

Analyzing commutation relations for the algebras $A_{3.i} \ (i=3,4, \ldots, 9)$
we obtain that the necessary conditions for $A^{10}_{2.1}$ to admit extension
to a realization of $A_{3.5}$ read as $r=0, \ x f' = -f$,\ of $A_{3.6}$ as $r=0,\
xf'=f,$ and of $A_{3.7}$ as $r=0, \ xf' = -qf \ (0<|q|<1).$ So $A^{10}_{2.1}$
gives rise to the following realizations:
\begin{eqnarray*}
A_{3.5} &:& e_1 = \partial_t, \ e_2 = x^{-1} u \partial_u, \ e_3 = t \partial_t
+x \partial_x;\\
A_{3.6} &:& e_1 = \partial_t, \ e_2 = x u \partial_u, \ e_3 = t \partial_t +x
\partial_x;\\
A_{3.7} &:& e_1 = \partial_t, \ e_2 = |x|^{-q} u \partial_u, \ e_3 = t
\partial_t +x \partial_x \ (0<|q|<1).
\end{eqnarray*}

If $e_3 =  \partial_x +r(t,x) \partial_u$, then
$$
[e_1, e_3] = r_t \partial_u, \ \
[e_2, e_3] = - f' u \partial_u -rf \partial_u.
$$
Analyzing commutation relations for $A_{3.i}$ $ (i=3,4,\ldots,9)$ we come to
conclusion that the realization $A^{10}_{2.1}$ cannot be extended
to a realization of the above three-dimensional Lie algebras.

The same conclusion holds true when $e_3 = r(t,x) \partial_u$
$(r_t \not =0$ or $r=1).$

Let $e_1 = f(x) u \partial_u, \ e_2 = \partial_t.$ If $e_3 = t
\partial_t + x \partial_x +r(t,x) \partial_u$ $(r_t \not =0$ or
$r=0),$ then it follows from commutation relations
$$
[e_1, e_3] = -(rf+xf'u) \partial_u, \ \
[e_2, e_3] = \partial_t+r_t \partial_u
$$
that the only possible extension of the realization $A^{10}_{2.1}$ is the
realization of $A_{3.5}$:
$$ \langle x^{-1} u \partial_u, \partial_t, t \partial_t+x\partial_x \rangle.$$
This realization coincide within notation with the already obtained
one.

Next, if $e_3 = \partial_x+r(t,x) \partial_u \ (r_t \not =0$ or $r=0),$ then
$$
[e_1, e_3] = -(f'u+rf) \partial_u, \ \
[e_2, e_3] = r_t \partial_u.
$$
Analyzing commutation relations for $A_{3.i}$ $ (i=3,4,\ldots,9)$ we come to
conclusion that the realization $A^{10}_{2.1}$ cannot be extended
to a realization of the above three-dimensional Lie algebras.

The same conclusion holds true for the case  $e_3 = r(t,x) \partial_u$
$(r_t \not =0$ or $r=1).$

Summing up the above considerations we see that the
realization $A^{10}_{2.1}$ can be extended to the following
realizations of non-decomposable three-dimensional solvable real
Lie algebras:
\begin{eqnarray*}
&& L^1 \sim A_{3.5}, \ \ L^1 = \langle \partial_t, x^{-1} u \partial_u, t
\partial_t +x \partial_x \rangle; \\
&& L^2 \sim A_{3.6}, \ \ L^2 = \langle \partial_t, xu \partial_u, t
\partial_t+x\partial_x \rangle; \\
&& L^3 \sim A_{3.7}, \ \ L^3 = \langle \partial_t, |x|^{-q} u \partial_u, t
\partial_t+x \partial_x \rangle \ (0<|q|<1).
\end{eqnarray*}

Solving the corresponding classifying equations yields the
following invariant equations:
\begin{eqnarray*}
L^1 &:& u_{tt} = u_{xx} -u^{-1} u^2_x -2 x^{-2} u \ln |u|+
x^{-2} u G(\omega), \ \ \omega = x u^{-1} u_x +\ln |u|; \\
L^2 &:& u_{tt} = u_{xx} -u^{-1} u^2_x + x^{-2} u G(\omega), \\
&& \omega = x u^{-1} u_x -\ln |u|; \\
L^3&:& u_{tt} = u_{xx} -u^{-1} u^2_x -q(q+1) x^{-2} u \ln |u|+u x^{-2}
G(\omega), \\
&& \omega = xu^{-1} u_x +q \ln |u| \ \ (0<|q|<1).
\end{eqnarray*}
Note that the algebras $L^1, L^2, L^3$ are maximal (in Lie's sense)
invariance algebras of the corresponding equations.

While classifying nonlinear equations invariant non-decomposable
three-dimensional solvable Lie algebras we discovered equations
whose maximal invariance algebras are four-dimensional. For example,
after extending the realization $A^9_{2.1}$ up to a realization of
the algebra $A_{3.3}$ we got the following realization of the
latter:
$$ \langle \partial_u, \partial_t, \partial_x+t \partial_u \rangle.$$
The corresponding invariant equation (\ref{1.9}) reads as
$u_{tt} = u_{xx} +G(u_x).$ However, the maximal invariance algebra
of the above equation is the four-dimensional Lie algebra $\langle \partial_t,
t \partial_u, \partial_u, \partial_x \rangle,$ which is a realization
of $A_{3.3} \oplus A_1$. Note also that we have obtained the above
invariant equation when classifying $A_{3.1}$-invariant equations.

By the above reason, we give below only those nonlinear invariant
equations whose maximal symmetry algebras are three-dimensional
non-decomposable solvable real Lie algebras.

\noindent
{\bf $A_{3.3}$-invariant equations}
\begin{eqnarray*}
A^1_{3.3} &=& \langle u \partial_u, \partial_t +k \partial_x, m \partial_t
+k^{-1} xu \partial_u \rangle \ \ (k>0, m\not  =0):\\[2mm]
&& F = -u^{-1} u^2_x +u G(\omega),\  \ \omega = x -kt +m k^2 u^{-1} u_x; \\[2mm]
A^2_{3.3} &=& \langle u \partial_u, \partial_x, m \partial_t+xu \partial_u
\rangle \ (m >0): \\[2mm]
&& F = -u^{-1} u^2_x +u G(\omega), \ \ \omega = t-m u^{-1} u_x; \\[2mm]
A^3_{3.3} &=& \langle |t|^{\frac{1}{2}} \partial_u, -|t|^{\frac{1}{2}} \ln |t|
\partial_u, t \partial_t +x \partial_x +\frac{1}{2}u \partial_u \rangle : \\
&& F = -\frac{1}{4} t^{-2} u +u^3_x G(\xi ,\omega), \ \xi = t x^{-1}, \omega =x
u^2_x; \\
A^4_{3.3} &=& \langle \partial_u, -t \partial_u, \partial_t +k \partial_x
\rangle \ (k \geq 0); \\[2mm]
&& F = G(\eta, u_x), \ \eta = x-kt.
\end{eqnarray*}

\noindent
{\bf $A_{3.4}$-invariant equations}
\begin{eqnarray*}
A^1_{3.4} &=& \langle |\eta|^{m-1} \partial_u, \partial_t +k \partial_x, t
\partial_t +x \partial_x+(mu +t |\eta|^{m-1} )  \partial_u\rangle \\[2mm]
&& (\eta = x-kt, \ \ k>0, m \not =1): \\[2mm]
&& F =(k^2-1) (m-1) (m-2) \eta^{-2} u -2 k (m-1) \eta^{m-2} \ln |\eta| \\[2mm]
&& + |\eta|^{m-2} G(\omega),  \ \ \omega = [\eta u_x -(m-1)u ]
|\eta|^{-m};\\[2mm]
A^2_{3.4} &=& \langle  \partial_u, -t  \partial_u,  \partial_t +k  \partial_x +u
 \partial_u \rangle \ (k\geq 0): \\[2mm]
&& F = e^t G(\eta, \omega), \ \ \eta = x-kt, \ \omega = e^{-t} u_x; \\[2mm]
A^{3}_{3.4} &=& \langle |t|^{\frac{1}{2}}  \partial_u, -|t|^{\frac{1}{2}} \ln
|t| \partial_u,t  \partial_t +x  \partial_x +\frac{3}{2} u  \partial_u \rangle:
\\
&& F = -\frac{1}{4} t^{-2} u +u^{-1}_x G(\xi,\omega), \ \xi = t x^{-1}, \ \
\omega = x^{-1} u^2_x; \\
A^4_{3.4} &=& \langle k x^{-1} u  \partial_u,  \partial_t -k x^{-1}
\ln |x| u
\partial_u, t  \partial_t +x  \partial_x \rangle \ (k>0): \\[2mm]
&& F = -3 kt x^{-3} u -2 x^{-2} u \ln |u| -u^{-1} u^2_x+ x^{-2} u G(\omega),\\[2mm]
&& \omega = x u^{-1} u_x +\ln |u| +kt x^{-1} ; 
\end{eqnarray*}
\begin{eqnarray*}
A^{5}_{3.4} &=& \langle \exp({kt x^{-1}})  \partial_u,  \partial_t +k x^{-1} u
\partial_u , t  \partial_t +x  \partial_x +(u+t \exp({kt x^{-1}}))  \partial_u
\rangle \ (k>0): \\[2mm]
&& F = k^2 x^{-4} u (t^2+x^2) +2 x^{-1} (ktx^{-1} +1) u_x \\
&& + 2 k \exp({kt x^{-1}}) x^{-1}\ln |x|+ x^{-1} \exp({ktx^{-1}}) G(\omega), \\[2mm]
&& \omega = \exp({-ktx^{-1}}) (u_x +kt x^{-2} u).
\end{eqnarray*}
{\bf $A_{3.5}$-invariant equations}
\begin{eqnarray*}
A^1_{3.5} &=& \langle |\eta|^{m-1} \partial_u, \partial_t +k \partial_x, t
\partial_t +x \partial_x+mu  \partial_u\rangle  \ (k>0, \ m \not =1)\\[2mm]
&& F =(k^2-1) (m-1) (m-2) \eta^{-2} u + |\eta|^{m-2} G(\omega), \\[2mm]
&& \omega = |\eta|^{-m} [\eta  u_x -(m-1)u ],  \eta=x -kt;\\[2mm]
A^2_{3.5} &=& \langle \partial_t, \partial_x, t \partial_t+x \partial_x \rangle
: \\[2mm]
&& F = u^2_x G(u); \\[2mm]
A^3_{3.5} &=& \langle \partial_t, \partial_x, t \partial_t +x \partial_x +mu
\partial_u \rangle \ (m \not =0): \\[2mm]
&& F = |u|^{1-\frac{2}{m}} G(\omega), \ \omega = |u_x|^m |u|^{1-m}; \\[2mm]
A^4_{3.5} &=& \langle \partial_t, \partial_x, t \partial_t+x
\partial_x+\partial_u \rangle : \\[2mm]
&& F = e^{-2u} G(\omega), \ \omega = e^u u_x; \\[2mm]
A^5_{3.5}& =  & \langle \partial_t, x^{-1} u \partial_u, t \partial_t
+x\partial_x \rangle : \\[2mm]
&& F = -u^{-1} u^2_x -2 x^{-2} u \ln |u| +x^{-2} u G(\omega), \\[2mm]
&& \omega = x u^{-1} u_x +\ln |u|; \\[2mm]
A^6_{3.5} &=& \langle \partial_t +k x^{-1} u \partial_u, \exp({kt x^{-1}})
\partial_u, t \partial_t +x \partial_x +u \partial_u \rangle \ (k>0) : \\[2mm]
&& F = k x^{-4} u [kt^2 -2t x +k x^2] + 2kt x^{-2} u_x +x ^{-1} \exp({ktx^{-1}})
G(\omega), \\[2mm]
&& \omega = \exp({-kt x^{-1}}) (u_x +kt x^{-2} u); \\[2mm]
A^7_{3.5} &=& \langle \varphi(t) \partial_u, \psi(t) \partial_u, \partial_x+u
\partial_u \rangle \ (\varphi' \psi -\varphi \psi' \not =0):\\[2mm]
&& F = \varphi^{-1} \varphi'' u +u_x G(t, \omega), \\[2mm]
&& \omega = e^{-x} u_x, \ \ \varphi'' \psi -\varphi \psi'' =0.
\end{eqnarray*}
{\bf $A_{3.6}$-invariant equations}
\begin{eqnarray*}
A^1_{3.6} &=& \langle \partial_t+k \partial_x, |\eta|^{m+1} \partial_u, t
\partial_t +x \partial_x +mu \partial_u \rangle  \ (k>0, \ m \not =-1):\\[2mm]
&& F =m(k^2-1)  (m+1) \eta^{-2} u + |\eta|^{m-2} G(\omega), \\[2mm]
&& \omega = |\eta|^{1-m} [ u_x -\eta^{-1} (m+1)u ],  \eta=x -kt;\\[2mm]
A^2_{3.6} &=& \langle \partial_t +m x^{-1} u \partial_u, xu \partial_u, t
\partial_t +x \partial_x \rangle \ (m \geq 0): \\[2mm]
&& F = -u^{-1} u^2_x -2m t x^{-3} u +x^{-2} u G(\omega), \\[2mm]
&& \omega = xu^{-1} u_x -\ln |u| +2 m t x^{-1}; 
\end{eqnarray*}
\begin{eqnarray*}
A^3_{3.6} &=& \langle \partial_t+k x^{-1} u \partial_u, \exp({ktx^{-1}}) \partial_u,
t \partial_t+x\partial_x-u \partial_u \rangle \ (k>0): \\[2mm]
&& F = x^{-4} [k^2 x^2 -2 ktx +k^2t^2] u +2kt x^{-2} u_x +x^{-3} \exp({ktx^{-1}})
G(\omega), \\[2mm]
&& \omega = \exp({-ktx^{-1}}) (x^2 u_x+kt u); \\[2mm]
A^4_{3.6} &=& \langle e^{-t} \partial_u, e^t \partial_u, \partial_t +k
\partial_x \rangle \ (k \geq 0): \\[2mm]
&& F = u +G(\eta, u_x), \ \eta = x-kt; \\
A^5_{3.6} &=& \langle |t|^{-\frac{1}{2}} \partial_u, |t|^{\frac{3}{2}}
\partial_u, t \partial_t+x \partial_x+\frac{1}{2}u \partial_u \rangle : \\
&& F = \frac{3}{4} t^{-2} u +|t|^{-\frac{3}{2}} G(\xi, \omega), \ \xi = t
x^{-1}, \ \ \omega = x^{-1} u^2_x.
\end{eqnarray*}
{\bf $A_{3.7}$-invariant equations}
\begin{eqnarray*}
A^1_{3.7} &=& \langle \partial_t +k \partial_x, |\eta|^{m-q} \partial_u, t
\partial_t +x \partial_x +mu \partial_u \rangle \\
&&  (k>0, m \not =q, \ 0<|q|<1): \\ [2mm]
&& F = (k^2-1) (m-q) (m-q-1) \eta^{-2} u +| \eta |^{m-2} G(\omega), \\ [2mm]
&& \omega = | \eta |^{1-m} [u_x -(m-q)\eta^{-1} u ], \ \eta = x-kt; \\ [2mm]
A^2_{3.7} &=& \langle \partial_t+k x^{-1} u \partial_u, \exp({ktx^{-1}}) \partial_u,
t \partial_t+x \partial_x+qu \partial_u \rangle \\
&&  (k>0, \ 0<|q|<1): \\[2mm]
&& F = [k^2x^{-2} +k^2x^{-4} t^2 -2k t x^{-3}] u+2kt x^{-2} u_x \\
&& +|x|^{q-2} \exp({ktx^{-1}}) G(\omega), \\ [2mm]
&& \omega = |x|^{1-q} \exp({-ktx^{-1}}) (u_x +ktx^{-2} u); \\
A^3_{3.7} &=& \langle |t|^{\frac{1}{2}q} \partial_u, |t|^{1-\frac{1}{2} q}
\partial_u, t \partial_t +x \partial_x +(1+\frac{1}{2}q)u \partial_u \rangle \
(q\not =0, \pm 1): \\
&& F = \frac{1}{4} q (q-2) t^{-2} u +|t|^{\frac{1}{2}(q-2)} G(\xi, \omega), \\
&& \xi = t x^{-1}, \ \omega = |t|^{-\frac{1}{2}q} u_x; \\
A^4_{3.7} &=& \langle \exp\left({\frac{1}{2}(q-1)t}\right)
\partial_u, \exp\left({\frac{1}{2}(1-q) t}\right)
\partial_u, \partial_t+k \partial_x+\frac{1}{2}(1+q) u \partial_u \rangle \\
&& \ (q\not =0, \pm 1; k \geq 0): \\
&& F = \frac{1}{4}(q-1)^2u +\exp\left({\frac{1}{2} (1+q)t}\right) G(\eta, \omega), \\
&& \eta = x -kt, \ \omega = \exp\left({-\frac{1}{2}(1+q) t}\right) u_x ; \\ [2mm]
A^5_{3.7} &=& \langle \partial_t +k x^{-1} u \partial_u, |x|^{-q} u \partial_u,
t \partial_t +x \partial_x \rangle \ (k \geq 0, q \not =0, \pm 1): \\[2mm]
&& F = -u^{-1} u^2_x -q (q+1) x^{-2} u \ln |u| +k(q-1)(q+2) t x^{-3} u \\[2mm]
&&+ u x^{-2} G(\omega), \ \ \omega = x u^{-1} u_x +q \ln |u| +k(1-q) t x^{-1}.
\end{eqnarray*}
{\bf $A_{3.8}$-invariant equations}
\begin{eqnarray*}
A^1_{3.8} &=& \langle \cos t \partial_u, -\sin t \partial_u, \partial_t +k
\partial_x \rangle \ (k\geq 0): \\[2mm]
&& F = -u +G(\eta, u_x), \eta = x-kt;
\end{eqnarray*}
\begin{eqnarray*}
A^2_{3.8} &=& \langle |t|^{\frac{1}{2}} \cos (\ln |t|) \partial_u,
-|t|^{\frac{1}{2}} \sin (\ln |t| ) \partial_u, t \partial_t+x
\partial_x+\frac{1}{2} u \partial_u \rangle : \\
&& F = -\frac{5}{4} t^{-2} u +|t|^{-\frac{3}{2}} G(\xi, \omega), \\
&& \xi = t x^{-1}, \ \omega = |t|^{\frac{1}{2}} u_x.
\end{eqnarray*}
{\bf $A_{3.9}$-invariant equations}
\begin{eqnarray*}
A^1_{3.9} &=& \langle \sin t \partial_u, \cos t \partial_u, \partial_t +k
\partial_x +qu \partial_u \rangle \ (k\geq 0, q>0): \\[2mm]
&& F = -u +e^{qt} G(\eta,\omega), \ \eta = x -kt, \ \ \omega = e^{-qt} u_x; \\
A^2_{3.9} &=& \langle |t|^{\frac{1}{2}} \sin (\ln|t|) \partial_u,
|t|^{\frac{1}{2}} \cos (\ln|t|) \partial_u, t \partial_t +x \partial_x
+(\frac{1}{2}+q) u \partial_u \rangle \\
&& \ (q\not = 0) :  F = -\frac{5}{4} t^{-2} u +|t|^{q-\frac{3}{2}} G(\xi,
\omega), \\
&& \xi = t x^{-1}, \ \ \omega = |t|^{\frac{1}{2}-q} u_x.
\end{eqnarray*}

\subsection{Complete group classification of equation (\ref{1.9})}

The aim of this subsection is finalizing group classification of
(\ref{1.9}). The majority of invariant equations obtained in the
previous subsection contain arbitrary functions of one variable.
So that we can utilize the standard Lie-Ovsyannikov approach in
order to complete their group classification.

\subsubsection{Equations depending on an arbitrary function of one variable.}

Note that equations belonging to the already investigated class of
(\ref{2.5}) are not considered.

As our computations show, new results could be obtained for the
equations
\begin{eqnarray}
u_{tt} &=& u_{xx} +uG(\omega), \ \ \omega = u^{-1} u_x, \label{2.27} \\
u_{tt} &=& u_{xx} +G(u_x) \label{2.28}
\end{eqnarray}
only. Below we give (without proof) the assertions describing their group properties.

\begin{tver} \label{tv2.5.1}
Equation (\ref{2.27}) admits wider symmetry group iff it is equivalent to the following equation
\begin{equation} \label{2.29}
u_{tt} = u_{xx} +m u^{-1} u^2_x \ \ \ \ (m \not =0, -1).
\end{equation}
The maximal invariance algebra of (\ref{2.29}) is the four-dimensional Lie algebra
$$ A_4 \sim A_{3.5} \oplus A_1, \ \ A_4 = \langle \partial_t, \partial_x, t
\partial_t +x \partial_x, u \partial_u \rangle.$$
\end{tver}

\begin{tver} \label{tv2.5.2}
Equation (\ref{2.28}) admits wider symmetry group iff it is
equivalent to one of the following PDEs:
\begin{eqnarray}
u_{tt} &=& u_{xx} +e^{u_x}; \label{2.31} \\
u_{tt} &=& u_{xx} +m \ln |u_x|, \ \ m>0; \label{2.32} \\
u_{tt} &=& u_{xx} +|u_x|^k, \ \ \ k \not =0,1. \label{2.33}
\end{eqnarray}
The maximal invariance algebras of the above equations are five-dimensional solvable Lie algebras listed below.
\begin{eqnarray*}
A^2_5 &=& \langle \partial_t, \partial_x, \partial_u, t \partial_u, t \partial_t
+x \partial_x+ (u-x) \partial_u \rangle; \\
A^3_5 &=& \langle \partial_t, \partial_x, \partial_u, t \partial_u, t \partial_t
+x \partial_x+ (2u+\frac{1}{2} m t^2) \partial_u \rangle; \\
A^4_5 &=& \langle \partial_t, \partial_x, \partial_u, t \partial_u, t \partial_t
+x \partial_x+ \frac{k-2}{k-1} u \partial_u \rangle.
\end{eqnarray*}
\end{tver}

Analyzing the remaining equations containing arbitrary functions
of one variable we come to conclusion that one of them can admit wider
invariance groups iff either
\begin{enumerate}
\renewcommand{\labelenumi}{\arabic{enumi})}
\renewcommand{\theenumi}{\arabic{enumi}}
\item it is equivalent to PDE of the form (\ref{2.5}), or
\item it is equivalent to PDE of the form (\ref{2.29}).
\end{enumerate}
Skipping the proof, we present two typical examples. We begin with
the equation
\begin{equation} \label{2.36}
u_{tt} = u_{xx} +u +G(u_x).
\end{equation}
This equation is invariant under the four-dimensional algebra
$\langle \partial_t, \partial_x, e^t \partial_u, e^{-t} \partial_u \rangle
$ isomorphic to $A_{3.6} \oplus A_1.$ Inserting $F = u +G(u_x)$ into
classifying equation (\ref{1.7}) yields the system of two equations for
the function $G$
$$
h' G' = -h'' -2 \lambda, \ \
[(h-\lambda) u_x +r_x ]G' -(h-2 \lambda) G = r_{tt} -r_{xx} -2 h' u_x -r.
$$
As we require $G'' \not =0$, it follows from the first equation
that $\lambda = h' =0$ and the second equation takes the form
$$ (h u_x + r_x) G' -h G = r_{tt} -r_{xx} -r.$$
Upon differentiating the above equation twice with respect to $u_x$ we get
$ (h u_x +r_x) G'' =0.$ As $G'' \not =0,$ relations $h = r_x =0$ hold.
Hence we conclude that the class of PDEs (\ref{2.36}) does not
contain equations admitting five-dimensional algebras.

The system of determining equations for symmetry group of
$A^2_{3.2}$-invariant equation
\begin{equation} \label{2.37}
u_{tt} = u_{xx} +u \eta^{-2} G(\omega), \ \ \eta = x-kt, \ \
\omega =\eta u^{-1}
u_x, \ \ k \geq 0,
\end{equation}
read as
\begin{eqnarray} \label{2.38}
&& (\eta^{-2} r \omega -\eta^{-1} r_x) G_{\omega} -\eta^{-2} r G = r_{tt}
-r_{xx}, \nonumber \\
&& [(\lambda_2 -k \lambda_1 ) \eta^{-3} \omega +\eta^{-1} h' ] G_{\omega}
-2(\lambda_2 -k \lambda_1) \eta^{-3} G =
-2 h' \eta^{-1} \omega -h ''.\nonumber
\end{eqnarray}
Differentiating the first equation with respect to $\omega$ yields
$$
(\eta^{-2} r \omega -\eta^{-1} r_x) G_{\omega \omega} =0,
$$
whence in view of inequality $ G_{\omega \omega}\not =0$ we get $r=0$.

Next, differentiating the second equation twice by $\omega$ we get
$$
[(\lambda_2 -lk \lambda_1) \eta^{-3} \omega +\eta^{-1} h'] G_{\omega
\omega\omega }=0,
$$
whence it follows that $ G_{\omega \omega\omega }=0.$ Indeed, if
this relation does not hold, we have $\lambda_2 = k \lambda_1, \
h' =0$ and operator (\ref{1.6}) takes the form
$$
\lambda (t \partial_t +x \partial_x ) +\lambda_1 (\partial_t + k \partial_x)
+C_1 u \partial_u, \ \ \lambda, \lambda_1 C_1 \in {\mathbb R}, \ \ k \geq 0.
$$
As the above operator contains at most three arbitrary constants it
cannot generate a four-parameter Lie transformation group.

By the above argument we can restrict our considerations to the
following class of functions $G:$
\begin{equation} \label{2.39}
G = A \omega^2 +B \omega +C, \ \ A \not =0, -1,\ B, C \in {\mathbb
R}.
\end{equation}

We can suppose that $A \not = -1$ in (\ref{2.39}) (since otherwise
(\ref{2.37}) belongs to the class of PDEs (\ref{2.5})). Inserting
(\ref{2.39}) into the second equation from (\ref{2.38}) yields
\begin{equation} \label{2.40}
2 (A+1) \eta^2 h' = B (\lambda_2 -k \lambda_1), \ \ \eta^2 B h'
+\eta^3 h'' = 2 C(\lambda_2 -k \lambda_1).
\end{equation}
If $k>0,$ then this system splits into the following equations (note
that $h = h(x)$):
$$
h' =0, \ \ B (\lambda_2 -k \lambda_1) = C(\lambda_2 -k \lambda_1) =0.
$$
Provided $|B|+|C| \not =0, $ there is no way for extending symmetry
of equation (\ref{2.37}). If, on the contrary, $B=C=0,$ then $F = A
u^{-1} u^2_x (A \not = 0, -1)$ and we obtain the equation equivalent
to (\ref{2.29}). Under $k =0$ system (\ref{2.40}) takes the form
\begin{eqnarray*}
&& 2 (A+1) x^2 h' = \lambda_2 B, \ \
 x^2 B h'+h'' x^3 = 2 \lambda_2 C.
\end{eqnarray*}
Hence
\begin{eqnarray*}
h&=& -\frac{1}{2} \lambda_2 (A+1)^{-1} B x^{-1} +C_1, \ \ C_1 \in
{\mathbb R}, \ \ C = \frac{B^2-2B}{4(A+1)}.
\end{eqnarray*}
In this case equation (\ref{2.37}) does admit additional symmetry operator
$$ \partial_x -\frac{B}{2(A+1)} x^{-1} u \partial_u$$
but the change of variables
$$ \bar t = t, \hskip 10mm \bar x = x, \hskip 10mm u = |x|^{\nu} v, \hskip 10mm
\nu = -\frac{B}{2(A+1)},$$
reduces it to the form (\ref{2.29}).

So equation (\ref{2.37}) admits wider symmetry group iff it is either
belongs to the class of (\ref{2.5}) or is equivalent to (\ref{2.29}).

To finalize the procedure of group classification of equations (\ref{1.9})
we need to consider invariant equations obtained in the previous section
that contain arbitrary functions of two variables.

\subsubsection{Classification of equations with arbitrary functions of
two variables.}

In the case under study the standard Lie-Ovsyannikov method is
inefficient and we apply our classifi\-ca\-ti\-on algorithm. In
order to do this we perform extension of three-dimensional solvable
Lie algebras to all possible realizations of four-dimensional
solvable Lie algebras. The next step will be to check which of the
obtained realizations are symmetry algebras of nonlinear equations
of the form (\ref{1.9}). In what follows we use the results of the
paper \cite{magda57}, where all inequivalent (within the action of
inner automorphism group) four-dimensional solvable abstract Lie
algebras are given.

We give full computation details for the case of $A_{3.6}$-invariant
equations. As shown in the previous subsection, there are two
inequivalent $A_{3.6}$-invariant equations, namely,
\begin{eqnarray*}
A^4_{3.6} &=& \langle e^{-t} \partial_u, e^t \partial_u, \partial_t
+k \partial_x \rangle \\ && (k \geq 0): F = u+ G(\eta, u_x), \ \eta
= x-kt; \\ A^5_{3.6} &=& \langle |t|^{-\frac{1}{2}} \partial_u, \
|t|^{\frac{3}{2}} \partial_u, t \partial_t +x \partial_x
+\frac{1}{2} u \partial_u \rangle : \\ && F = \frac{3}{4} t^{-2} u
+|t|^{-\frac{3}{2}} G(\xi, \omega),  \ \xi = t x^{-1}, \ \omega =
x^{-1} u^2_x.
\end{eqnarray*}
According to \cite{magda57} the algebra $A_{3.6}$ is the subalgebra
of the following four-dimensional solvable Lie algebras: $2 A_{2.2},
A_{3.6} \oplus A_1; A_{4.2} (q=-1); A_{4.8} (q=-\frac{1}{2}).$

{Algebra $2A_{2.2}.$}
The algebra $2A_{2.2} = \langle e_1, e_2, e_3, e_4  \rangle$ is
determined by the following commutation relations (note that we
give non-zero relations only):
$$
[e_1, e_2] = [e_1, e_4] = [e_2, e_3] = [e_2, e_4]=0, \ \
[e_1, e_2] = e_2, [ e_3, e_4]=e_4.
$$
It contains a subalgebra $A_{3.6} = \langle e_1 -e_3, e_2, e_4
\rangle.$ That is why, we can choose as $ e_1 -e_3, e_2, e_4 $ the
basis operators of the realization of $A_{3.6}$. Next, we take as
$e_1+e_3$ an arbitrary operator of the form (\ref{1.6}) and require
for the commutation relations
\begin{eqnarray}
&& [e_1 -e_3, e_1+e_3] =0, \ \ [e_1+e_3, e_2] = e_2, \ \ [e_1 +e_3, e_4] = e_4
\label{2.41}
\end{eqnarray}
to hold.

\underline{Realization $A^4_{3.6}.$} \ In this case
\begin{eqnarray*}
&& e_1 -e_3 = -\partial_t -k \partial_x, \ \  e_2= e^{-t} \partial_u, \ \ e_4 =
e^t \partial_u, \\
&& e_1 +e_3  = (\lambda t +\lambda_1) \partial_t +(\lambda x +\lambda_2)
\partial_x + (h u +r) \partial_u.
\end{eqnarray*}
It follows from (\ref{2.41}) that
$$
\lambda = \lambda_1 =0, \ \ r =\gamma = \gamma(\eta), \ \ h=-1.
$$
Using the change of variables
$$ \bar t = t, \ \ \bar x = x, \ \ v = u+\Lambda (\eta), $$
where $\Lambda =\Lambda(\eta)$ is a solutions of equation $
\lambda_2 \Lambda' +\Lambda = -\gamma,$ we simplify the operator
$e_1 +e_3$ to become
$$
e_1 +e_3 = \alpha \partial_x - u \partial_u, \ \ \alpha \in {\mathbb R}.
$$

Requiring invariance under the above operator yields that
$\alpha \not =0$ (otherwise $G$ would be linear in $u_x$).
With this condition we rewrite the invariant equation to become
$$
G = \exp({-\alpha^{-1} \eta}) H(\omega), \hskip 10mm \omega = \exp({\alpha^{-1}
\eta}) u_x.
$$

Thus we arrive at the following realization of the algebra $2A_{2.2}$:
$$ \langle e^{-t} \partial_u, e^t \partial_u, \partial_t +k \partial_x, \alpha
\partial_x -u \partial_u \rangle \ (k \geq 0, \ \alpha \not =0).$$
This algebra is admitted by the equation
$$
u_{tt} = u_{xx} +u +\exp({-\alpha^{-1} \eta}) G(\omega), \ \eta = x-kt, \ \omega
= \exp({\alpha ^{-1} \eta}) u_x.
$$
If the function $G \ (G_{\omega \omega} \not =0)$ is arbitrary, then
the obtained realization is the maximal symmetry algebra of the
equation under study. What is more, no $G$ exists such that the above
equation admits a wider invariance algebra.

\underline{Realization $A^5_{3.6}.$} In this case
\begin{eqnarray*}
&& e_1 -e_3 = -t \partial_t -x \partial_x -\frac{1}{2} u \partial_u, \ \
 e_2 = |t|^{-\frac{1}{2}} \partial_u, \\
&& e_1 +e_3 = (\lambda_t +\lambda_1) \partial_t +(\lambda_x +\lambda_2)
\partial_x +(h u +r) \partial_u, \ \   e_4 = |t|^{\frac{3}{2}} \partial_u.
\end{eqnarray*}
It follows from commutation relations (\ref{2.41}) that $\lambda_1
= \lambda_2 = \lambda =0,$ $h = -1, \ r=|t|^{\frac{1}{2}} \gamma (\xi),
\ \xi = t x^{-1}.$

Making the change of variables
$$ \bar t = t, \ \ \bar x = x, \ \ v = u-|t|^{\frac{1}{2}}\gamma (\xi) $$
we get $r=0$ in $e_1 +e_3$. Consequently, without loss of generality we
can choose $e_1 +e_3 = -u \partial_u.$ Requiring for $A^5_{3.6}$-invariant
equation to admit the operator $e_1+e_3$ yields the equation
$2 \omega G_\omega =-G,$ whence $G = |\omega|^{-\frac{1}{2}} H(\xi)$.
Consequently, the function $F$ is linear in the variable $u_x$.
This means that $A^5_{3.6}$ does not admit extension to a realization
$2A_{2.2}$ that can be a symmetry algebra of an equation of the form
(\ref{1.9}).

\underline{Algebra $A_{3.6} \oplus A_1$.} What we need to do here is
to supplement the set of operators $e_1, e_2, e_3 $ forming the basis
of $A_{3.6}$  by the operator $e_4$ of the form (\ref{1.6}) and verify
the commutation relations
\begin{equation} \label{2.42}
[e_1, e_4] = [e_2, e_4] = [e_3, e_4] =0
\end{equation}
\underline{Realization $A^4_{3.6}$.} It follows from (\ref{2.42})
that $h = \lambda = \lambda_1 =0, \ \ r = \gamma(\eta), \ \eta = x-kt$
in the operator $e_4$ so that
$$
e_4 = \alpha \partial_x +\gamma (\eta) \partial_u, \ \ \alpha \in {\mathbb
R}.
$$
If $\alpha \not =0,$ then $e_4$ is equivalent to $\partial_x.$
Hence we get two possible realizations of the algebra $A_{3.6} \oplus A_1$:
\begin{eqnarray*}
&& \langle  e^{-t} \partial_u, \ e^t \partial_u, \partial_t, \partial_x \rangle
;\\
&& \langle  e^{-t} \partial_u, \ e^t \partial_u, \partial_t+k \partial_x,
\gamma(\eta) \partial_u \rangle .
\end{eqnarray*}
Analyzing the above realizations we come to conclusion that the
second one cannot be invariance algebra of a nonlinear equation
of the form (\ref{1.9}). The first realization is the maximal (if $G$
is an arbitrary function) invariance algebra of the equation (\ref{2.36}).

\underline{Realization $A^5_{3.6}.$ } It follows from (\ref{2.42})
that $\lambda_1 = \lambda_2 = h = \lambda =0,$
$$
r = |t|^{\frac{1}{2}} \gamma(\xi) , \ \ \xi = t x^{-1},
$$
so that the operator $e_4$ necessarily takes the form
$e_4 = |t|^{\frac{1}{2}} \gamma(\xi) \partial_u$. As the
straightforward verification shows thus obtained realization
cannot be invariance algebra of a nonlinear equation of the
form (\ref{1.9}).

\underline{Algebra $A_{4.2} \ (q=-1)$.} We need to supplement the set of operators $e_1, e_2, e_4$ forming the basis of $A_{3.6}$ by the operator $e_3$ of the form (\ref{1.6}) so that the following commutation relation hold
\begin{equation} \label{2.43}
[e_1, e_3] = [e_2, e_3] =0, \ \ \  [e_3, e_4] =e_2 +e_3.
\end{equation}
\underline{Realization $A^4_{3.6}$.} In this case
\begin{eqnarray*}
e_1 &=& e^{-t} \partial_u, \ \ e_2 = e^t \partial_u, \ \
e_4 = -\partial_t -k \partial_x
\end{eqnarray*}
and it follows from (\ref{2.43}) that coefficients of $e_3$ satisfy equations $h = \lambda = \lambda_1=\lambda_2 =0$, the function $r$ being a solution of the equation
$$
r_t +k r_x = r +e^t.
$$
This realization cannot be invariance algebra of nonlinear equation
of the form (\ref{1.9}).

\underline{Realization $A^5_{3.6}$.} In this case
\begin{eqnarray*}
e_1 &=& |t|^{-\frac{1}{2}} \partial_u, \  \ e_2 = |t|^{\frac{3}{2}} \partial_u,
\ \
e_4 = -t \partial_t -x \partial_x-\frac{1}{2} u \partial_u.
\end{eqnarray*}
It follows from (\ref{2.43}) that the coefficients of the operator
$e_3$ satisfy equations $\lambda = \lambda_1 = \lambda_2 =h =0$
and the function $r$ is a solution of the equation
$$
t r_t +x r_x = \frac{3}{2} r +|t|^{\frac{3}{2}}.
$$
Further analysis shows that the so obtained realization cannot
be invariance algebra of a nonlinear equation of the form (\ref{1.9}).

\underline{Algebra $A_{4.8} \ (q=-\frac{1}{2}).$} We need to
supplement the set of operators $e_1, e_3, e_4$ forming the
basis of $A_{3.6}$ by the $e_2$ of the form (\ref{1.6}) in order
to satisfy the commutation relations
\begin{equation} \label{2.44}
[e_1, e_2] =0, \ \  [e_2, e_3] =e_1, \ \ \  [e_2, e_4] =e_2.
\end{equation}
\underline{Realization $A^4_{3.6}$.} In this case
\begin{eqnarray*}
e_1 &=& e^{-t} \partial_u, \ \ e_3 = e^t \partial_u, \ \
e_4 = \frac{1}{2}\partial_t +\frac{1}{2}k \partial_x
\end{eqnarray*}
and the second commutation relation yields the false equality $1=0$.

\underline{Realization $A^5_{3.6}$.} In this case
\begin{eqnarray*}
e_1 &=& |t|^{-\frac{1}{2}} \partial_u, \  \ e_2 = |t|^{\frac{3}{2}} \partial_u,
\ \
e_4 = -t \partial_t -x \partial_x-\frac{1}{2} u \partial_u
\end{eqnarray*}
and again the second commutation relation from (\ref{2.44}) cannot
be satisfied.

So that there are no extensions of the realization of $A_{3.6}$ to
a realization of the algebra $A_{3.8} \ (q=-\frac{1}{2})$.

The remaining equations containing arbitrary functions of two
variables are handled in a similar way. The results can be summarized
as follows
\begin{enumerate}
\renewcommand{\labelenumi}{\arabic{enumi})}
\renewcommand{\theenumi}{\arabic{enumi}}
\item if the functions contained in the equations under study
are arbitrary, then the corresponding realizations are their maximal
invariance algebras, and
\item except for equation (\ref{2.28}), all the equations
in question do not allow for extension of their symmetry.
\end{enumerate}

Below we give the complete list of invariant equations obtained
through group analysis of equations with arbitrary functions of
two variables.

\subsubsection{ Equations invariant under four-dimensional solvable
Lie algebras.}

{$A_{2.2} \oplus 2 A_1$-invariant equations }
\begin{eqnarray*}
1) && \langle \partial_x, \partial_t+u \partial_u, e^t \partial_u, e^{-t}
\partial_u \rangle :\ \
 F = u +e^t G(\omega), \ \ \omega = u^{-t} u_x; \\
2) && \langle \frac{1}{2k} (\partial_t +k \partial_x ), e^{x+kt} \partial_u,
e^{\eta} \partial_u , \partial_x +u \partial_u \rangle \ (k>0, \eta = x-kt): \\
&& F = (k^2-1) u +e^{\eta} G(\omega), \ \ \omega = e^{-\eta} (u_x -u ).
\end{eqnarray*}

{$2A_{2.2}$-invariant equations}
\begin{eqnarray*}
1) && \langle \partial_t +\epsilon u \partial_u, \partial_x, e^{x+kt}
\partial_u, e^{x-kt} \partial_u \rangle \ (\epsilon =0,1; k>0) : \\
&& F = (k^2-1) u +e^{\epsilon t} G(\omega), \ \ \omega = e^{-\epsilon t
}(u_x-u); \\
2) && \langle \alpha \partial_x-u \partial_u, \partial_t+k \partial_x, e^{-t}
\partial_u, e^{t} \partial_u \rangle \ (k\geq 0, \alpha >0): \\
&& F = u+\exp({-\alpha^{-1} \eta}) G(\omega), \ \eta = x-kt, \ \omega =
\exp({\alpha^{-1} \eta}) u_x.
\end{eqnarray*}

{$A_{3.3} \oplus A_1$-invariant equations}
\begin{eqnarray*}
1) && \langle \partial_t, \partial_x,\partial_u,t \partial_u \rangle :
F = G(u_x).
\end{eqnarray*}

{$A_{3.4} \oplus A_1$-invariant equations}
\begin{eqnarray*}
1) && \langle \partial_u, \partial_x, t \partial_t +x \partial_x+(u+x)
\partial_u, t \partial_u \rangle :\\
&& F = t^{-1} G(\omega), \ \ \omega = u_x-\ln|t|; \\
2) && \langle \partial_t +u \partial_u , \partial_x, t \partial_u ,  \partial_u
\rangle : \ \
 F = e^{t} G(\omega), \ \ \omega = e^{-t} u_x; \\
3) && \langle x^{-1} \partial_u, \partial_x-x^{-1} (u +\ln |x|) \partial_u, t
\partial_t+x \partial_x ,  tx^{-1} \partial_u \rangle : \\
&& F = 2x^{-1} u_x +x^{-2} +t^{-1} x^{-1} G(\omega), \ \omega = xu_x +u -\ln
|tx^{-1}|.
\end{eqnarray*}

{$A_{3.5} \oplus A_1$-invariant equations}
\begin{eqnarray*}
1) && \langle \partial_x, \partial_u, t \partial_t+x \partial_x +u \partial_u, t
\partial_u \rangle :\ \
 F = t^{-1} G(u_x); \\
2) && \langle x^{-1}\partial_u,  \partial_x -x^{-1} u \partial_u,t\partial_t+x
\partial_x, tx^{-1}\partial_u \rangle : \\
&& F = -2 x^{-2} u +2t^{-1}(u_x +x^{-1} u)\ln |t(u_x +x^{-1}u)| \\
&& +t^{-1} (u_x +x^{-1} u)  G(\omega), \ \ \omega = x u_x +u  .
\end{eqnarray*}

{$A_{3.6} \oplus A_1$-invariant equations}
\begin{eqnarray*}
1) && \langle \partial_x, t\partial_u, t \partial_t+x \partial_x, \partial_u
\rangle :\ \
 F = t^{-2} G(\omega), \ \omega = t^{-1} u_x; \\
2) && \langle \partial_t, \partial_x, e^{t} \partial_u,e^{-t}\partial_u \rangle
: \ \
 F = u+G(u_x).
\end{eqnarray*}

{$A_{3.7} \oplus A_1$-invariant equations}
\begin{eqnarray*}
1) && \langle \exp\left({-\frac{1}{2} (1-q) t}\right) \partial_u,
\exp\left({\frac{1}{2} (1-q)t}\right)\partial_u,  \partial_t+\frac{1}{2}(1+q) u\partial_u, \partial_x \rangle \\
&& \ (q\not =0,\pm 1): \ F = \frac{1}{4} (1-q)^2 u +
\exp\left({\frac{1}{2}(1+q) t}\right)
G(\omega), \\
&& \omega = \exp\left({-\frac{1}{2} (1+q) t}\right) u_x; \\
2) && \langle \partial_x, |t|^{\frac{1}{2} (1-q)} \partial_u,
|t|^{\frac{1}{2}(1+q)} \partial_u,
t\partial_t+x \partial_x +\frac{1}{2} (1+q) u \partial_u  \rangle \\
&& (q\not =0, \pm 1):  F = \frac{1}{4} (q^2-1) t^{-2} u +|t|^{\frac{1}{2}(q-3)}
G(\omega), \\
&& \omega = |t|^{\frac{1}{2}(1-q)} u_x;\\
3) && \langle |t|^{-\frac{1}{q}} |\xi|^{\frac{q+1}{2q}} \partial_u, \partial_x
-\frac{1+q}{2q} x^{-1} u \partial_u, -q(t \partial_t +x \partial_x),
|\xi|^{\frac{1+q}{2q}} \partial_u \rangle \  \\
&&(q\not =0, \pm 1): F = \left [ \frac{1-q^2}{4q^2} (t^{-2} +x^{-2})\right ]
u+\frac{1+q}{q} x^{-1} u_x \\
&& +t^{-2} |\xi| ^{\frac{1+q}{2q}} G(\omega) ,\ \ \xi = t x^{-1}, \ \
 \omega = |\xi|^{\frac{q-1}{2q}} \left [ x u_x +\frac{q+1}{2q} u \right ].
\end{eqnarray*}

{$A_{3.8} \oplus A_1$-invariant equations}
\begin{eqnarray*}
1) && \langle \sin t \partial_u, \cos t \partial_u,\partial_t, \partial_x
\rangle :
F =-u +  G(u_x).
\end{eqnarray*}

{$A_{3.9} \oplus A_1$-invariant equations}
\begin{eqnarray*}
1) && \langle \sin t \partial_u, \cos t \partial_u,\partial_t+qu \partial_u,
\partial_x \rangle (q>0):\\
&&  F =  -u +e^{qt} G(\omega), \ \ \omega = e^{-qt} u_x.
\end{eqnarray*}

{$A_{4.1}$-invariant equations}
\begin{eqnarray*}
1) && \langle \partial_u,-t \partial_u,\partial_x, \partial_t-tx \partial_u
\rangle :\ \
 F =  G(\omega), \ \ \omega = u_x +\frac{1}{2} t^2;\\
2) && \langle \partial_u,-t \partial_u,\alpha \partial_x+\frac{1}{2} t^2
\partial_u, \partial_t+kx \partial_x \rangle \ (k \geq 0, \alpha >0): \\
&& F = \alpha^{-1}(x-kt) + G(u_x).
 \end{eqnarray*}

{$A_{4.2}$-invariant equations}
\begin{eqnarray*}
1) && \langle |t|^{1-\frac{1}{2}q} \partial_u , |t|^{\frac{1}{2}q} \partial_u,
\partial_x, t \partial_t +x \partial_x +\left [ \left (1+\frac{1}{2} q \right) u
+x |t| ^{\frac{1}{2}q} \right ] \partial_u \rangle \\
&&  (q \not =0,1): F = \frac{1}{4} q (q-2) t^{-2} u +|t|^{\frac{1}{2}(q-3)}
G(\omega), \\
&&  \omega = |t|^{\frac{1}{2}(1-q)} u_x -2 |t|^{\frac{1}{2}}; \\
2) && \langle \partial_x, \sqrt{|t|} \partial_u, \sqrt{|t|} \ln |t| \partial_u,
t \partial_t +x \partial_x +\left(q+\frac{1}{2}  \right ) u \partial_u \rangle
\\
&&  (q\not =0): F = -\frac{1}{4} t^{-2} u +|t|^{q-\frac{3}{2}} G(\omega), \ \
 \omega =|t|^{\frac{1}{2}-q} u_x.
 \end{eqnarray*}

{$A_{4.3}$-invariant equations }
\begin{eqnarray*}
1) && \langle \partial_x, |t|^{\frac{1}{2}} \partial_u ,- |t|^{\frac{1}{2}}\ln
|t|  \partial_u,  t \partial_t +x \partial_x +\frac{1}{2}  u  \partial_u \rangle
:\\
&& F = -\frac{1}{4}  t^{-2} u +|t|^{-\frac{3}{2}} G(\omega), \ \ \omega =
|t|^{\frac{1}{2}} u_x ; \\
2) && \langle \partial_x,t \partial_u, \partial_u, t \partial_t +x \partial_x
\rangle :  \ \
 F = t^{-2} G(\omega), \ \omega =t u_x;\\
3) && \langle e^{kt} \partial_u, \partial_t +ku \partial_u, \beta \partial_x +t
e^{kt} \partial_u, e^{-kt} \partial_u \rangle \ (k \not =0, \ \beta >0): \\
&& F = k^2 u +2 k \beta^{-1} x e^{kt} +e^{kt}G(\omega)  , \ \ \omega = e^{-kt}
u_x;\\
4) && \langle e^{x+kt} \partial_u, e^{\eta} \partial_u, \alpha (\partial_x +u
\partial_u) +2k t e^{\eta} \partial_u, -\frac{1}{2k} (\partial_t +k \partial_x)
\rangle \\
&& (\alpha \not =0, \ k>0):\
 F = (k^2-1) u -4 k^2 \alpha^{-1} \eta e^{\eta} +e^{\eta} G(\omega), \\
&& \omega = e^{-\eta} (u_x -u),  \eta = x-kt.
\end{eqnarray*}

{$A_{4.4}$-invariant equations }
\begin{eqnarray*}
1) && \langle |t|^{\frac{1}{2}} \partial_u, -|t|^{\frac{1}{2}}\ln |t|
\partial_u, \partial_x, t \partial_t +x \partial_x +\left [\frac{3}{2} u -x
|t|^{\frac{1}{2}} \ln |t| \right ] \partial_u \rangle :\\
&& F = \frac{1}{4} t^{-2} u +|t|^{-\frac{1}{2}} G(\omega), \ \omega =
|t|^{-\frac{1}{2}} u_x +\frac{1}{2} \ln^2|t|.
\end{eqnarray*}

{$A_{4.5}$-invariant equations}
\begin{eqnarray*}
1) && \langle \partial_x, |t|^{m-\alpha} \partial_u,|t|^{1-m+\alpha}
\partial_u, t \partial_t +x \partial_x +mu  \partial_u \rangle\\
&& (m \not = \frac{1}{2}(1+\alpha), \frac{1}{2} +\alpha; \alpha \not =0): \\
&& F = (m-\alpha) (m-\alpha-1) t^{-2} u +|t|^{m-2} G(\omega), \ \omega =
|t|^{1-m} u_x .
\end{eqnarray*}

{$A_{4.6}$-invariant equations}
\begin{eqnarray*}
1) && \langle \partial_x, |t|^{\frac{1}{2}} \sin (q^{-1} \ln |t|) \partial_u,
|t|^{\frac{1}{2}} \cos (q^{-1} \ln |t|) \partial_u, qt \partial_t +qx \partial_x
\\
&& \left(\frac{1}{2} q+p \right) u \partial_u \rangle  \  (q\not =0, p \geq 0):
\\
&& F=-\left(\frac{1}{4} +q^{-2} \right) t^{-2} u+
|t|^{q^{-1}\left(p-\frac{3}{2}q \right)} G(\omega), \ \
\omega = |t|^{q^{-1}\left(\frac{1}{2}q-p \right)} u_x.
\end{eqnarray*}

{$A_{4.7}$-invariant equations}
\begin{eqnarray*}
1) && \langle \partial_u, -t \partial_u,\partial_t+k \partial_x, t \partial_t +x
\partial_x +\left(2u -\frac{1}{2}t^2 \right) \partial_u \rangle  \  (k \geq 0):
\\
&& F=-\ln |\eta| +G(\omega), \ \
\omega = \eta^{-1}u_x, \ \ \eta = x-kt.
\end{eqnarray*}

{$A_{4.8}$-invariant equations}
\begin{eqnarray*}
1) && \langle \partial_t +\epsilon u \partial_u,\partial_x,  e^x \partial_u, t
e^x \partial_u \rangle \ (\epsilon =0; 1): \\
&& F = -u +e^{\epsilon t} G(\omega), \ \ \omega = e^{-\epsilon t} (u_x -u); \\
2) && \langle |x|^{m-q} \partial_u, \partial_t, t |x|^{m-q} \partial_u, t
\partial_t+x\partial_x+mu\partial_u \rangle \ (q\not =0, \ m \in {\mathbb R}):
\\
&& F = -(m-q)(m-q-1) x^{-2} u +|x|^{m-2} G(\omega), \\
&& \omega =|x|^{1-m} [u_x-(m-q) x^{-1} u]; \\
\end{eqnarray*}
\begin{eqnarray*}
3) && \langle \partial_t +k \partial_x, \partial_u, t \partial_u,
t\partial_t+x\partial_x+qu\partial_u \rangle \ (k>0, q \in {\mathbb R}):\\
&& F = |\eta|^{q-2} G(\omega), \ \omega = |\eta|^{1-q} u_x, \ \ \eta = x-kt; \\
4) && \langle x^{-1} \partial_u, \partial_t +\partial_x-x^{-1} u \partial_u,
tx^{-1} \partial_u, t \partial_t +x \partial_x \rangle : \\
&& F = 2x^{-1} u_x +x^{-1} (t-x)^{-1} G(\omega), \ \ \omega =xu_x +u;\\
5) && \langle \partial_u, -t \partial_u, \partial_t+k \partial_x +u \partial_u,
\alpha \partial_x+u \partial_u \rangle \ (\alpha \not =0, k \geq 0): \\
&& F = \exp({\alpha^{-1} \eta +t}) G(\omega), \ \
\omega = \exp({-\alpha^{-1} \eta -t}) u_x, \ \eta = x-kt.
\end{eqnarray*}

{$A_{4.10}$-invariant equations}
\begin{eqnarray*}
1) && \langle \sin t \partial_u, \cos t \partial_u, \partial_x +u \partial_u,
\partial_t+k \partial_x \rangle (k \geq 0): \\
&& F = -u +e^{\eta} G(\omega), \ \ \omega = e^{-\eta} u_x, \ \eta = x-kt.
\end{eqnarray*}
 In the above formulas $G= G(\omega)$ is an arbitrary function satisfying the
condition $F_{u_x u_x} \not =0$.

\section{Symmetry reduction and solutions of nonlinear wave equations}
\setcounter{section}{7}

Among various applications of Lie symmetry groups the most prominent
and remarkable one is a possibility to construct exact solutions of
nonlinear PDEs. The basic idea is reducing multi-dimensional differential
equations to ordinary differential equations via special ansatzes
(invariant solutions). A regular (but not the only) way to derive those
ansatzes is to utilize symmetry group admitted by the equation under study
(for more details see, e.g., \cite{magda1, magda2}). Though the obtained
ordinary differential equations are, as a rule, nonlinear, they possess
in many cases a residual symmetry allowing for constructing their general
or particular solutions. Inserting the latter into the corresponding
ansatz yields the exact solution of initial nonlinear multi-dimensional
PDE. This method is often referred to in the literature as symmetry
reduction of PDEs.

We apply the symmetry reduction approach to derive the families of
exact solutions of nonlinear wave equations (\ref{1.9}) having the
richest symmetry properties.

To perform reduction of PDEs (\ref{1.9}) to ordinary differential
equations we need to obtain all inequivalent one-dimensional
subalgebras of the symmetry algebras of the equations under study.
What is more, basis operators of the said one-dimensional algebras
$$
\tau(t,x,u) \partial_t +\xi(t,x,u) \partial_x +\eta(t,x,u) \partial_u,
$$
have to obey the following restriction \cite{magda1}:
\begin{equation} \label{2.45}
|\tau| +|\xi| \not =0
\end{equation}
in some open domain $\Omega$ of the space $V = {\mathbb R}^2
\times {\mathbb R}^1$
of independent ${\mathbb R}^2 = \langle t,x \rangle$ and dependent
${\mathbb R}^1= \langle u \rangle$ variables.

As we proved in the previous sections, equations
\begin{eqnarray}
u_{tt} &=& u_{xx} -u^{-1} u^2_x; \label{2.46} \\
u_{tt} &=& u_{xx} +e^{u_x}; \label{2.47} \\
u_{tt}&=& u_{xx} +m \ln |u_x| \ (m >0); \label{2.48} \\
u_{tt} &=& u_{xx} +|u_x|^k \ (k \not =0, 1). \label{2.49}
\end{eqnarray}
enjoy the highest symmetry properties amongst PDEs of the form (\ref{1.9}).

The first step of symmetry reduction algorithm is classifying
one-dimensional subalgebras of the invariance algebras of the above equations
taking into account constraint (\ref{2.45}). To classify one-dimensional
subalgebras we utilize the method suggested in \cite{magda57} and the lists
of one-dimensional subalgebras of four-dimensional subalgebras obtained
in \cite{magda57}.

Equation (\ref{2.46}) admits the algebra
$$
A^1_5 = (A_{3.3} \oplus A_1) + \hskip -3.4mm \supset  \langle e_5 \rangle,
$$
where $A_{3.3} = \langle e_1, e_2, e_3 \rangle =
\langle u \partial_u, \partial_x, xu \partial_u \rangle,$ $
A_1 =\langle e_4  \rangle = \langle \partial_t \rangle,
e_5 =  t \partial_t +x \partial_x.$

In what follows we will need the commutation relations of the basis operators
of the algebra $A_{3.3} \oplus A_1$ with the operator $e_5$:
\begin{eqnarray*}
&& [e_1, e_5 ] =0, \ \ [e_2, e_5] = e_2, \ \
 [e_3, e_5 ] =-e_3, \ \ [e_4, e_5] = e_4.
\end{eqnarray*}

According to \cite{magda57} one-dimensional subalgebras of $A_{3.3}
\oplus A_1$ defined within action of inner auto\-mor\-phism group of
this algebra read as $\langle e_1 \rangle,$ $\langle e_1+\alpha e_4
\rangle, $ $\langle e_4 \rangle,$ $\langle e_2 \rangle,$$\langle
e_2+\alpha e_4 \rangle,$ $\langle e_3 \rangle,$ $\langle e_3+\alpha
e_4 \rangle,$ $\langle e_2+\beta e_3 \rangle,$ $\langle e_2+\beta
e_3 +\alpha e_4  \rangle \ (\alpha, \beta \not =0).$ The above
subalgebras can be further simplified by using transformation group
generated by the operator $e_5$. For example, using the
Campbell-Hausdorff formula we transform $e_1 +\alpha e_4$ as
follows:
$$ \exp (\theta e_5) (e_1 +\alpha e_4) \exp(-\theta e_5) = e_1 +\alpha
e^{\theta} e_4.$$
Consequently, putting $\theta = -\ln|\alpha|$ we simplify $e_1 +\alpha
e_4$ to become $e_1 \pm e_4$. Similarly, we prove that we can put $\alpha =\pm 1$
in $e_3 +\alpha e_4$ and $ \beta = \pm 1$ in $e_2 +\beta e_3$,
$e_3 +\beta e_3 +\alpha e_4$.

To complete classification of one-dimensional subalgebras we have to
describe all inequivalent subalgebras with non-zero projection on the
basis operator $e_5$, i.e., subalgebras of the form
\begin{equation} \label{2.50}
\Lambda = e_5 +\alpha_1 e_1 +\alpha_2 e_2 +\alpha_3 e_3 +\alpha_4 e_4, \ \
\alpha_1, \alpha_2, \alpha_3, \alpha_4 \in {\mathbb R}.
\end{equation}
Utilizing the automorphism $\exp(\theta e_4)$ with properly chosen $\theta$
we have $\alpha_4 =0$ in (\ref{2.50}). Next, applying transformation
$\exp(\theta_1 e_2 +\theta_2 e_3)$ to operator (\ref{2.50}) reduces it to one
of the following operators: $e_5, e_5 +\alpha e_1 \ (\alpha \not =0).$

So the list of one-dimensional subalgebras of the five-dimensional algebra
$A^1_5$ determined up to the action of inner automorphism group is exhausted by
the following algebras:
$\langle e_1 \rangle, $ $\langle e_1\pm e_4 \rangle, $ $\langle e_4 \rangle,$
$\langle e_2 \rangle,$ $\langle e_2 +\alpha e_4 \rangle,$ $\langle e_3 \rangle,$
$\langle e_3\pm  e_4 \rangle,$ $\langle e_2\pm  e_3 \rangle,$ $\langle e_2\pm
e_3 +\alpha e_4  \rangle, \langle e_5 \rangle, \langle e_5 +\alpha e_1 \rangle
\ (\alpha\not =0).$ By direct verification we prove that the basis operators
of the algebras $\langle e_1 \rangle, \langle e_3 \rangle$ do not satisfy
condition (\ref{2.45}).

Finally, we make use of the fact that the discrete groups of
transformations
\begin{eqnarray*}
&& \bar t = -t, \ \ \bar x = x, \ \ v=u;\\
&& \bar t = t, \ \ \bar x = -x, \ \ v=u;\\
&& \bar t = -t, \ \ \bar x = -x, \ \ v=u,
\end{eqnarray*}
also belong to the equivalence group of (\ref{2.46}). Using the
above transformations enables to further simplify the optimal system
of inequivalent subalgebras
\begin{eqnarray} \label{2.51}
&& \langle e_1+e_4 \rangle, \ \ \langle e_4 \rangle,\ \ \langle e_2 \rangle,\ \
\langle e_2+\alpha e_4 \rangle,\ \langle e_3+  e_4 \rangle,\\
&& \langle e_2\pm  e_3 \rangle,\ \ \langle e_2\pm  e_3 +\alpha e_4  \rangle,
\langle e_5 \rangle, \langle e_5 +\alpha e_1 \rangle  \ (\alpha> 0). \nonumber
\end{eqnarray}

The second step of the method of symmetry reduction is constructing the
complete set of invariants $f(t,x,u)$ for each inequivalent one-dimensional
subalgebra. As a typical example, we consider the case of subalgebra
$\langle e_1 +e_4 \rangle.$ To construct its invariants we need
to integrate the first-order PDE
$$(e_1 +e_4)\circ  F(t,x,u)=0$$
or
$$ u F_u+F_t =0.$$
The complete set of first integrals of the above equation
reads as $ \omega_1 =x, \ \ \omega_2 = e^{-t} u.$
Hence we get the general form of invariant solution (ansatz)
$ \omega_2 = \varphi(\omega_1).$
Solving this equation with respect to $u$ we finally have
\begin{equation} \label{2.52}
u = e^t \varphi(x).
\end{equation}

Inserting (\ref{2.52}) into (\ref{2.46}) yields ordinary differential equation
for unknown function $\varphi$
$$
\varphi'' -\varphi^{-1} (\varphi')^2-\varphi =0,
$$
which is easily integrated
$$
\varphi = \exp\left[ \frac{1}{2} x^2 +C_1 x +C_2 \right ], \ \ C_1, C_2 \in
{\mathbb R}.
$$
Inserting the so obtained expression for $\phi$ into ansatz (\ref{2.52})
yields the explicit form of invariant solution of equation (\ref{2.46})
$$
u = \exp\left [ t +\frac{1}{2} x^2 +C_1 x +C_2 \right], \ \ C_1, C_2 \in
{\mathbb R}.
$$

Analysis of the remaining subalgebras from (\ref{2.51}) yields the following results.
Invariant solutions of (\ref{2.46}) that correspond to the algebras $\langle e_4 \rangle,$
$\langle e_2 \rangle, $ $\langle e_2 +\alpha e_4 \rangle
(\alpha>0) $ read as
\begin{eqnarray*}
u&=& \exp[C_1 x+C_2], \\
u&=& C_1 t+C_2, \\
u&=& C_2 |t-\alpha x +C_1|^{1-\alpha^2},
\end{eqnarray*}
where $C_1, C_2 \in {\mathbb R}.$

The ansatz corresponding to the algebra $\langle e_3 +e_4 \rangle$ has the
form
$$
u = e^{tx} \varphi(x).
$$
It reduces (\ref{2.46}) to ordinary differential equation
$$
\varphi'' -\varphi^{-1}(\varphi')^2-x^2\varphi =0,
$$
whose general solution has the form
$$
\varphi = \exp \left[\frac{1}{12} x^4 +C_1 x+C_2 \right], \ \ C_1, C_2 \in
{\mathbb R}.
$$
Inserting the above expression into the corresponding ansatz
we get another family of exact solutions of nonlinear wave equation
(\ref{2.46})
$$
u = \exp \left[tx +\frac{1}{12} x^4 +C_1 x+C_2 \right], \ \ C_1, C_2 \in
{\mathbb R}.
$$
A similar analysis applied to the algebra $\langle e_2 +\epsilon e_3 \rangle \
(\epsilon = \pm 1)$ yields two families of exact solutions (\ref{2.46}):
\begin{eqnarray*}
u&=& C_1 \exp\left({-\frac{1}{2}x^2}\right) \cos(t+C_2), \ \ \epsilon =-1; \\
u&=& C_1 \exp\left({\frac{1}{2}x^2}\right)\cosh(t+C_2), \ \ \epsilon =1;
\end{eqnarray*}
where $C_1, C_2 \in {\mathbb R}.$

The ansatz invariant under the algebra $\langle e_2 +\epsilon e_3 +\alpha e_4 \rangle \
(\epsilon = \pm 1; \alpha>0)$ reads as
$$
u = \exp\left({\frac{1}{2} \epsilon x^2}\right)\varphi(\eta), \ \ \eta = t-\alpha x.
$$
It reduces PDE (\ref{2.46}) to equation
$$
(\alpha^2 -1) \varphi'' -\alpha^2 \varphi^{-1} (\varphi')^2 +\epsilon \varphi
=0,
$$
whose general solution is given by one of formulas below
\begin{eqnarray*}
u&=& \exp(C_1 \pm \eta), \ \ \alpha=1, \epsilon =1; \\
u&=& C_2\left[ \cos \left(C_1 +\frac{\eta}{\alpha^2-1}\right)
\right]^{1-\alpha^2}, \ \ \alpha>0, \ \alpha\not =1,\ \   \epsilon =-1; \\
u&=& C_2\left[ \cosh \left(C_1 +\frac{\eta}{1-\alpha^2}\right)
\right]^{1-\alpha^2}, \ \ \alpha>0, \ \alpha\not =1,\ \   \epsilon =1;
\end{eqnarray*}
where $C_1, C_2 \in {\mathbb R}.$ The corresponding invariant solutions
of (\ref{2.46}) have the form
\begin{eqnarray*}
u&=& \exp\left[C_1+\frac{1}{2} x \pm (t-x) \right]; \\
u&=& C_2\exp\left({-\frac{1}{2}x^2}\right) \left[ \cos \left(C_1 +\frac{t -\alpha
x}{\alpha^2-1}\right) \right]^{1-\alpha^2}, \ \ \alpha>0, \ \alpha\not =1; \\
u&=& C_2 \exp\left({-\frac{1}{2}x^2}\right) \left[ \cosh \left(C_1 +\frac{t-\alpha
x}{1-\alpha^2}\right) \right]^{1-\alpha^2}, \ \ \alpha>0, \ \alpha\not =1;
\end{eqnarray*}
where $C_1, C_2 \in {\mathbb R}.$

The algebras $\langle e_5 \rangle, \langle e_5 +\alpha e_1 \rangle \
(\alpha>0)$ give rise to the so called auto-model (\cite{magda1})
solutions of (\ref{2.46}). Solution invariant under $\langle e_5 \rangle$
reads as
\begin{eqnarray*}
u &= &\exp \left \{ \int^{\xi} \left( C_1 (\eta^2-1) +\frac{1}{4} (\eta^2-1) \ln
\left| \frac{1+\eta}{1-\eta}\right| -\frac{1}{2} \eta \right]^{-1} d \eta+C_2
\right \}, \\
&&  \xi= tx^{-1}, \ \ C_1, C_2 \in {\mathbb R}.
\end{eqnarray*}

Solution invariant under the algebra $\langle e_5+\alpha e_1 \rangle \
(\alpha>0)$ has the form
$$
u = |t|^{\alpha} \varphi(\xi), \ \ \xi = tx^{-1}.
$$
Inserting this expression into (\ref{2.46}) yields ordinary
differential equation
$$
\xi^2(\xi^2-1) \varphi'' -\xi^4 \varphi^{-1} (\varphi')^2 +2 \xi (\xi^2
-\alpha) \varphi' -\alpha (\alpha-1) \varphi =0.
$$
If $\alpha =1,$ then the general solution of the above equation
is of the form
\begin{eqnarray*}
 \varphi& =& \exp \left \{ \int^{\xi} \left[C_1 \eta^2 +\frac{1}{2} \eta \ln
\left|\frac{1+\eta}{1-\eta}\right| -\eta\right ]^{-1} d \eta +C_1 \right\}, \\
&&  \xi = tx^{-1}, \ \ C_1, C_2 \in {\mathbb R}.
\end{eqnarray*}
Provided $\alpha>0, \alpha \not =1,$ the change of variables
$$
\varphi = \exp f, \ \ f = f(\xi)
$$
reduces the equation for $\phi$ to become
$$
\xi^2(\xi^2-1) f'' -\xi^2(f')^2 +2 \xi(\xi^2-\alpha) f' -\alpha(\alpha-1)
=0.
$$
Now making the change of variables, $f'=g, g=g(\xi)$ we arrive at the
Riccatti equation
\begin{equation} \label{2.53}
\xi^2(\xi^2-1)g' = \xi^2 g^2-2\xi(\xi^2-\alpha) g +\alpha(\alpha-1).
\end{equation}

In view of the above we have the two families of exact solutions of nonlinear
wave equation (\ref{2.46}):
\begin{eqnarray*}
u&=&  t\exp   \left \{ \int^{\xi} \left[C_1 \eta^2 +\frac{1}{2} \eta \ln
\left|\frac{1+\eta}{1-\eta}\right| -\eta\right ]^{-1} d \eta +C_2 \right\}, \\
&&  \xi = tx^{-1}, \ \ C_1, C_2 \in {\mathbb R}; \\
u&=&  |t|^{\alpha} \exp   \left [ \int^{\xi} g(\eta) d \eta +C_1 \right],
\end{eqnarray*}
where $\alpha>0, \ \alpha\not=1, \ \ \xi = tx^{-1},  C_1 \in {\mathbb R},$
and the function $g= g(\xi)$ is a solution of (\ref{2.53}).

Now we turn to nonlinear wave equation (\ref{2.47}). Its maximal symmetry
algebra is $A^2_5 =(A_{3.3} \oplus A_1 \rangle +\hskip -4.1mm \supset
\langle e_5 \rangle$, where $A_{3.3} = \langle e_1, e_2, e_3 \rangle=
\langle \partial_u, \partial_t, t \partial_u \rangle,$ $A_1 =
\langle e_4 \rangle = \langle \partial_x \rangle,$ $ e_5 = t \partial_t
+x\partial_x +(u-x) \partial_u.$ What is more, the basis operators of
the algebra $A_{3.3} \oplus A_1$ obey the following commutation
relations with $e_5$:
$$
[e_1, e_5] = e_1, \ \ [e_2, e_5] = e_2, \ \ [e_3, e_5] =0, \ \
[e_4, e_5] = e_4 -e_1.
$$
Using these and taking into account condition (\ref{2.45}) we obtain the
following system of optimal one-dimensional subalgebras of the algebra
$A^2_5$:
\begin{eqnarray*}
&& \langle e_4 \rangle, \ \langle e_2 \rangle, \ \langle e_2+\alpha e_4 \rangle,
\ \langle e_3 +\alpha e_4 \rangle, \langle e_2 \pm e_3 \rangle,\\
&&  \langle e_2 +\beta e_3 +\alpha e_4 \rangle, \ \langle e_5 \rangle, \
\langle e_5+\alpha e_3  \rangle \ (\alpha>0, \beta \not =0).
\end{eqnarray*}

Skipping the intermediate computations we give the final result,
the families of exact solutions of nonlinear equation (\ref{2.47}):
\begin{eqnarray*}
&& \langle e_4 \rangle : u = C_1 t+C_2, \ \ C_1, C_2 \in {\mathbb  R}; \\
&& \langle e_2 \rangle : u =(x+ C_1)[1-\ln|x+C_1|]+C_2, \ \ C_1, C_2 \in
{\mathbb  R}; \\
&& \langle e_2 +\alpha e_4 \rangle : u =(x-\alpha t) \ln|1-\alpha^2|\\
&& -(x-\alpha t+C_1)[\ln|x-\alpha t+C_1|-1]+C_2, \\
&& \ \ \ \ \alpha>0, \alpha\not =1, \ \  C_1, C_2 \in {\mathbb  R}; \\
&& \langle e_3 +\alpha e_4 \rangle : u =\alpha^{-1} tx+\alpha^2
\exp({\alpha^{-1}t}) +C_1 t+C_2, \ \ \alpha>0,  C_1, C_2 \in {\mathbb  R};
\end{eqnarray*}
\begin{eqnarray*}
&& \langle e_2 +\epsilon e_3 \rangle : u =\frac{1}{2} \epsilon t^2 +\varphi(x),
\ \ \epsilon=\pm1, \ \varphi' = y(x), \\
&& y-\ln|\epsilon -e^{y}|= \epsilon x+C_1, \ \ C_1 \in {\mathbb R};\\
&& \langle e_2 +\beta e_3 +e_4 \rangle : u = \frac{1}{2}\beta t^2+(x-t) \ln
|\beta| +C_1, \ \ \beta \not =0, \ \ C_1 \in {\mathbb R};\\
&& \langle e_2 +\beta e_3+\alpha e_4  \rangle : u =\frac{1}{2} \beta t^2
+\varphi(\eta), \ \ \eta = x-\alpha t, \ \ \varphi = y'(\eta), \\
&& y-\ln|\beta-e^y|=\beta(1-\alpha^2)^{-1} \eta +C_1, \ \ \alpha>0, \alpha
\not=1, \beta \not =0, C_1 \in {\mathbb  R};\\
&& \langle e_5 \rangle : u =-x\ln|x| +C_1 t+x, \ \ \ C_1 \in {\mathbb R};\\
&& \langle e_5 +\alpha e_3 \rangle : u =(\alpha t-x)\ln|x| +x \varphi(\xi), \ \
\xi = tx^{-1}, \ \alpha>0, \\
&& (\xi^2-1) \varphi''+\exp[-\xi \varphi'+ \varphi+\alpha \xi -1] =1+\alpha \xi.
\end{eqnarray*}

Equation (\ref{2.48}) admits the algebra $A^3_5 = (A_{3.5} \oplus
A_1) + \hskip -3.8mm \supset \langle e_5 \rangle,$ and $A_{3.5} =
\langle e_1, e_2, e_3 \rangle = \langle \partial_u, \partial_t,
t\partial_u \rangle,$ $A_1 = \langle e_4 \rangle = \langle
\partial_x \rangle,$ $ e_5 = t \partial_t+x
\partial_x+\left (2u+\frac{1}{2} mt^2 \right ) \partial_u.$ Its optimal
system of one-dimensional subalgebras has the form
$$
\langle e_1+e_4 \rangle, \ \langle e_4 \rangle, \ \langle e_2 \rangle, \ \langle
e_2 +\alpha e_4 \rangle, \langle e_3 +\alpha e_4 \rangle, \langle e_5 \rangle \
(\alpha>0).
$$
Below we give the list of exact solutions of (\ref{2.48}) invariant
under the above subalgebras.
\begin{eqnarray*}
&& \langle e_1 +e_4 \rangle : u = x+C_1 t +C_2, C_1, C_2 \in {\mathbb R} ;\\
&& \langle e_4 \rangle : u = C_1 t +C_2, C_1, C_2 \in {\mathbb R} ;\\
&& \langle e_2 \rangle : u = \varphi(x), \ \ \varphi'= f(x), \ \int
\frac{df}{\ln |f|}=-mx +C_1, \ C_1 \in {\mathbb R};\\
&& \langle e_2+\alpha e_4  \rangle : u = x- t +C_1, C_1\in {\mathbb R}, {\rm \
if \ } \alpha=1;\\
&& u = \varphi (\eta), \ \eta = x-\alpha t, \ \varphi' = f(\eta), \\
&&  \int \frac{df}{\ln |f|} = -\frac{m}{1-\alpha^2} (x-\alpha t) +C_1, \ C_1 \in
{\mathbb R}, {\rm \ if \ } \alpha>0, \alpha \not =1;\\
&& \langle e_3+\alpha e_4  \rangle : u = \alpha^{-1} tx +\frac{1}{2} m t^2
(1+\ln \alpha) +\frac{1}{2} mt^2 \left (\ln |t| -\frac{1}{2} \right) \\
&& + C_1 t+C_2, \alpha>0,  C_1, C_2 \in {\mathbb R}.\\
&& \langle e_5 \rangle : u =\frac{1}{2} m^2 \ln |t| +t^2 \varphi, \
{\rm where}\ \varphi =
\varphi(\xi) \ (\xi = tx^{-1}) {\rm \  satisfy \ equation \ } \\
&& \xi^2 (\xi^2 -1) \varphi'' +2 \xi(\xi^2-2) \varphi' -2 \varphi +m \ln |\xi^2
\varphi'| -\frac{3}{2}m =0.
\end{eqnarray*}

Note that the particular solution of reduced equation $\varphi = -\xi^{-1}
\exp({\frac{3}{2}})$ gives rise to the following invariant solution of
(\ref{2.48}):
$$
u = \frac{1}{2} mt^2 \ln |t| -tx \exp({\frac{3}{2}}).
$$

Equation (\ref{2.49}) is invariant under the algebra $A^4_5 =
(A_{3.5} \oplus A_1) +\hskip -4.8mm \supset  \langle e_5 \rangle,$
and $A_{3.5} = \langle e_1, e_2, e_3 \rangle = \langle
\partial_u,\partial_t, t \partial_u \rangle, $ $A_1 = \langle e_4
\rangle = \langle \partial_x \rangle,$ $e_5 = t \partial_t +x
\partial_x +\frac{k-2}{k-1} u \partial_u, \ k\not =0;1.$ The optimal system of
one-dimensional subalgebras of this algebra reads as $\langle e_5 \rangle,
\langle e_1 +e_4 \rangle, \ \langle e_4 \rangle, \langle
e_2 \rangle, \langle e_2 +\alpha e_4 \rangle,$ $\langle e_3 +e_4 \rangle,
$$\langle e_2 \pm e_3 \rangle, $ $\langle e_2 \pm e_3 +\alpha e_4\rangle \
(\alpha>0)$. If $k=2,$ then the above list should include also the
algebra $\langle e_5 +\alpha e_1\rangle\ (\alpha>0)$.

Below we give exact solutions of nonlinear wave equation (\ref{2.49}) invariant
under the above algebras.
\begin{eqnarray*}
&& \langle e_1+e_4 \rangle: u = \frac{1}{2} t^2 +x +C_1 t +C_2, \ \
C_1, C_2 \in
{\mathbb R};\\
&& \langle e_4 \rangle : u = C_1 t +C_2, \ \ C_1, C_2 \in {\mathbb R}; \\
&& \langle e_2 \rangle : u = (2-k)^{-1} (C_1 +(k-1)x)^{\frac{2-k}{1-k}} +C_2,
C_1, C_2 \in {\mathbb R},\\
&&  {\rm \ if \ } k \not =2; \\
&& \ \ \ \ u = C_2 \ln |x-C_1|, \ C_1, C_2 \in {\mathbb R}, {\rm \ if \ } k
=2; \\
&& \langle e_2+\alpha e_4 \rangle: u =C, \ \ C \in {\mathbb R}, {\rm \ if \ }
\alpha  =1; \\
&& \ \ \ \ u = (1-\alpha^2) \ln |C_1 -x +\alpha t| +C_2, \ C_1, C_2
\in {\mathbb R}, \\
&&  {\rm \ if \ } k =2, \alpha>0, \alpha \not =1;\\
&& \ \ \ \ u = \frac{1-k}{2-k}
\left(\frac{1-k}{\alpha^2-1}\right)^{\frac{1}{1-k}} \left| x-\alpha t +C_1
\right |^{\frac{2-k}{1-k}} +C_2,  \ C_1, C_2 \in {\mathbb R}, \\
&&  {\rm \ if \ } k \not =0,1,2; \alpha>0, \alpha \not =1;\\
&& \langle e_3+e_4 \rangle: u =tx +t (\ln |t| -1) +C_1 t +C_2,\
C_1, C_2  \in {\mathbb R}, \\
&& {\rm \ if \ } k  =-1; \\
&& \ \ \ \ u = tx -\ln |t| +C_1t +C_2,  \ C_1, C_2  \in {\mathbb R},
{\rm \ if \ } k  =-2; \\
&& \ \ \ \ u = tx +(k^2+3k+2)^{-1} |t|^{k+2}  +C_1t +C_2,  \ C_1, C_2  \in
{\mathbb R},\\
&&  {\rm \ if \ } k  \not =0,1,-1,-2;\\
&& \langle e_2+\epsilon e_3 \rangle: u =\frac{1}{2} \epsilon t^2 \varphi, \
\varphi = \varphi(x)=\int^x f(\eta) d \eta +C_2,\\
&&  {\rm \ where \ } f  {\rm \ is\ defined \ by \ }\\
&& \int \frac{df}{\epsilon -|f|^k} = \eta +C_1,   C_1, C_2  \in {\mathbb R}, \
\epsilon = \pm 1;\\
&& \langle e_2+\epsilon e_3+\alpha e_4  \rangle: u =\frac{1}{2} \epsilon t^2
+\varphi, \ \varphi = \varphi(\eta) = \int^{\eta} f(z) dz +C_2, \\
&& \eta = x-\alpha t, {\rm \ where \ } f  {\rm \ is \ defined \ by \ }\\
&& \int \frac{df}{\epsilon -|f|^k} = (1-\alpha^2)^{-1} z +C_1, C_1, C_2 \in
{\mathbb R}, \ \epsilon = \pm 1, \ \alpha>0;\\
&& \langle e_5 \rangle : u = |t|^{\frac{k-2}{k-1}} \varphi(\xi), \ \xi =
tx^{-1}, {\rm \ where \ } \varphi {\rm \ is \ defined \ by \ }\\
&& \xi^2(\xi^2 -1) \varphi '' +2 \xi\left (\xi^2-\frac{k-2}{k-1}\right) \varphi'
+(-1)^k \xi^{2k} |\varphi'|^k +\\
&& +\frac{k-2}{(k-1)^2} \varphi =0, \ \  {\rm \ if \ } k \not =0,1,2;\\
&& u = \int^{\xi} \left [C_1 (1-\eta^2) +\frac{1}{4} (1-\eta^2) \ln \left
|\frac{1+\eta}{1-\eta}\right| -\frac{1}{2} \eta \right]^{-1} d \eta +C_2, \\
&&  C_1, C_2 \in {\mathbb R}, \xi = tx^{-1}, {\rm \ if \ } k=2; \\
&& \langle e_5+\alpha e_1  \rangle : u =\alpha \ln |t| +\varphi, \ \varphi =
\varphi(\xi)=\int^{\xi} f(\eta) d \eta +C, \\
&&  \xi = tx^{-1}, C \in {\mathbb R}, \ f =f(\eta) {\rm \ is \ a
\  solution \ of \ Riccatti \ equation} \\
&& \eta^2 (\eta^2-1) f' +\eta^4 f^2+ 2 \eta^3 f +\alpha =0, \alpha>0, {\rm \
if \ } k=2.
\end{eqnarray*}

\newpage
\section{Concluding remarks}
\setcounter{section}{8}

Let us briefly summarize the results obtained in this paper.

We prove that the problem of group classification of the general
quasi-linear hyperbolic type equation (\ref{1.1}) reduces to
classifying equations of more specific forms
\begin{eqnarray*}
{\rm I.} && u_{tt} = u_{xx} + F(t,x,u,u_x), \ \ F_{u_x u_x} \not =0; \\
{\rm II.} && u_{tt} = u_{xx} + g(t,x,u) u_x +f(t,x,u), \ \ g_u \not =0; \\
{\rm III.} && u_{tx} = g(t,x) u_x +f(t,x,u), \ g_x \not =0, \ f_{uu} \not =0; \\
{\rm IV.} && u_{tx} = f(t,x,u), \ \ f_{uu} \not =0.
\end{eqnarray*}
The cases of PDEs that are essentially nonlinear in $u_x$ (the class
of PDEs I) and either linear in $u_x$ or do not depend on $u_x$ (the
classes II - IV) need to be considered separately.

If we denote as ${\cal DE}$ the set of PDEs II -- III, then the results
of application of our algorithm for group classification  of
equations I -- IV can be summarized as follows.
\begin{enumerate}
\renewcommand{\labelenumi}{\arabic{enumi})}
\renewcommand{\theenumi}{\arabic{enumi}}
\item We perform complete group classification of the class ${\cal DE}$.
We prove that the Liouville equation has the highest symmetry properties
among equations from ${\cal DE}$. Next, we prove that the only
equation belonging to this class and admitting the four-dimensional
invariance algebra is the nonlinear d'Alembert equations. It is
established that there are twelve inequivalent equations from ${\cal DE}$
invariant under three-dimensional Lie algebras. We give the lists of
all inequivalent equations from ${\cal DE}$ that admit one- and
two-dimensional symmetry algebras.
\item We have studied the structure of invariance algebras admitted
by nonlinear equations from the class I. It is proved, in particular,
that their invariance algebras are necessarily solvable.
\item We perform complete group classification of nonlinear equations
from the class of PDEs I. We prove that the highest symmetry algebras
admitted by those equations are five-dimensional and construct all
inequivalent classes of equations invariant with respect to five-dimensional
Lie algebras. We also construct all inequivalent equations of the form
I admitting one-, two-, three- and four-dimensional Lie algebras.
\end{enumerate}

The results of group classification of the class of nonlinear
wave equations (\ref{1.1}) are utilized for constructing their
explicit solutions. Namely, we perform symmetry reduction of all
equations (\ref{1.1}) admitting five-dimensional symmetry algebras
to ordinary differential equations and constructed multi-parameter
families of their exact solutions.

\end{document}